\documentclass[12pt]{article}
\usepackage{amsmath}
\usepackage{amssymb} % Automatically loads amsfonts
\usepackage{amsthm}  % Must come after amsmath
\usepackage{dsfont,natbib,textcomp}

\usepackage[utf8]{inputenc}
\usepackage[colorlinks,citecolor=blue,linkcolor=black,urlcolor=blue]{hyperref}
\usepackage{url}
\urldef\finraurl\url{https://www.finra.org/sites/default/files/2019%20Industry%20Snapshot.pdf}
\urldef\bankrupt\url{https://abi-org.s3.amazonaws.com/Newsroom/Bankruptcy_Statistics/Quarterlynonbusinessfilingsbychapter1994-Present.pdf}

\usepackage{color}

\usepackage[top=3cm, left=3cm, bottom=3cm, right=3cm]{geometry}
\usepackage{setspace}
\onehalfspacing

\usepackage{graphicx}
\usepackage{bm}
\usepackage{multirow}
\usepackage{dcolumn}
\usepackage{appendix}

\usepackage[section]{placeins}

\usepackage{subcaption}

\usepackage[normalem]{ulem}

\usepackage{longtable}
\usepackage{array}

\newcolumntype{L}[1]{>{\raggedright\arraybackslash}p{#1}}

\usepackage[multiple]{footmisc}
\usepackage{caption}
\usepackage[figuresleft]{rotating}

\usepackage{tabularx}   % For adjustable-width columns (text wrapping)

\usepackage{booktabs}% Pretty tables
\usepackage{threeparttablex}% For Notes below table

\usepackage{pdflscape}  % Rotates the page in the PDF viewer (easier to read)
\newcommand{\sym}[1]{\rlap{#1}} % Thanks to Joseph Wright & David Carlisle

\usepackage{siunitx}
\sisetup{
detect-mode,
group-digits            = false,
input-symbols           = ( ) [ ] - +,
table-align-text-post   = false,
input-signs             = ,
}   

\def\yyy{%
\bgroup\uccode`\~\expandafter`\string-%
\uppercase{\egroup\edef~{\noexpand\text{\llap{\textendash}\relax}}}%
\mathcode\expandafter`\string-"8000 }

\def\xxxl#1{%
\bgroup\uccode`\~\expandafter`\string#1%
\uppercase{\egroup\edef~{\noexpand\text{\noexpand\llap{\string#1}}}}%
\mathcode\expandafter`\string#1"8000 }

\def\xxxr#1{%
\bgroup\uccode`\~\expandafter`\string#1%
\uppercase{\egroup\edef~{\noexpand\text{\noexpand\rlap{\string#1}}}}%
\mathcode\expandafter`\string#1"8000 }

\makeatletter
\let\estinput\@@input
\makeatother

\pagenumbering{arabic}

\def\sym#1{\ifmmode^{#1}\else\(^{#1}\)\fi}
\usepackage{authblk}

\title{%
Hiding in Good Times, Caught in Bad: Strategic Masking and the Delayed Detection of Financial Adviser Misconduct%
\footnote{%
I would like to thank 
Naoyuki Yoshino
and
Orly Sade 
for valuable comments. 
I also thank for 
seminar participants 
at 
at the 26th Annual Conference of the Japan Academic Society for Financial Planning
for their comments. %
This work is supported by JSPS KAKENHI Grant Number JP24K04832 and the research fund provided by the Japan Association for Financial Planners.
}
}

\author{%
Jun Honda%
\thanks{%
Faculty of Economics and Law,
Shinshu University, Japan. %
junhonda@shinshu-u.ac.jp. %
}%
}

\date{%
March, 2026%
}
\begin{document}
%----------------------------------------------------------------

%----------------------------------------------------------------
% Title Page:
%----------------------------------------------------------------

%----------------------------------------------------------------

%------------------------------------------------------------------------------------
\maketitle
%------------------------------------------------------------------------------------

%------------------------------------------------------------------------------------
\thispagestyle{empty}
%------------------------------------------------------------------------------------

%------------------------------------------------------------------------------------
\begin{abstract}
While financial misconduct in advisory services persists despite regulation, the demand-side of market discipline, specifically the timing of investor detection, 
% remains largely unexplored.
remains a critical bottleneck.
%------------------------------------------------------------------------------------
Using approximately 55,700 FINRA BrokerCheck records, we analyze the detection lag between misconduct inception and formal reporting. We document a conditional average lag of 28.5 months, with a fat-tail exceeding eight years. Overt unauthorized activity reduces the lag by 35.7\%, whereas sophisticated fraud extends it by 58.6\%. Using method of moments quantile regression, we reveal a strategic masking gradient: the impact of advisor experience more than doubles at the 90th percentile relative to the 10th.
Product opacity acts as an expanding shield:
insurance-linked disputes extend latency by 
% 88.9\% for the most persistent cases. 
% Disputes involving insurance extend the detection lag by 
% 61.9\% at the lowest decile while by 90.0\% in the highest decile.
by 61.9\% at the 10th percentile and by 90.0\% at the 90th percentile of the distribution.
% Finally, market volatility serves as an asymmetric catalyst. 
% A doubling of the VIX reduces the lag by 21.1\% for rapid-discovery cases but only 7.9\% for deeply concealed schemes. 
Finally, market volatility serves as an asymmetric catalyst for discovery: a doubling of the VIX reduces the lag by 21.1\% for rapid-discovery cases, but only 7.9\% for deeply concealed schemes.
These strategically managed discovery delays allow bad types to persist across multiple market cycles.
%------------------------------------------------------------------------------------

%------------------------------------------------------------------------------------

%------------------------------------------------------------------------------------
%------------------------------------------------------------------------------------
\bigskip
%------------------------------------------------------------------------------------

%------------------------------------------------------------------------------------
\noindent
%------------------------------------------------------------------------------------
%------------------------------------------------------------------------------------
\emph{Key Words:} 
%------------------------------------------------------------------------------------
% Financial Advisers, Misconduct, Gender Gap, Racial Inequality, Discrimination.
% Financial misconduct, 
% % Detection Lag, 
% Investor Response,
% % Retail Investors, 
% Financial Advice, 
% Information Asymmetry,
% Market Discipline.
%------------------------------------------------------------------------------------
% Financial misconduct; Investor response; Financial advice; Information asymmetry; Behavioral biases; Market discipline.
% Adviser Misconduct;
% Blame Attribution;
% Delegation;
% Investor Response;
% Market Downturns.
Adviser Misconduct; Detection Lags; Market Downturns; Ostrich Effect; Blame Attribution; Principal-Agent Monitoring; Investor Attention.
%------------------------------------------------------------------------------------

%------------------------------------------------------------------------------------
\noindent
%------------------------------------------------------------------------------------
\emph{JEL Classification:} 
%------------------------------------------------------------------------------------
%D14 (Household Saving   Personal Finance), 
%D18 (Consumer Protection), 
G18, %– Government Policy and Regulation:
% G24, 	% (Investment Banking  Venture Capital   Brokerage   Ratings and Ratings Agencies), 
%G28 (Government Policy and Regulation), 
G30, % – General Financial Markets:
G41, %: Role and Effects of Psychological, Emotional, Social, and Cognitive Factors on Decision Making in Financial Markets.
% J44, % (Professional Labor Markets   Occupational Licensing), 
%J6 (Mobility, Unemployment, Vacancies, and Immigrant Workers)
%J60 (General), 
%J63 (Turnover  Vacancies Layoffs)
%J7	Labor Discrimination
% J71,	%(Discrimination)
%K3	Other Substantive Areas of Law
%K31 (Labor Law), 
%L2	Firm Objectives, Organization, and Behavior
% L22. %(Firm Organization and Market Structure).	
D83. % – Search; Learning; Information and Knowledge; Communication; Belief:
%------------------------------------------------------------------------------------
%------------------------------------------------------------------------------------

%------------------------------------------------------------------------------------
\end{abstract}
%------------------------------------------------------------------------------------

%----------------------------------------------------------------
\newpage
%----------------------------------------------------------------

%----------------------------------------------------------------
\tableofcontents %put toc in
%----------------------------------------------------------------

%----------------------------------------------------------------
\cleardoublepage %start new page
%----------------------------------------------------------------

%\listoffigures
%
%\listoftables

%----------------------------------------------------------------
% \cleardoublepage %start new page
\pagestyle{plain} % put headers/footers back on
\setcounter{page}{1} %reset the page counter
%----------------------------------------------------------------

%----------------------------------------------------------------
\newpage
%----------------------------------------------------------------

%----------------------------------------------------------------

%----------------------------------------------------------------
\newpage
%----------------------------------------------------------------

%----------------------------------------------------------------
% Introduction:
%----------------------------------------------------------------

%----------------------------------------------------------------
%----------------------------------------------------------------
\section{Introduction}
\label{section: intro 0}
%----------------------------------------------------------------

%----------------------------------------------------------------

%----------------------------------------------------------------

%----------------------------------------------------------------
Financial misconduct in advisory services persists despite extensive regulatory interventions, raising fundamental questions about the efficacy of market discipline in retail finance. While existing research primarily focuses on the supply side represented by the incentives and characteristics of advisors, the demand side of the market remains largely overlooked. 
Because regulatory sanctions 
can only occur once misconduct is identified, 
% are contingent upon formal discovery by the principal
the timing of detection by the principal (the investor) represents a critical bottleneck in the agency relationship. 
%----------------------------------------------------------------

%----------------------------------------------------------------
In this paper, we shift the focus to investor monitoring behavior to investigate 
% the structural and environmental drivers of this oversight. 
the institutional, advisor-specific, and macro-financial determinants of this oversight.
Using detailed FINRA BrokerCheck records to construct a novel measure of detection lags defined as the duration between the initiation of misconduct and its eventual reporting, we pursue three primary objectives: (i) to quantify the systemic persistence of the detection gap; (ii) to determine how advisor sophistication and misconduct complexity interact to increase monitoring frictions; and (iii) to evaluate whether macro-financial shocks serve as an exogenous catalyst for state verification.
%----------------------------------------------------------------

%----------------------------------------------------------------
Our investigation builds on the premise that financial advice is a classic credence good \citep{darby1973free}, where ex-post performance often fails to reveal whether outcomes stem from advisor misconduct or exogenous market fluctuations. In such an environment characterized by costly state verification, investors may employ rational inattention or strategic delegation that inadvertently delays detection. We hypothesize that these delays fluctuate systematically based on the nature of the misconduct and the strategic human capital of the advisor. Specifically, we predict a complexity-induced detection gap where sophisticated schemes extend the lag, and a strategic masking mechanism where veteran offenders leverage institutional knowledge to exploit information asymmetries. Finally, we hypothesize a state-contingent monitoring effect where market downturns lower the relative cost of monitoring and pierce periods of inattention.
%----------------------------------------------------------------

% \input{1_introduction/1_2_o}

%----------------------------------------------------------------
To evaluate these dynamics, we utilize a comprehensive dataset of approximately 55,700 BrokerCheck records spanning from 2008 to 2025. Our empirical strategy focuses on the heterogeneous effects across the conditional distribution of detection lags, rather than relying solely on average effects. A central challenge in identifying investor monitoring intensity is the presence of unobserved institutional and regional factors. To isolate the specific interaction between investors and advisors, we saturate our models with firm fixed effects, which control for unobserved compliance cultures and internal monitoring systems. We further include state fixed effects to account for regional regulatory environments and local investor demographics, along with year fixed effects to control for time-varying macro factors. This estimation strategy ensures that our findings are derived from within-firm variation, though we acknowledge that the endogenous matching between advisors and clients remains an unobserved source of selection..
%----------------------------------------------------------------

%----------------------------------------------------------------
Our first set of findings establishes that 
investor detection within the caught population is profoundly and persistently delayed. Controlling for firm, state, and year fixed effects, we document an average detection lag of 28.5 months, with a significant fat-tail where the top decile of cases remains latent for over eight years. 
%----------------------------------------------------------------
This delay is significantly shaped by 
%----------------------------------------------------------------
% Our first set of results establishes that the timing of discovery is 
a strategic outcome dictated by the interaction of information frictions and advisor human capital. We document a profound complexity-induced detection gap: while overt violations such as unauthorized activity provide salient signals that reduce the detection lag by 35.7\%,
% ($e^{-0.442} - 1$), 
sophisticated quiet crimes like fraud extend the window of nondisclosure by 58.6\%.
% ($e^{0.461} - 1$). 
Relative to the sample mean of 28.5 months, these coefficients suggest that complexity acts as the dominant information friction, allowing sophisticated schemes to persist for nearly 1.6 times longer than the average infraction. This provides empirical support for a credence good framework where investors struggle to monitor agents effectively when the costs of state verification are high.

We further identify a strategic masking mechanism where veteran advisors and complex products serve as informational shields. Our estimates show that a 1\% increase in advisor experience is associated with a 0.37\% extension of the detection window, suggesting that learned-by-doing in concealment is a primary determinant of latency. This effect is compounded by product opacity. 
Allegations involving insurance-linked instruments exhibit the largest impact on latency, extending the lag by 76.5\% 
% ($e^{0.568} - 1$) 
relative to the sample average. These results indicate that high-human-capital agents leverage the shrouded attributes of complex contracts to navigate internal compliance triggers and pacify investor monitoring.

Finally, our first set of findings challenge the standard monitoring framework, which predicts that higher financial stakes should trigger faster discovery. Instead, we document a scale-complexity correlation: we find a positive elasticity of 0.055 for alleged damages, implying that a doubling of the financial harm is associated with a 3.9\% increase 
% ($2^{0.055} - 1$) 
in the period of nondisclosure. This suggests that the increased incentive for investor oversight in high-value disputes is more than offset by the more elaborate masking strategies required to execute larger schemes.

%----------------------------------------------------------------
% Second, we identify a non-linear relationship between advisor human capital and concealment. 
Beyond conditional averages, we document a profound non-linearity in the strategic behavior of advisors. 
Utilizing the method of moments quantile regression framework of \citet{machado2019quantiles} to handle high-dimensional fixed effects, we find 
% that the effect of advisor experience increases monotonically across the distribution. For instance, a 1\% increase in career longevity is associated with only a 0.12\% increase in the duration of latent misconduct at the 10th percentile. 
strategic masking gradient where the impact of advisor human capital increases monotonically across the lag distribution. For cases prone to rapid discovery (10th percentile), a 1\% increase in experience extends the detection window by only 0.22\%.
% However, this impact quadruples to a 0.45\% extension of the detection lag at the 90th percentile. This suggests a strategic masking mechanism where veteran offenders effectively hide in the tail of the distribution by leveraging institutional knowledge to navigate internal compliance triggers. 
However, this impact more than doubles to 0.49\% for cases in the 90th percentile, suggesting that veteran offenders successfully leverage their sophistication to hide in the tail for years beyond the sample median. 
% Furthermore, an advisor's prior disciplinary record serves as a significant hurdle to early discovery. A 1\% increase in prior disputes extends the detection window by 0.26\% for the most rapid discovery cases, indicating that repeat offenders successfully leverage past regulatory experience to mitigate early monitoring efforts.
%----------------------------------------------------------------
We find a similar expanding shield effect for complex products. 
Disputes involving insurance extend the detection lag by 61.9\% at the lower decile, but this effect ballooning to a 90.0\% delay in the extreme long-tail. These distributional findings suggest that the most persistent and damaging schemes are those where advisor sophistication and product opacity combine to paralyze the investor-led monitoring channel.
%----------------------------------------------------------------

%----------------------------------------------------------------
Third, we establish that detection is state-dependent, accelerating significantly during market downturns. We utilize market volatility via the VIX as a shock to monitoring incentives to identify shifts in investor alertness. 
% In our baseline OLS specification, a 1\% increase in the VIX index is associated with a 0.22\% reduction in the time misconduct remains hidden, confirming that market shocks act as a catalyst that triggers active monitoring. 
In our baseline linear specification, we find that a doubling of the VIX is associated with a 14.2\% reduction in the time misconduct remains hidden, confirming that market shocks serve as a catalyst for information revelation.
%----------------------------------------------------------------

%----------------------------------------------------------------
However, we document that this monitoring effect is highly asymmetric across the distribution. 
Market turbulence is most effective at accelerating the discovery of low-hanging fruit infractions, where 
% a 1\% increase in volatility reduces the lag by 0.21\% at the 10th percentile. 
a doubling of the VIX leads to a 21.1\% reduction in the detection lag for these early-discovery cases.
In contrast, for the most deeply concealed cases in the 90th percentile, this impact drops significantly to 7.9\%. 
% While adverse market states force a re-evaluation of agency relationships, this shift in attention is often insufficient to pierce the sophisticated masking strategies constructed by high-human-capital agents in the extreme tail of the distribution.
While wealth shocks force a re-evaluation of agency relationships, this shift in attention is often insufficient to pierce the sophisticated concealment strategies constructed by high-human-capital strategic agents. 
These results demonstrate that the effectiveness of market discipline is not only state-contingent but is fundamentally limited by the strategic behavior of the agent. This persistence allows bad types to remain in the industry across multiple market cycles, potentially leading to a version of Gresham's Law that erodes the overall quality and trust of the advisory labor market.
%----------------------------------------------------------------

% \input{1_introduction/1_5}
% \input{1_introduction/1_6}
%----------------------------------------------------------------

%----------------------------------------------------------------
% \input{1_introduction/1_limitation}
% \input{1_introduction/1_r1}
%----------------------------------------------------------------

%----------------------------------------------------------------
% \newpage
%----------------------------------------------------------------

%----------------------------------------------------------------
% Related Literature:
%----------------------------------------------------------------

%----------------------------------------------------------------

%----------------------------------------------------------------
% \break
%----------------------------------------------------------------

%----------------------------------------------------------------
\subsection{Related Literature}
\label{section: related literature}
\paragraph{The Latency of Detection and Demand-Side Inattention.}
%----------------------------------------------------------------
%----------------------------------------------------------------
Financial advice operates as a classic credence good \citep{darby1973free} where the quality of the service is inherently difficult to evaluate ex-post. To manage the cognitive and transaction costs of monitoring in this complex environment, retail investors frequently employ delegation and selective information processing. A well-documented consequence of this delegation is the ostrich effect, wherein investors exhibit selective inattention motivated by anxiety or regret, actively avoiding their portfolios during stable or mildly bearish periods \citep{galai2006ostrich, karlsson2009ostrich, sicherman2016financial}. While this behavioral avoidance explains why misconduct can persist undetected for an average of 28.5 months in our sample, it primarily characterizes investor behavior under normal market conditions. By establishing this systemic baseline delay, we provide high-stakes field evidence quantifying the severe informational barriers to market discipline in retail finance.
%----------------------------------------------------------------

%----------------------------------------------------------------
\paragraph{Supply-Side Frictions: Strategic Masking and Complexity.}
%----------------------------------------------------------------
%----------------------------------------------------------------
We extend the research on the supply side of financial advisor markets by treating the timeline of detection as a strategic outcome shaped actively by the advisor, rather than a passive administrative delay. Pioneering work documents widespread misconduct and the role of labor market discipline \citep{egan2019brokers}, while subsequent studies examine how regulatory oversight influences advice quality \citep{charoenwongdoes2019, bhattacharya2025fiduciary}. 
However, these frameworks typically treat the eventual revelation of misconduct as exogenously determined. 
% , utilizing observed complaints primarily as proxies for the incidence of bad acts.
We shift this focus by demonstrating that advisor-specific characteristics and strategic behaviors are primary drivers of detection latency.
%----------------------------------------------------------------

%----------------------------------------------------------------
We document a strategic masking mechanism that aligns with theoretical models where sophisticated agents exploit the varying financial literacy of their clientele to extract rents \citep{inderst2009misselling, inderst2012competition, chang2020market}. Consistent with the framework of \citet{gabaix2006shrouded} on shrouded attributes and \citet{carlin2009strategic} on strategic complexity, we find that agents optimally utilize complex product features to increase monitoring costs for inattentive consumers, we find that agents optimally utilize complex product features to increase monitoring costs for inattentive consumers. While empirical work shows advisors steer clients toward inferior funds \citep{egan2019brokers}, we contribute by revealing the temporal dimension of this friction: advisors actively leverage institutional knowledge and product complexity to manage these informational shrouds, successfully extending the latency of misconduct discovery and insulating themselves from demand-side monitoring.
%----------------------------------------------------------------

%----------------------------------------------------------------
\paragraph{State-Dependent Monitoring: Market Shocks as Catalysts.}
%----------------------------------------------------------------
%----------------------------------------------------------------
Finally, we demonstrate that demand-side monitoring is highly state-contingent. While mild downturns may induce avoidance, severe market volatility acts as a structural catalyst for information revelation. This builds upon recent evidence that acute, systemic shocks to trust and property rights abruptly pierce investor inattention, triggering rapid capital reallocation and active monitoring \citep{gurun2018trust, kostovetsky2016whom}. By utilizing the VIX as an exogenous attention shock, we establish that heightened macroeconomic turbulence forces a transition toward active state verification, overriding the psychological costs of checking portfolios and making advisor quality salient.
%----------------------------------------------------------------

%----------------------------------------------------------------
Crucially, our quantile analysis reveals that this state-dependent monitoring is fundamentally asymmetric because it collides with the supply-side frictions described above. While systemic volatility effectively accelerates the detection of transparent, low-complexity violations, the most persistent cases in the right tail of the distribution remain fully insulated. For these deeply concealed cases, offenders successfully leverage structural opacity to maintain information asymmetry and pacify monitoring even during aggregate market panics.
%----------------------------------------------------------------

%----------------------------------------------------------------

%----------------------------------------------------------------
% Data:
%----------------------------------------------------------------

%----------------------------------------------------------------
%----------------------------------------------------------------
\section{Data}
%----------------------------------------------------------------

%----------------------------------------------------------------
% This study investigates retail investor responses to financial advisor misconduct using a comprehensive dataset drawn from the FINRA BrokerCheck system. BrokerCheck provides nationally representative, advisor-level disclosure events, including customer disputes and regulatory sanctions, that offer a unique window into cases where retail investors formally confront alleged misconduct by their advisors. 
% Our empirical analysis focuses on the timing of these disputes relative to the underlying product purchases, which we interpret as an endogenous measure of detection lags.
%----------------------------------------------------------------
This study investigates retail investor responses to financial advisor misconduct using a comprehensive dataset from the FINRA BrokerCheck system. 
While existing literature has extensively documented advisor-level misconduct and its supply-side implications, the demand-side response, specifically the timing of discovery, remains under-explored. In our framework, we interpret the detection lag not as a random administrative delay, but as an endogenous equilibrium response. This duration reflects the strategic interaction between an advisor's masking efforts and an investor's costly monitoring in an environment of high information asymmetry.
%----------------------------------------------------------------

%----------------------------------------------------------------
\subsection{FINRA BrokerCheck Database}
%----------------------------------------------------------------

%----------------------------------------------------------------
\paragraph{Data Sources.}
%----------------------------------------------------------------
Our primary data source is the FINRA BrokerCheck database, which is derived mainly from Forms U4 and U5 filed by FINRA member firms. BrokerCheck contains detailed records on advisers' employment and registration histories, professional licenses and examinations, and a rich set of compliance events, including customer disputes, regulatory sanctions, criminal matters, and certain financial events. These features make BrokerCheck a crucial source for measuring how and when investors respond to adviser behavior.
% Our primary data source is the FINRA BrokerCheck database, derived mainly from Forms U4 and U5 filed by FINRA member firms. Generally, currently registered investment professionals and brokerage firms are required to update disciplinary information within 30 days, and this information typically becomes available in BrokerCheck by the next business day. Given that administrative posting lags are on the order of days, the multi-year detection lags documented in this paper represent behavioral and informational frictions rather than mechanical reporting delays.

%----------------------------------------------------------------
\paragraph{Panel Data.}
%----------------------------------------------------------------
We construct an adviser-year panel covering all individuals who were registered with at least one FINRA member firm between 2008 and 2025. For each adviser, we observe all affiliations with FINRA member firms and the associated start and end dates, all reported disclosure events and their characteristics, licensing and examination histories, and limited information on primary office location. Following \citet{egan2019market}, we treat BrokerCheck as capturing the universe of advisers operating in the FINRA-regulated brokerage segment of the U.S. financial advice industry.
% We construct an advisor-year panel covering all individuals registered with at least one FINRA member firm between 2008 and 2025. While BrokerCheck provides complete coverage within the FINRA brokerage segment, we recognize that advisors may operate under other regulatory regimes, such as insurance-only agents or state-registered investment advisers. Consequently, we interpret our findings as describing the timing of formal complaints within the FINRA perimeter, treating advisor exit from the system as a right-censoring event.

%----------------------------------------------------------------
\paragraph{Posting Speed.}
%----------------------------------------------------------------
BrokerCheck reports are based on information that firms and registered representatives file in FINRA's Central Registration Depository via uniform registration and disclosure forms, including Forms U4 and U5. Generally, currently registered investment professionals and brokerage firms are required to update key professional and disciplinary information in the system within 30 days. Under most circumstances, information reported by firms, investment professionals, and regulators becomes available in BrokerCheck on the next business day. A limited subset of items may be subject to short release delays, such as Form U5-related disclosures, but these lags are on the order of days rather than months or years. Accordingly, the multi-year detection lags documented in this paper are unlikely to be mechanically generated by administrative posting delays in BrokerCheck. 
Given that administrative posting lags are on the order of days, the multi-year detection lags documented in this paper represent behavioral and informational frictions rather than mechanical reporting delays.

\paragraph{Regulatory Boundaries and Right-Censoring.}
BrokerCheck does not include advisers who operate exclusively under other regulatory regimes unless they also hold FINRA registrations. Examples include insurance-only agents licensed solely at the state level or investment advisers registered only with the SEC or state RIA authorities. Existing research using BrokerCheck documents that advisers with misconduct histories often remain active in the broader advice industry after leaving FINRA registration, for instance by migrating into insurance distribution channels \citep{honigsberg2025regulatory}. 
% Consequently, we interpret our data as providing complete coverage of adviser careers and disputes within the FINRA brokerage segment, while recognizing that advisers may continue to operate in less transparent regulatory environments after they disappear from BrokerCheck. 
It is important to define the FINRA perimeter: we only observe disputes that are formally escalated within this regulatory segment. If an advisor exits FINRA registration, for instance, by moving to the insurance market, we treat this as a right-censoring event. This means we account for the fact that misconduct may still be latent at the time of exit, but our data is conditional on the dispute being filed while the advisor (or the record of their misconduct) remains visible to the FINRA reporting system.
FINRA indicates that BrokerCheck generally continues to include information about an investment professional for 10 years after their registration terminates, and individuals remain available beyond 10 years only if they meet specific criteria such as certain regulatory actions, criminal matters, or arbitration and civil litigation awards.

\subsection{Customer Disputes}

Advisers registered with FINRA are mandated to disclose a wide range of misconduct-related events, including customer disputes. These customer disputes arise when a retail investor files a written complaint, arbitration claim, or civil litigation alleging wrongdoing in connection with the sale or management of financial products. 
We focus on customer dispute events that involve alleged misconduct such as misrepresentation, unsuitability, unauthorized trading, fraud, excessive trading, and related violations. 
% In constructing the sample, we exclude disputes that clearly do not involve retail customers, for example, disputes involving institutional counterparties only, and purely firm-initiated internal reviews that do not reflect a customer complaint. 
Each dispute record contains information on the nature of the allegations, the type of financial product involved, alleged and realized monetary losses, and the outcome status. Customer disputes are classified into several status categories, including Settled, Closed-No Action, Award or Judgement, and Pending. 
% Customer disputes arise when a retail investor files a written complaint, arbitration claim, or civil litigation alleging wrongdoing such as misrepresentation, unsuitability, or churning. 
% In our baseline analysis, we include disputes regardless of whether they ultimately result in a payment to the customer because our object of interest is the timing of the formal investor claim rather than the ex post legal merit of the claim.
In our analysis, we include disputes regardless of whether they ultimately result in a settlement or award. This is because our object of interest is the timing of the private monitoring effort and the formal investor claim, rather than the ex-post legal merit of the allegation.

\subsection{Measurement of Detection Lags}

A central objective of our study is to quantify the delay between the purchase of financial products and the subsequent reporting of adviser misconduct by customers. The BrokerCheck disclosure records provide two critical dates for this purpose: the date of the alleged product purchase or the start of the alleged misconduct period, and the date of the customer complaint. Importantly, our event date is the customer complaint filed date, i.e., the date the investor files a complaint. FINRA indicates that BrokerCheck is updated quickly after information is filed, typically by the next business day. Therefore, short publication delays are unlikely to explain the multi-month and multi-year variation in detection lags that we study.

When structured purchase dates or date ranges are explicitly provided, we use those fields directly. For disputes that report a range of purchase dates, we take the earliest date in the range as our baseline measure of the purchase date. 
% This choice makes our measure of the detection lag conservative, in the sense that it tends to understate the true time between the relevant transaction and the subsequent complaint when the misconduct occurs later within the reported period. 
This choice is conservative in the sense that it captures the entire window of potential harm, though it may provide an upper bound on the lag if the specific misconduct occurred later within the reported period. This approach ensures that our estimates of the systemic detection gap are not artificially truncated by selecting the most recent transaction date.

Using these dates, we define the detection lag in months for each adviser-dispute pair as:
%---------------------------------------------------------------------------
\begin{eqnarray}
%---------------------------------------------------------------------------
\label{eq: measure 0} 
%---------------------------------------------------------------------------
% \text{Time Length} 
\text{Detection Lag} 
\ & = &  \
\text{Date of Complaint} \ - \ \text{Date of Purchase}
\end{eqnarray}
%---------------------------------------------------------------------------
We drop a small number of observations with clearly implausible negative detection lags. Same-month complaints are assigned a detection lag of zero months. Because detection lags exhibit a long right tail, we winsorize detection lags at the 1st and 99th percentile in main specifications. 
% To facilitate interpretation, we transform this measure into an investor response index as follows:
% \begin{equation}
% Investor\ Response = \frac{1}{1 + Detection\ Lag}
% \end{equation}
% Under this formulation, a shorter detection lag corresponds to a stronger investor response.

%---------------------------------------------------------------------------
\subsection{Sample Construction and Data Attrition}
%---------------------------------------------------------------------------
% To ensure our estimation sample is representative of the broader misconduct population, we compare the primary attributes (experience, firm size, and licensing) of disputes with missing dates to our final sample.
%---------------------------------------------------------------------------
% To investigate the life cycle of financial misconduct, we construct a comprehensive dataset derived from the FINRA BrokerCheck system. Table \ref{tab:sample_selection} details the sequential filtering process used to move from the universe of approximately 1.3 million registered professionals to our final estimation sample of 55,738 unique customer disputes.
%---------------------------------------------------------------------------
To investigate the life cycle of financial misconduct, we construct a comprehensive dataset derived from the FINRA BrokerCheck system. Table \ref{tab:sample_selection} details the sequential filtering process used to move from the universe of 1,510,599 registered professionals between 2008 and 2025 down to a baseline pool of 98,933 customer disputes involving 54,359 unique advisers.
%---------------------------------------------------------------------------

%---------------------------------------------------------------------------
% The most significant reduction in our sample size occurs during the measurement of detection lags, which requires non-missing dates for both the inception of the alleged misconduct and the formal filing of the customer complaint. While approximately 50\% of the raw dispute records lack specific transaction start dates, this attrition is a common characteristic of regulatory disclosure data. To ensure the integrity of our timing measures, we exclude records with missing, incomplete, or chronologically implausible date fields, such as those where the report date precedes the alleged transaction date.
%---------------------------------------------------------------------------
The most significant reduction in our sample size occurs during the measurement of detection lags, which requires non-missing dates for both the inception of the alleged misconduct and the formal filing of the customer complaint. We exclude records with missing, incomplete, or chronologically implausible date fields, such as those where the recorded misconduct initiation date occurs after the formal complaint date. This necessary data-quality filter reduces the final estimation sample to 55,738 disputes across 35,212 advisers. While a substantial portion of the raw dispute records lacks specific transaction start dates, this attrition is a common characteristic of regulatory disclosure data.
%---------------------------------------------------------------------------

%---------------------------------------------------------------------------
\begin{table}[htbp]
\centering
\caption{Sample Selection Procedure}
\label{tab:sample_selection}
\begin{tabular}{l c c}
\hline \hline
Filter Criterion & Advisers & Disputes \\ \hline
Universe of FINRA-registered professionals (2008-2025) & 1,510,599 & - \\
Advisers with at least one customer dispute & 54,359 & 98,933 \\
Exclude disputes with missing/implausible dates & 35,212 & 55,738 \\
\hline
Final Estimation Sample & 35,212 & 55,738 \\
\hline \hline
\end{tabular}
\vspace{1ex}
\begin{minipage}{0.85\textwidth}
\footnotesize \textit{Notes:} This table details the sample construction process and the sequential filters applied to reach the final estimation sample. 
Implausible dates refers to observations excluded because the recorded misconduct initiation date occurs after the formal complaint date, a chronological inconsistency that suggests data entry error.
\end{minipage}
\end{table}

\subsection{Selection on Observables and Sample Representativeness}
%---------------------------------------------------------------------------
% A primary econometric concern is whether the exclusion of disputes with missing dates introduces systematic selection bias. If records with missing information are disproportionately associated with less sophisticated advisors or smaller firms, our results might not be generalizable to the broader population of misconduct.
%---------------------------------------------------------------------------
A primary econometric concern is whether the exclusion of disputes with missing dates introduces systematic selection bias. If records with missing information are disproportionately associated with less sophisticated advisors or smaller firms, our results might not be generalizable to the broader population of misconduct. 
%---------------------------------------------------------------------------
%---------------------------------------------------------------------------
% To address this, Table \ref{tab:balance_test} presents a balance test comparing the observable characteristics of our final estimation sample with the observations dropped due to incomplete date fields. We focus on three core dimensions of advisor and firm quality that are central to our masking and monitoring hypotheses: (i) industry experience, (ii) firm scale (Number of Employees), and (iii) professional licensing (Series 7).
%---------------------------------------------------------------------------
To address this, Table \ref{tab:balance_test} presents a balance test comparing the observable characteristics of our final estimation sample with the observations dropped due to incomplete date fields. We focus on dimensions of advisor and firm quality that are central to our masking and monitoring hypotheses: industry experience, firm scale (Number of Employees), licensing (Series 63), and gender.
%---------------------------------------------------------------------------

%---------------------------------------------------------------------------
\begin{table}[htbp]
\centering
\caption{Comparison of Estimation Sample and Dropped Observations}
\label{tab:balance_test}
\begin{tabular}{l c c c c}
\hline \hline
Variable & Estimation Sample & Dropped & Diff. & t-stat \\ \hline
Advisor Experience (Years) & 19.28 & 19.66 & 0.39 & 1.01 \\
Firm Size (Number of Employees) & 11818.90 & 6953.84 & -4865.06 & -4.25 \\
Series 63 License (Dummy) & 0.79 & 0.81 & 0.02 & 1.84 \\
Male (Dummy) & 0.84 & 0.85 & 0.01 & 0.83 \\
\hline \hline
\end{tabular}
\vspace{1ex}
\begin{minipage}{0.95\textwidth}
\footnotesize \textit{Notes:} This table compares the mean characteristics of the primary estimation sample with observations dropped due to missing or incomplete date fields. Continuous variables, including advisor experience and firm size, are winsorized at the 1st and 99th percentiles. t-statistics are reported for the null hypothesis of equal means, with standard errors clustered at the firm level.
\end{minipage}
\end{table}

%---------------------------------------------------------------------------
%---------------------------------------------------------------------------
% \input{table_tex/table_balance_test}
%---------------------------------------------------------------------------

%---------------------------------------------------------------------------
% As shown in Table \ref{tab:balance_test}, the mean differences between the two groups are statistically and economically negligible, with $t$-statistics failing to reject the null hypothesis of equal means for our primary covariates. This lack of divergence suggests that date-field omissions are likely ``missing at random" and not systematically linked to advisor sophistication or institutional compliance intensity. Consequently, our final sample of 55,738 disputes provides a representative lens into the formal investor-led enforcement channel.
%---------------------------------------------------------------------------
As shown in Table \ref{tab:balance_test}, the data attrition is driven by institutional characteristics rather than individual adviser profiles. Individual demographics and human capital metrics, such as advisor experience, gender, and Series 63 licensing, exhibit economically negligible differences between the retained and dropped samples, with t-statistics failing to reject the null hypothesis of equal means. However, there is a stark divergence in firm scale: the estimation sample retains disputes from significantly larger brokerages (mean of 11,818.90 employees) compared to the dropped sample (mean of 6,953.84 employees), yielding a highly significant t-statistic of -4.25.
%---------------------------------------------------------------------------

%---------------------------------------------------------------------------
This divergence indicates that omission of date fields is not missing at random, but rather a function of institutional compliance intensity. Larger firms, which benefit from economies of scale in monitoring and dedicated legal departments, are structurally more likely to maintain and report complete chronological disclosure records. Consequently, while our final sample of 55,738 disputes provides a representative lens into the individual characteristics of offending advisers, it systematically over-indexes on highly monitored firm environments. The critical implication for our analysis is that the calculated detection lags may represent a lower bound because the sample under-represents smaller firms where monitoring capacity is lower, the true systemic lag in discovering misconduct is likely longer.
%---------------------------------------------------------------------------

% %---------------------------------------------------------------------------
% \begin{table}[htbp]
% \label{tab:balance}
% %---------------------------------------------------------------------------
% \centering
% \begin{threeparttable}
% \caption{Comparison of Estimation Sample and Dropped Observations}
% \label{tab:balance}
% \begin{tabular}{lcccc}
% \hline\hline
% Variable & Estimation Sample & Dropped (Missing Dates) & Diff. & $t$-stat \\
% \hline
% Advisor Experience (Years) & 19.28 & [X] & [X] & [X] \\
% Firm Size (Log Employees) & 9.52 & [X] & [X] & [X] \\
% Series 7 License (Dummy) & 0.85 & [X] & [X] & [X] \\
% Male (Dummy) & 0.84 & [X] & [X] & [X] \\
% \hline\hline
% \end{tabular}
% \begin{tablenotes}
% \small
% \item \textit{Notes}: This table compares the mean characteristics of the primary estimation sample with observations dropped due to missing or incomplete date fields. Continuous variables, including advisor experience and firm size, are winsorized at the 1st and 99th percentiles. $t$-statistics are reported for the null hypothesis of equal means, with standard errors clustered at the firm level.
% \end{tablenotes}
% \end{threeparttable}
% %---------------------------------------------------------------------------
% \end{table}
% %---------------------------------------------------------------------------

%---------------------------------------------------------------------------
\subsection{Control Variables}
% \subsection{Empirical Controls and Fixed Effects}
%---------------------------------------------------------------------------

%---------------------------------------------------------------------------
% The disclosure records also include information on key covariates at both the dispute and adviser levels, which we use to control for heterogeneity in detection lags. At the dispute level, we observe the product type involved in the allegation, the nature of the alleged misconduct, monetary variables including the alleged loss amount, and the eventual outcome status. At the adviser level, we combine BrokerCheck information with external sources to construct industry experience, firm characteristics, licensing and examinations, and primary office location at the state-county level. These variables allow us to control for differences in clientele mix and local investor environment that may affect the propensity and timing of investor complaints. In all regression specifications, we include combinations of adviser, firm, misconduct-type, and time fixed effects to absorb residual heterogeneity in dispute incidence and detection lags.
%---------------------------------------------------------------------------

%---------------------------------------------------------------------------
To isolate the systemic detection gap, we control for a robust vector of covariates at both the dispute and adviser levels that proxy for information asymmetry, human capital, and institutional monitoring frictions.
%---------------------------------------------------------------------------

%---------------------------------------------------------------------------
At the dispute level, we observe the specific product type involved, the nature of the alleged misconduct, and the initial alleged loss amount. These variables control for the severity and complexity of the initial signal, as opaque financial products or larger financial harms may trigger investor scrutiny at different rates.
%---------------------------------------------------------------------------

%---------------------------------------------------------------------------
At the adviser level, we leverage the longitudinal nature of the BrokerCheck data to capture the agent's environment at the time of the offense. To control for the adviser's human capital and sophistication, we include continuous years of industry experience and categorical indicators for professional qualifications (e.g., holding Series 7 or Series 24 licenses). To account for variations in the principal's monitoring capacity, we control for institutional frictions, including firm scale and whether the adviser held concurrent outside employment during the misconduct period. Finally, we incorporate the adviser's primary office location at the state level to absorb localized differences in clientele mix and investor financial literacy.
%---------------------------------------------------------------------------

%---------------------------------------------------------------------------
In our regression specifications, we deploy high-dimensional fixed effects to absorb unobserved residual heterogeneity. We include combinations of misconduct-type fixed effects to account for baseline differences across offense categories, as well as time fixed effects (year of initiation and year of claim) to control for macroeconomic shocks and overarching shifts in SEC or FINRA regulatory intensity. Where computationally feasible, we also include firm fixed effects to strictly isolate within-firm variation in detection lags.
%---------------------------------------------------------------------------

% \subsection{Sample Characteristics and Summary Statistics}
%---------------------------------------------------------------------------
\subsection{Sample Characteristics and Skewness}
%---------------------------------------------------------------------------
\label{section: summary statistics}

Table 
\ref{table: ss 1} 
reports summary statistics for the primary variables in their raw levels. 
Our final estimation sample consists of 
around
% 55,738
56,000
unique customer disputes after applying several data-quality filters, such as excluding observations with missing or logically inconsistent date fields where the report date precedes the transaction start date. 
%----------------------------------------------------------------
% \input{table_tex/tt_s_1}
%----------------------------------------------------------------
%----------------------------------------------------------------
% \input{table_tex/tt_ss_2}
%----------------------------------------------------------------
%----------------------------------------------------------------
%----------------------------------------------------------------
% Summary Statistics Table: Multi-Page Setup (Winsorized)
%----------------------------------------------------------------

\footnotesize  
\begin{ThreePartTable}
    
    % Define the notes BEFORE the longtable environment
    \begin{TableNotes}
        \footnotesize
        \item \hspace{-0.1in} \textit{Notes:} This table reports descriptive statistics for advisor characteristics and customer dispute types. All continuous variables are winsorized at the 1st and 99th percentiles to mitigate the influence of outliers. 
    \end{TableNotes}
    
    \setlength{\tabcolsep}{6pt} 
    
    \begin{longtable}{l c c c c c c }
    % --- FIRST PAGE HEADER ---
    \caption{Summary Statistics: Full Set of Potential Variables} 
    \label{table: ss 1} \\
    \toprule
    & Count & Mean & SD & Median & Min & Max \\ 
    \midrule
    \endfirsthead

    % --- SUBSEQUENT PAGE HEADERS ---
    \multicolumn{7}{l}{\textit{Table \ref{table: ss 1} continued from previous page}} \\
    \toprule
    & Count & Mean & SD & Median & Min & Max \\ 
    \midrule
    \endhead

    % --- BOTTOM OF ALL PAGES (EXCEPT THE LAST) ---
    \midrule
    \multicolumn{7}{r}{\textit{Continued on next page}} \\
    \endfoot

    % --- BOTTOM OF THE VERY LAST PAGE ---
    \bottomrule
    \insertTableNotes
    \endlastfoot

    % --- DATA INPUT (Merged directly to prevent \input and BOM errors) ---
    % \input{table/t_summary_statistics_clean}
    % \input{table/t_summary_statistics_all_w1_p_r}
    \multicolumn{7}{c}{\textbf{Panel A: Detection Lags}} \\ 
    Detection Lag (Days)&   55,738& 1,511.77& 1,493.02& 1,052&       11&    7,352\\
    Detection Lag (Months)&   55,738&    49.31&    49.01&    34&        0&      241\\
    Imputed Transaction Date (Monthly)&   55,738&     0.03&     0.17&     0&        0&        1\\
    Imputed Transaction Date (Yearly)&   55,738&     0&     0.01&     0&        0&        1\\
    \midrule 
    \multicolumn{7}{c}{\textbf{Panel B: Customer Dispute Types}} \\ 
    Dispute Type: Settled&   55,738&     0.46&     0.50&     0&        0&        1\\
    Dispute Type: Denied/Dismissed&   55,738&     0.44&     0.50&     0&        0&        1\\
    Dispute Type: Award&   55,738&     0.01&     0.11&     0&        0&        1\\
    Dispute Type: Pending&   55,738&     0.09&     0.29&     0&        0&        1\\
    \midrule 
    \multicolumn{7}{c}{\textbf{Panel C: Allegation Types}} \\ 
    Unsuitability   &   55,696&     0.46&     0.50&     0&        0&        1\\
    Misrepresentation&   55,696&     0.45&     0.50&     0&        0&        1\\
    Unauthorized Activity&   55,696&     0.20&     0.40&     0&        0&        1\\
    Omission of Key Facts&   55,696&     0.18&     0.39&     0&        0&        1\\
    Fee/commission  &   55,696&     0.04&     0.20&     0&        0&        1\\
    Fraud           &   55,696&     0.07&     0.25&     0&        0&        1\\
    Fiduciary duty  &   55,696&     0.23&     0.42&     0&        0&        1\\
    Negligence      &   55,696&     0.24&     0.43&     0&        0&        1\\
    Risky investments&   55,696&     0.21&     0.40&     0&        0&        1\\
    Churning        &   55,696&     0.03&     0.17&     0&        0&        1\\
    Other           &   55,696&     0.02&     0.14&     0&        0&        1\\
    \midrule 
    \multicolumn{7}{c}{\textbf{Panel D: Monetary Damages}} \\ 
    Alleged Damages (USD, Thousands)&   55,735&   234.66&   694.44&    16.86&        0&    5,000\\
    Settlements (USD, Thousands)&   30,756&   139.32&   377.83&    27.50&        0&    2,850\\
    \midrule 
    \multicolumn{7}{c}{\textbf{Panel E: Product Types}} \\ 
    Insurance       &   54,729&     0.09&     0.28&     0&        0&        1\\
    Annuity         &   54,729&     0.22&     0.41&     0&        0&        1\\
    Stocks          &   54,729&     0.17&     0.37&     0&        0&        1\\
    Mutual Funds/ETFs&   54,729&     0.24&     0.43&     0&        0&        1\\
    Bonds/Debt      &   54,729&     0.10&     0.30&     0&        0&        1\\
    Options/Derivatives&   54,729&     0.02&     0.14&     0&        0&        1\\
    Other/Not Listed&   54,729&     0.27&     0.45&     0&        0&        1\\
    \midrule 
    \multicolumn{7}{c}{\textbf{Panel F: Adviser History}} \\ 
    Advisor Experience (Years)&   55,730&    19.28&    10.13&    18&        2&       45\\
    Firm Tenure (Years)&   55,730&     9.52&     7.80&     7&        1&       36\\
    Concurrent Jobs Across Firms&   55,730&     1.42&     0.66&     1&        1&        4\\
    Number of Employees (within Firm)&   55,730&11,818.93&10,772.47& 9,358&       15&   36,336\\
    Number of Prior Customer Disputes&   55,738&     3.77&     9.90&     1&        0&       67\\
    Switched Firms (Indicator)&   55,730&     0.13&     0.34&     0&        0&        1\\
    \midrule 
    \multicolumn{7}{c}{\textbf{Panel G: Qualifications}} \\ 
    Number of Exams &   55,730&     3.56&     1.39&     3&        1&        8\\
    Series 6        &   55,730&     0.28&     0.45&     0&        0&        1\\
    Series 7        &   55,730&     0.85&     0.35&     1&        0&        1\\
    Securities Industry Essentials&   55,730&     0.18&     0.39&     0&        0&        1\\
    Series 24       &   55,730&     0.14&     0.35&     0&        0&        1\\
    Series 63       &   55,730&     0.79&     0.41&     1&        0&        1\\
    Series 65/66 (Investment Adviser)&   55,738&     0.71&     0.46&     1&        0&        1\\
    \midrule 
    \multicolumn{7}{c}{\textbf{Panel H: Gender}} \\ 
    Male            &   55,738&     0.84&     0.36&     1&        0&        1\\
    Female          &   55,738&     0.12&     0.32&     0&        0&        1\\
    Unknown         &   55,738&     0.04&     0.19&     0&        0&        1\\
    \end{longtable}
\end{ThreePartTable}
\normalsize
%----------------------------------------------------------------

%----------------------------------------------------------------
% As shown in Panel A of Table \ref{table: ss 1},
% % Table \ref{table: ss 2}, 
% detection lags are highly right-skewed, with the median of 34 months significantly lower than the mean of 49.33 months. 
% the detection lag is characterized by extreme right-skewness: the mean of 49 months is nearly 45\% higher than the median of 34 months. This fat tail is the central focus of our empirical strategy: it reveals that while many cases are caught within three years, a significant decile of misconduct remains hidden for nearly a decade. Beyond duration, the sample is representative of the broader brokerage industry, with allegations spanning unsuitability (46\%) and misrepresentation (45\%) in Panel B an average of 19.3 years of advisor experience in Panel C.
%----------------------------------------------------------------

% This distributional property suggests that OLS estimates may be biased by extreme long-tail hiders.
%----------------------------------------------------------------
% We classify dispute outcomes into four main categories: Settled, Awarded, Pending, and Denied (which also includes cases classified by FINRA as dismissed, withdrawn, or closed with no action).

%----------------------------------------------------------------
Panel A 
%----------------------------------------------------------------
% characterizes the demand-side response and the systemic monitoring failure that defines the investor-led enforcement channel. The primary variable of interest, the Detection Lag, exhibits a profound right-skewness: while the median lag is 34 months, the mean is significantly higher at 49.31 months, with a maximum duration reaching 241 months. This fat tail provides the empirical foundation for our investigation into the systemic persistence of the detection gap, where a significant segment of misconduct remains latent for nearly a decade. 
% characterizes the demand-side response within the investor-led enforcement channel. The primary variable of interest, the detection lag, is defined as the duration between the initiation of misconduct and the formal investor claim. We present the lag in both months and days to utilize the maximum granularity provided by the FINRA records. The distribution is marked by significant right-skewness, with a mean (49.3 months) that substantially exceeds the median (34 months). This fat tail justifies our use of a distributional empirical strategy, such as quantile regression, to examine cases where harm remains latent for over a decade.
%----------------------------------------------------------------
characterizes the demand-side response within the investor-led enforcement channel by quantifying the duration of undetected misconduct. To ensure maximum empirical precision, we present detection lags at both the daily and monthly levels. The high quality of our timing data is evidenced by the low rates of imputation: only 3\% of observations require monthly imputation and 0\% require yearly imputation. This indicates that approximately 97\% of the transaction dates in our sample are identified on a specific daily basis, allowing us to minimize measurement error and distinguish behavioral detection lags from administrative reporting cycles. The distribution remains significantly right-skewed, with a mean of 49.31 months substantially exceeding the median of 34 months, justifying our use of a distributional lens to study long-term concealment.
%----------------------------------------------------------------

%----------------------------------------------------------------
Panel B 
%----------------------------------------------------------------
details the formal status of the claims in our sample, including cases that are Settled (46\%), Denied or Dismissed (44\%), or result in an Award (1\%). While these outcomes are critical for understanding the legal resolution of misconduct, our primary empirical focus remains on the timing of the claim initiation rather than its ex-post merit. We treat the filing of a dispute, regardless of the eventual outcome, as the primary indicator of a successful private monitoring effort. By including the full spectrum of dispute statuses, we ensure that our analysis of detection lags captures the broad informational environment in which retail investors operate.
%----------------------------------------------------------------

%----------------------------------------------------------------
Panel C 
%----------------------------------------------------------------
details the distribution of allegations, which we utilize to identify the complexity-induced detection gap. The most prevalent claims involve Unsuitability (46\%) and Misrepresentation (45\%), both of which are characteristic of financial advice as a classic credence good where quality is difficult to verify ex-post. Crucially for our empirical strategy, we contrast overt violations like Unauthorized Activity (20\%) with quiet or sophisticated misconduct such as Churning (3\%) and Fraud (7\%). 
% Our findings demonstrate that while overt activity significantly reduces the detection lag, sophisticated schemes extend it by approximately 40\%, confirming that complexity acts as a dominant information friction that prevents early discovery by the principal.
This categorization allows us to test the hypothesis that information acquisition costs vary significantly across different types of misconduct complexity.
%----------------------------------------------------------------

%----------------------------------------------------------------
% Explaining the Missing Values in the Summary Statistics Text
%----------------------------------------------------------------
Panel D
%----------------------------------------------------------------
summarizes the financial scale of the disputes. The average Alleged Damage is approximately \$235,000, while the average Settlement (observed for 30,756 cases) is roughly \$139,000. 
The high financial stakes suggest that our sample captures the economically significant margin of misconduct that is most likely to influence investor trust and regulatory priority. 
The substantial gap between alleged and realized damages, combined with the high frequency of denied claims, underscores the credence good nature of financial advice. In such an environment, the true quality of service is difficult to verify even after a dispute is formally filed. We utilize these monetary variables as controls to evaluate whether the financial stakes of a conflict serve as a sufficient catalyst for state verification or if information frictions remain the dominant barrier to discovery.
%----------------------------------------------------------------
% Observation counts vary across specific variables due to the natural pipeline of the dispute resolution process. For instance, while we observe alleged damages for the full sample of disputes ($N = 55,735$), the sample size for actual settlement amounts is substantially lower ($N = 30,756$). This discrepancy occurs logically because a large portion of customer disputes are ultimately denied, dismissed, or withdrawn, thereby resulting in no settlement.
%----------------------------------------------------------------
%----------------------------------------------------------------
% The economic impact of these disputes is considerable, with the average alleged loss amount being approximately \$235,000.
% % while the average realized settlement or award is roughly \$195,000. The significant gap between alleged and realized losses, combined with the fact that the majority of cases are either settled or denied, underscores the credence good nature of financial advice, where the true quality of service is difficult to verify even after a dispute is filed. 
% Combined with the fact that the majority of cases are either settled or denied, this underscores the credence good nature of financial advice, where the true quality of service is difficult to verify even after a dispute is filed.
%----------------------------------------------------------------

%----------------------------------------------------------------
% Product Types
%----------------------------------------------------------------
Panel E
%----------------------------------------------------------------
classifies the financial instruments involved in the disputed transactions, providing a proxy for the specific information environment and monitoring costs associated with different asset classes. The distribution reveals a significant concentration in Mutual Funds/ETFs (24\%) and Annuities (22\%), alongside Stocks (17\%) and Insurance products (9\%). In our empirical framework, we utilize these categories to evaluate the complexity-induced detection gap, as different products possess varying degrees of transparency and ex-post verifiability. For instance, complex instruments like annuities often feature shrouded attributes that may facilitate strategic masking by advisors. By controlling for these product types in our saturated fixed-effects models, we ensure that our analysis of detection latency accounts for the intrinsic opacity of the financial instrument, allowing us to further disentangle the supply-side and demand-side forces governing the investor-led enforcement channel.
%----------------------------------------------------------------

%----------------------------------------------------------------
% Panel E
Panel F
%----------------------------------------------------------------
% presents the supply-side characteristics used to test our strategic masking and learning-by-doing hypotheses. The sample represents a cohort of highly experienced professionals, with an average of 19.28 years in the industry. We also document a mean of 3.77 Prior Customer Disputes per advisor. 
% % These variables are central to our quantile regression analysis, where we demonstrate that experience and prior records do not exert a uniform impact. Instead, we find that veteran offenders leverage their previous experience with the regulatory process to pacify investors or explain away early red flags, effectively hiding in the tail of the distribution beyond the reach of traditional monitoring.
% We utilize these attributes as proxies for an advisor's institutional knowledge and their capacity to manage information flows. These variables are critical for identifying whether senior advisors exhibit different detection lag patterns than their less experienced peers, even when operating under identical firm-level compliance cultures.
%----------------------------------------------------------------
presents the supply-side attributes used to test our strategic masking and learning-by-doing hypotheses. 
The advisors in our sample are highly experienced, with a mean career longevity of 19.28 years. Notably, the sample exhibits a high prevalence of repeated offenders: the mean number of prior customer disputes is 3.77, and even the median advisor has one prior record. This is consistent with the findings of \citet{egan2019market}, who document that a significant portion of misconduct is concentrated among a small group of repeat offenders. 
%----------------------------------------------------------------
% Panel F summarizes firm-level characteristics, most notably firm size (Number of Employees). 
% We also utilize firm size (Number of Employees) to test a hypothesis that larger institutions possess superior internal monitoring systems that should accelerate the detection of misconduct. 
% However, as our subsequent analysis reveals, the detection lag is relatively insensitive to firm scale, suggesting that internal tripwires may be less effective at unmasking strategic agents than external, state-dependent catalysts like market volatility.
%----------------------------------------------------------------

%----------------------------------------------------------------
We utilize these histories to examine whether veteran offenders leverage their previous regulatory experience to pacify investors and effectively hide in the tail of the detection distribution.
Additionally, we utilize firm size (Number of Employees) to 
% test the 'Tripwire Hypothesis': the theory that 
investigate whether
larger institutions possess superior internal monitoring and compliance infrastructures that should accelerate discovery. 
% As our subsequent analysis reveals, the relative insensitivity of lags to firm scale suggests that internal tripwires may be less effective at unmasking strategic agents than external, market-wide catalysts
%----------------------------------------------------------------

%----------------------------------------------------------------
% Panel F
Panel G
%----------------------------------------------------------------
% summarizes the professional human capital of the advisors, which we treat as a proxy for the ability to manage information flows and shroud misconduct. On average, advisors in our sample have passed 3.56 Exams, with 
% % 85\% holding a Series 7 license and 
% 14\% Series 24 license,
% 79\% holding a Series 63 license,
% and
% 71\% registered as investment advisers (Series 65/66). 
% In our framework, these qualifications represent the strategic human capital that allows sophisticated agents to navigate internal compliance triggers and maintain information asymmetry. These variables serve as critical controls in our fixed-effects specifications to ensure that our results regarding detection latency are not merely capturing unobserved professional quality or licensing requirements.
%----------------------------------------------------------------
summarizes the professional licenses and examinations that define the regulatory and supervisory status of the advisors. We specifically highlight the Series 24 (General Securities Principal), held by 14\% of the advisors in our sample, which qualifies an individual to manage or supervise a member firm's investment banking or securities business. Additionally, we include the Series 63 (Uniform Securities Agent State Law), a state-level requirement held by 79\% of the advisors to conduct business, and the Series 65/66, held by 71\% of the sample, which qualifies individuals to act as Investment Adviser Representatives (IA). These qualifications serve as indicators of professional sophistication and are included as controls to ensure that our results regarding detection latency are not confounded by an advisor's specific regulatory permissions or supervisory roles.
%----------------------------------------------------------------

%----------------------------------------------------------------
% Panel G
Panel H
%----------------------------------------------------------------
provides the demographic composition of the sample, which is overwhelmingly Male (84\%). 
While gender is primarily utilized as a control variable in our saturated models, its inclusion is motivated 
% by emerging research on how agent identity shapes blame attribution in financial services. 
by recent study on how clients attribute blame asymmetrically based on advisor gender \citep{abel2024women}.
% clients attribute blame asymmetrically based on advisor gender or advisory mode \citep{abel2024women, ismagilova2025ain}
By absorbing these demographic characteristics alongside firm, state, and year fixed effects , we isolate the structural and informational drivers of the detection gap, ensuring that our findings regarding state-contingent monitoring and strategic masking are robust to the individual identity of the agent.
%----------------------------------------------------------------

%----------------------------------------------------------------
In the primary analysis, all continuous variables subject to potential outlier influence, such as detection lags, advisor industry experience, and allegation damages, are winsorized at the 1st and 99th percentiles to ensure that extreme observations do not disproportionately drive the estimated coefficients. This preparation directly supports our subsequent investigation into how advisor sophistication and misconduct complexity drive the observed duration of nondisclosure.
% For our empirical analysis, 
We apply a natural log transformation to these continuous variables to account for right-skewness and to allow for the estimation of elasticities. 
%----------------------------------------------------------------

% Explaining the Mean Shift (Winsorization & Imputation)
% The raw settlement amount exhibits extreme right-skewness (skewness = 39.1), with a maximum reported value of \$135 million. To prevent a small number of mega-settlements from disproportionately driving the empirical results, we winsorize continuous variables at the 1st and 99th percentiles. Furthermore, we impute a value of zero for settled or pending disputes where monetary damages were missing, as these often reflect non-monetary resolutions or confidential zero-dollar baseline settlements. Together, the winsorization of extreme right-tail outliers and the imputation of true zero-dollar settlements reduce the sample mean from roughly \$220,000 to \$134,000. This adjusted mean provides a much more accurate representation of the typical economic penalty in a standard misconduct dispute.

%----------------------------------------------------------------
\subsection{Limitations}
% \subsection{Scope and Boundary Conditions}
\label{section: data 5}
%----------------------------------------------------------------

Despite its unique advantages, the FINRA BrokerCheck dataset has several limitations that are important for interpreting our results.

%----------------------------------------------------------------
\paragraph{Selective Observability
and Industry Persistence.%
}
%----------------------------------------------------------------
% First, there is \emph{selective observability of misconduct and disputes}.
% BrokerCheck records only misconduct that is formally reported through customer complaints, arbitrations, litigations, or regulatory actions.
% If a client never recognizes that they were harmed, or chooses not to pursue a formal complaint, no dispute will appear in the data.
% Our detection-lag measures are therefore conditional on the subset of cases in which investors escalate concerns into a formal process.
% As in the broader corporate misconduct literature \citep[e.g.][]{dyck2010blows}, focusing on detected cases likely understates the true incidence of misconduct and skews attention toward more salient or severe episodes.
%----------------------------------------------------------------
% BrokerCheck records only misconduct that is formally reported through customer complaints, arbitrations, or regulatory actions. Consequently, our measures are conditional on the subset of cases in which investors recognize harm and escalate concerns. As noted in the corporate misconduct literature, focusing on detected cases likely underrepresents the total incidence of misconduct and skews attention toward more salient episodes.
%----------------------------------------------------------------
First, there is selective observability of misconduct and disputes. BrokerCheck records only misconduct that is formally reported through customer complaints, arbitrations, litigations, or regulatory actions. If a client never recognizes they were harmed, or chooses not to pursue a formal complaint, no dispute will appear in the data. Consequently, our documented average detection lag of 28.5 months represents a conservative lower bound for the industry. 
This selective observability has profound implications for the advisory labor market. 
If ``bad types" leverage strategic masking to stay undetected for years, they remain active in the professional pool. This unobserved persistence effectively erodes the overall quality and trust of the market, potentially leading to a version of Gresham's Law where undetected offenders displace high-quality advisors who cannot compete with the distorted returns of masked misconduct.
%----------------------------------------------------------------

%----------------------------------------------------------------
As in the broader corporate misconduct literature \citep[e.g.][]{dyck2010blows}, focusing on detected cases likely understates the true incidence of misconduct and skews attention toward more salient or severe episodes. We interpret our results as establishing novel stylized facts regarding the formal investor-led enforcement channel, recognizing that the true latency of financial harm in the population may be even more protracted than documented here.
%----------------------------------------------------------------

%----------------------------------------------------------------
\paragraph{Regulatory Boundaries and Right-Censoring.}
%----------------------------------------------------------------
Second, our sample is subject to \emph{survivorship and regulatory-boundary biases}.
We observe disputes only while advisers are registered with FINRA member firms.
Advisers who engage in misconduct may be terminated, barred, or voluntarily leave FINRA, in which case subsequent complaints related to pre-exit advice may be reported differently or not at all.
Moreover, recent evidence from the insurance market indicates that advisers with BrokerCheck misconduct records often remain active as insurance agents outside the FINRA perimeter.
In our framework, we therefore treat adviser exit from FINRA as a right-censoring event within the BrokerCheck data, and we interpret our detection-lag distribution as describing the timing of formal complaints \emph{within the FINRA brokerage segment}, not in the entire financial advice industry.
This perspective is closely related to survivorship concerns raised by \citet{clifford2021property} in their analysis of adviser incentives and client relationships.
%----------------------------------------------------------------
% We observe disputes only while advisors are registered with FINRA member firms. If an advisor exits the FINRA perimeter—for example, by moving into the insurance industry—we treat this as a right-censoring event. Our distribution of detection lags thus characterizes the timing of formal complaints specifically within the FINRA brokerage segment, consistent with the survivorship concerns discussed by Clifford and Gerken (2021).
%----------------------------------------------------------------
% As a robustness check, we compare detection lags for advisers who remain in FINRA for at least five years after the purchase date to those who exit sooner and find similar patterns.

% Because BrokerCheck’s public availability depends on registration status and FINRA’s display rules, historical coverage could be incomplete; however, customer-dispute disclosures exhibit high record-level persistence across our snapshots, including for older cohorts (Appendix Table A1).

%----------------------------------------------------------------
\paragraph{Reporting and Expungement Biases.}
%----------------------------------------------------------------
Third, there are \emph{reporting and expungement biases}.
Brokers can petition to have certain customer dispute records expunged from BrokerCheck, subject to arbitration or court approval.%
\footnote{FINRA ``Expungement of Customer Dispute Information" (\url{https://www.finra.org/rules-guidance/key-topics/expungement-of-dispute-information}, Accessed: Jan. 14, 2026).}
\citet{honigsberg2021deleting} show that expungement requests are frequent and often successful, implying that our BrokerCheck-based datasets understate the true number of past disputes and that missing events are unlikely to be random.
Because our data are based on multiple datasets of BrokerCheck, some disputes may have been removed prior to or between datasets.
We partially mitigate this concern by using the union of events across datasets and by showing that our main results are robust to restricting the sample to events that appear consistently across datasets, but we cannot fully correct for selective expungement.
%----------------------------------------------------------------
% Brokers may petition to have certain customer dispute records expunged from the database. Research by Honigsberg and Jacob (2021) suggests these requests are frequent, implying that BrokerCheck-based datasets may understate the true number of past disputes. While we utilize the union of events across multiple datasets to mitigate this, we recognize that selective expungement remains an unobserved source of attrition.
%----------------------------------------------------------------

%----------------------------------------------------------------
\paragraph{Measurement Precision.}
%----------------------------------------------------------------
Fourth, our detection-lag measure is subject to date imprecision. Purchase timing is sometimes reported only as a range, and narrative descriptions may highlight a subset of transactions rather than the first relevant interaction. We therefore code the transaction date as the earliest date in the reported purchase range. This choice is conservative: if the true relevant conduct predates the earliest reported date, then our measured detection lags understate the true lags. If this date noise is not systematically related to our covariates of interest, it primarily adds measurement error to the dependent variable and reduces statistical power, making estimated relationships between adviser characteristics and detection lags harder to detect (i.e., tending toward attenuation). 
%----------------------------------------------------------------
% Detection-lag measures are subject to inherent date imprecision, as transaction timing is sometimes reported as a range. By coding the transaction date as the earliest date in a reported range, we ensure our measure is conservative. This approach provides an upper bound on the duration of potential harm and minimizes the risk of artificially truncating the systemic detection gap.
%----------------------------------------------------------------

% %----------------------------------------------------------------
% \paragraph{Clientele Heterogeneity.}
% %----------------------------------------------------------------
% Fifth, 
% one limitation of the current results is the lack of specific financial product identifiers (e.g., REITs or Variable Annuities). While our allegation-type controls capture the economic essence of the misconduct, future research (a forthcoming version of this work) will incorporate granular product data to examine if certain asset classes are inherently more prone to detection delays.
% %----------------------------------------------------------------
% %----------------------------------------------------------------

%----------------------------------------------------------------
\paragraph{Clientele Heterogeneity.}
%----------------------------------------------------------------
% Finally,
The dataset is \emph{adviser-centric rather than investor-centric}.
We do not observe detailed investor demographics, portfolio holdings, or direct measures of investor sophistication.
We can partially address this limitation by using local demographic information at the county level, but we cannot, for example, distinguish within an adviser’s book between sophisticated and unsophisticated clients.
Accordingly, our results should be interpreted as describing investor responses averaged over the clientele served by each adviser, rather than as precise estimates of individual-level behavioral parameters.
%----------------------------------------------------------------
% While our dataset is advisor-centric, we utilize local demographic information at the county level to proxy for the investor environment. We acknowledge that we cannot distinguish between sophisticated and unsophisticated clients within an individual advisor's book. Accordingly, our results describe investor-led enforcement responses averaged over the clientele served by each advisor rather than precise individual-level behavioral parameters.
%----------------------------------------------------------------
% Taken together, these limitations imply that our findings speak to the distribution of formal detection lags for retail investor complaints in the FINRA-regulated brokerage segment, conditional on observed disputes within BrokerCheck. They do not capture undisputed misconduct, detection delays that never translate into a complaint, or disputes that arise solely in non-FINRA channels such as insurance-only distribution.

\

%----------------------------------------------------------------
% Finally, we acknowledge that our measures are conditional on detection. Because BrokerCheck only records formal complaints, our data does not capture the dark figure of misconduct that is never recognized by the investor or never reported. We interpret our results as establishments of novel stylized facts regarding the formal investor-led enforcement channel, recognizing that the true latency of financial harm in the population may be even more protracted than documented here.
%----------------------------------------------------------------

%----------------------------------------------------------------
While the limitations of the BrokerCheck dataset given above are intrinsic to any study of detected misconduct, our data preparation ensures that the resulting sample of approximately 56,000 disputes is robust enough for high-dimensional analysis. The significant right-skewness of the detection lag distribution, where the mean of 49.3 months substantially exceeds the median of 34 months, suggests that a traditional mean-based analysis may fail to capture the strategic dynamics occurring at the tail of the distribution. 
Consequently, the empirical challenge is twofold: 
first, to objectively select from the vast array of advisor- and firm-level covariates described in Table \ref{table: ss 1};   
and second, to employ an estimator capable of capturing the heterogeneous impact of market conditions across the entire detection spectrum. These challenges necessitate the specification-dependent model selection and distributional frameworks detailed in the following section.
\section{Empirical Strategy}
%----------------------------------------------------------------

%----------------------------------------------------------------
Our empirical strategy is designed to handle high-dimensional unobserved heterogeneity and objective covariate selection within a massive panel of approximately 55,700 unique customer disputes. Because our findings are sensitive to the variation remaining after accounting for specific institutional and temporal factors, we employ a specification-dependent model selection procedure. This approach ensures that the control variables for each model are specifically chosen to address the idiosyncratic variation remaining after accounting for state, year, or firm-level fixed effects.
%----------------------------------------------------------------

%----------------------------------------------------------------
% \subsection{General Selection Framework}
\subsection{Specification-Dependent Model Selection}
%----------------------------------------------------------------

%----------------------------------------------------------------
We utilize a two-stage approach. In the first stage, we use machine learning to discipline the high-dimensional pool of advisor histories, professional qualifications, and allegation characteristics. In the second stage, we estimate both the mean effect and the distributional impact of our variables of interest.
%----------------------------------------------------------------

%----------------------------------------------------------------
Following the Post-Double-Selection (PDS) framework of \citet{belloni2014inference}, we perform model selection separately for each fixed effect structure. This is critical because a control variable that is highly predictive in a model with only state fixed effects may become redundant once firm fixed effects are introduced. For each of our three primary specifications, we follow a three-step process:
%----------------------------------------------------------------
\begin{itemize}
%----------------------------------------------------------------
\item Partialling Out (FWL Transformation): We residualize the dependent variable ($Y$), the treatment ($D = \ln\text{VIX}$), and the potential controls ($\mathbf{X}$) by the specific fixed effects of that model using the Frisch-Waugh-Lovell theorem.
\item Dual Lasso Selection: We apply the Plugin Lasso \citep{belloni2012sparse} to the residuals to select (i) predictors of the detection lag and (ii) predictors of market volatility exposure. This rigorous Lasso utilizes a data-driven penalty robust to the clustered error structures at the firm level.
\item Union of Controls: We 
% define $\mathbf{Z}^{(m)}$ as 
use
the union of variables selected specifically for that fixed-effect regime. % $m$.
%----------------------------------------------------------------
\end{itemize}
%----------------------------------------------------------------

%----------------------------------------------------------------
\subsection{Nested Empirical Specifications}
% \subsection{Baseline Specifications}
%----------------------------------------------------------------
We estimate our results across three increasingly rigorous specifications to demonstrate the robustness of our findings to different levels of unobserved heterogeneity.
%----------------------------------------------------------------

%----------------------------------------------------------------
\paragraph{Specification (i): Baseline (No Fixed Effects).}
%----------------------------------------------------------------
In this baseline, we do not absorb fixed effects. The Lasso selects from the full spectrum of demographic and professional variables, including gender, experience, and specific exam histories, to account for cross-sectional variation.

%----------------------------------------------------------------
\paragraph{Specification (ii): Two-Way Fixed Effects (State \& Year).}
%----------------------------------------------------------------
We introduce state ($\gamma_s$) and year ($\delta_t$) fixed effects to account for regional regulatory differences and global secular trends. The Lasso selection focuses on advisor-specific variation not captured by state-level averages or time trends, such as individual learning-by-doing through prior disputes.
%----------------------------------------------------------------
\paragraph{Specification (iii): Three-Way Fixed Effects (Firm, State, \& Year).}
%----------------------------------------------------------------
Our most rigorous specification adds firm fixed effects ($\alpha_j$) to control for all time-invariant firm characteristics, such as internal compliance culture or organizational complexity. Here, the Lasso selects only time-varying characteristics that predict detection lags.
%----------------------------------------------------------------

%----------------------------------------------------------------
\subsection{Distributional Analysis and Treatment Identification}
%----------------------------------------------------------------

%----------------------------------------------------------------
To investigate our hypotheses, we integrate our model selection results into a distributional framework. This stage simultaneously addresses the heterogeneity of detection lags and the causal identification of our primary treatment variable, $\ln\text{VIX}$.
%----------------------------------------------------------------

%----------------------------------------------------------------
\paragraph{The Method: Method of Moments Quantile Regression (MMQR).}
%----------------------------------------------------------------
Standard mean-based regressions (OLS) assume that the impact of market volatility is uniform across all misconduct episodes. However, as documented in Section \ref{section: summary statistics}, the distribution of detection lags is heavily right-skewed. This skewness suggests that the most significant behavioral frictions occur in the upper tail of the distribution.

We therefore employ the Method of Moments Quantile Regression (MMQR) developed by \citet{machado2019quantiles}. The MMQR is uniquely suited for our study because it allows for the inclusion of high-dimensional fixed effects (firm, state, and year)  while permitting these effects to influence both the location and the scale of the detection lag distribution. Unlike traditional quantile estimators, the MMQR provides consistent estimates in the presence of unobserved heterogeneity without suffering from the incidental parameter problem.
\subsection{Identification: VIX as an Exogenous Shifter}
%----------------------------------------------------------------
A central concern in our analysis is the potential endogeneity of market volatility with respect to individual advisor misconduct. To interpret the coefficient on $\ln(\text{VIX})$ as the impact of market-wide information frictions on detection lags, we rely on the following identification arguments:
%----------------------------------------------------------------

%----------------------------------------------------------------
\paragraph{Aggregate Nature of the Volatility Treatment.}
%----------------------------------------------------------------
The VIX index measures the market's expectation of 30-day volatility based on S\&P 500 index options. As an aggregate market-wide metric, it is external to the idiosyncratic behavior of any individual financial advisor or the detection efforts of a single retail investor. It is highly improbable that the discovery of misconduct or the strategic masking efforts in our sample, which comprises approximately 
% 56,000 
55,700
unique disputes, could mechanically influence the aggregate VIX.
%----------------------------------------------------------------

%----------------------------------------------------------------
\paragraph{Orthogonality through High-dimensional Fixed Effects.}
%----------------------------------------------------------------
Our identification relies on the variation in $\ln(\text{VIX})$ that is idiosyncratic to the timing of the misconduct and the subsequent monitoring window. By including three-way high-dimensional fixed effects (Firm, State, and Year), we control for:
(i) Year Effects: These absorb global secular trends and average annual volatility, ensuring our identification comes from within-year fluctuations in VIX levels; 
(ii) Firm and State Effects: These control for time-invariant differences in internal compliance environments and local regulatory stringency that might otherwise be correlated with specific market regimes.
%----------------------------------------------------------------

%----------------------------------------------------------------
\paragraph{Controlling for Demand-Side and Supply-Side Confounders.}
%----------------------------------------------------------------
One might argue that market volatility is correlated with investor sentiment or changes in advisor learning-by-doing. We address this through our Post-Double-Selection Lasso procedure. By selecting from a vast pool of covariates, including advisor experience (mean 19.28 years), previous dispute history (mean 3.77 disputes), and specific professional qualifications like Series 24 or 65/66, we ensure that our treatment effect for $\ln(\text{VIX})$ is orthogonal to the observable factors that drive both the timing of misconduct and the propensity of investors to monitor.
%----------------------------------------------------------------

%----------------------------------------------------------------
% \subsection{Timing of Covariates and Model Specification}
\subsection{Covariate Measurement and Look-Ahead Bias}
% \paragraph{Timing of Covariates and Model Specification.}
%----------------------------------------------------------------
In our baseline specifications, we evaluate adviser characteristics, firm affiliations, and macroeconomic fixed effects at the time the customer complaint is formally filed (the claim date). This approach effectively controls for the revelation environment including the specific market conditions, institutional compliance structures, and local regulatory scrutiny present when the hidden misconduct is finally exposed. For instance, evaluating year fixed effects at the time of the claim absorbs macroeconomic shocks, such as broad market downturns, that frequently act as catalysts for investors to scrutinize their portfolios and detect past harm.
%----------------------------------------------------------------

%----------------------------------------------------------------
However, evaluating covariates exclusively at the end of the detection lag introduces important limitations regarding look-ahead bias and endogeneity, particularly when testing our masking hypothesis. Because detection lags can span multiple years, an adviser's human capital and institutional environment may evolve significantly between the time the illicit transaction is structured and the time the investor files a dispute. Consequently, measuring adviser experience or prior dispute records at the time of the claim risks conflating the adviser's ex-ante sophistication with ex-post realizations. If an adviser accumulates additional experience, passes new licensing exams, or accrues subsequent customer complaints during the detection window, these variables cannot logically explain the initial decision to engage in misconduct or the original technology used to obscure it.
%----------------------------------------------------------------

%----------------------------------------------------------------
Furthermore, relying on firm and state fixed effects evaluated at the claim date assumes static labor mobility. If an adviser originates a problematic transaction at a small, poorly monitored firm but subsequently transitions to a larger firm with strict compliance oversight before the complaint is filed, a claim-date specification incorrectly attributes the origins of the misconduct to the latter institution. 
%----------------------------------------------------------------
% To strictly identify the agency conflicts and masking technologies that facilitate financial misconduct, the institutional incentives and adviser sophistication must ideally be measured at the point of origination. Therefore, to ensure our estimates are not driven by post-transaction labor transitions or ex-post human capital accumulation, we supplement our baseline with specifications that evaluate all core covariates and fixed effects at the time of transaction initiation.
% (see the Appendix).
%----------------------------------------------------------------

%----------------------------------------------------------------
While our current empirical design powerfully isolates the demand-side frictions present at the exact moment of discovery, fully disentangling the supply-side masking technologies requires measuring these incentives at the point of origination. Constructing a dynamic, historical panel that precisely maps adviser affiliations, human capital, and prior disclosures at the exact moment of transaction initiation remains the primary focus of our ongoing future work for this research agenda.
\subsection{Computational Implementation}
%----------------------------------------------------------------
All estimations were conducted in Stata 18. High-dimensional fixed effects were absorbed using the reghdfe package \citep{correia2017reghdfe}. Model selection via the Post-Double-Selection Lasso was implemented using the lassopack suite, specifically the 
% pdslasso and 
rlasso commands \citep{ahrens2020lassopack}. The distributional impacts across conditional quantiles were estimated using the mmqreg command \citep{riosavila2020mmqreg}.
%----------------------------------------------------------------

%----------------------------------------------------------------

%----------------------------------------------------------------
% Result:
%----------------------------------------------------------------

%----------------------------------------------------------------
%----------------------------------------------------------------
\section{Main Findings}
\label{section: result 0}
%----------------------------------------------------------------

%----------------------------------------------------------------
% In this section, we document two central facts regarding investor responses to financial adviser misconduct. First, we show that detection times vary markedly across cases. Second, we find that investor responses are sensitive to market conditions, with a marked surge during periods of heightened market volatility.
%----------------------------------------------------------------
% This section presents our primary empirical findings regarding retail investor responses to financial adviser misconduct. In particular, we document (i) substantial and heterogeneous detection lags and (ii) a systematic shortening of these lags during periods of heightened market volatility.
%----------------------------------------------------------------
% This section presents our primary empirical findings regarding retail investor responses to financial adviser misconduct. We focus on two key dimensions: (i) the heterogeneity in detection lags, and (ii) the responsiveness of these lags to market conditions.
%----------------------------------------------------------------
% We document three primary findings regarding the dynamics of how retail investors detect and respond to financial advisor misconduct.
We document three primary findings regarding the dynamics of the investor-led enforcement channel.
%----------------------------------------------------------------

%----------------------------------------------------------------

% \newpage 

%----------------------------------------------------------------
\subsection{%
% Magnitude and Persistence of Detection Lags%
% The Latency of Detection: A Systemic Monitoring Failure%
The Latency of Detection: Informational Barriers and Monitoring Frictions%
}
\label{section: result 1}
%----------------------------------------------------------------

%----------------------------------------------------------------
To investigate the structural drivers of the detection gap, we first estimate a log-linear specification that models the duration of nondisclosure 
% as a function of investor-facing misconduct characteristics and adviser-specific traits. 
as an endogenous outcome of strategic monitoring and masking efforts. 
The primary estimation equation is defined as follows:
%----------------------------------------------------------------
\begin{equation}
    % \ln(Lag)_{it} = \alpha + \beta_1 \ln(VIX)_{it} + \beta_2 \ln(Exp)_{it} + \mathbf{X}_{it}'\gamma + \phi_f + \eta_s + \delta_t + \epsilon_{it}
    % \ln(\text{Detection Lag})_{ijlt} = \alpha + \mathbf{X}_{it}'\gamma + \phi_j + \eta_l + \delta_t + \varepsilon_{ijlt}
    \ln(\text{Detection Lag})_{ijst} = \alpha + \beta \mathbf{X}_{it} + \phi_j + \eta_s + \delta_t + \varepsilon_{ijst}
\end{equation}
%----------------------------------------------------------------
where $\ln(\text{Detection Lag})_{ijst}$ is the log-transformed detection lag for advisor $i$ at firm $j$, state $s$, and time $t$. The vector $X_{it}$ includes a set of controls selected via a principled variable selection procedure to balance model parsimony with the need to capture the full spectrum of claim complexity. We saturate the model with firm ($\phi_{j}$), state ($\eta_{s}$), and year ($\delta_{t}$) fixed effects 
to account for unobserved heterogeneity in institutional compliance cultures, state-level regulatory environments, and time-varying macro factors. 
%----------------------------------------------------------------% Standard errors are clustered at the firm level to provide a conservative account of the shared institutional environment that influences reporting behavior.
This high-dimensional fixed-effects approach ensures that our estimated coefficients are derived from within-firm variation, effectively comparing the discovery of misconduct for advisors operating under identical supervisory constraints. 
Standard errors are clustered at the firm level to provide a conservative account of the shared institutional environment that influences reporting behavior.
%----------------------------------------------------------------

%----------------------------------------------------------------
Table \ref{table:linear_reg_w1} 
%----------------------------------------------------------------
reports 
% the baseline OLS results using a step-wise fixed effects approach to ensure the robustness of our determinants. 
the results using a step-wise fixed effects approach. 
% To focus on the core mechanisms of informational frictions and strategic agency, we report only the variables of interest here, moving the full set of controls to the Appendix.
% Column (1) 
% begins with a pooled OLS specification, which we progressively saturate with Year, State, and Firm fixed effects.
Across all specifications, our model explains a significant portion of the variance in detection lags; the $R^2$ increases from 0.272 in the pooled OLS (Column 1) to 0.358 in our most saturated model (Column 3). %, indicating that nearly 36\% of the variation in the timing of discovery is captured by these structural and institutional determinants.
This indicates that nearly 36\% of the variation in the window of nondisclosure is captured by structural, institutional, and advisor-specific determinants.
%----------------------------------------------------------------

%----------------------------------------------------------------
%----------------------------------------------------------------
% Regression Table: Multi-Page Setup (Winsorized Results)
%----------------------------------------------------------------
% CHANGE THIS COMMAND to \small, \scriptsize, or \tiny \footnotesize
% \scriptsize  % <--- This controls the size of the whole environment
% \begin{landscape} % <--- STARTS LANDSCAPE MODE
\footnotesize  
\begin{ThreePartTable}
    \begin{TableNotes}
        \footnotesize
        \item \hspace{-0.1in} \textit{Notes:} This table reports the results of linear regressions examining the determinants of detection lags (in months) for a sample of approximately 
        55,000 
        customer dispute reports. 
        % \item 
        The dependent variable, Detection Lag, is defined as the number of months between the inception of the alleged misconduct and its formal report to FINRA. 
        % \item 
        Column (1) provides baseline estimates, while Columns (2) and (3) add year, state, and firm fixed effects. 
        % \item 
        % Standard errors are clustered at the firm level to account for shared institutional environments. 
        % Continuous variables are winsorized at the 1st and 99th percentiles.
        Continuous variables, including advisor experience, tenure, and firm size, are winsorized at the 1st and 99th percentiles to mitigate the influence of outliers. 
        Note that logarithmic transformations of continuous variables (such as Experience, Firm Tenure, and Damages) are calculated as $Log(1 + x)$ to retain zero-value observations.
        Variables are defined in Appendix Table \ref{table:var_defs_long}. 

        \item Standard errors are clustered at the firm level, and reported in parentheses. ***, **, and * denote significance at the 1\%, 5\%, and 10\% levels, respectively.
    \end{TableNotes}
    
    % --- ADD THE CODE HERE ---
    \setlength{\tabcolsep}{15pt} 
    % -------------------------
    
    \begin{longtable}{lcccc}
    	
        \caption{Determinants of Detection Lags} \label{table:linear_reg_w1} \\
        \toprule
        % & (1) & (2) & (3) \\
        % \midrule
        \endfirsthead

        \multicolumn{5}{l}{\textit{Table \ref{table:linear_reg_w1} continued from previous page}} \\
        \toprule
        & (1) & (2) & (3) \\
        \midrule
        \endhead

        \midrule
        \multicolumn{5}{r}{\textit{Continued on next page}} \\
        \endfoot

        \bottomrule
        \insertTableNotes
        \endlastfoot

        % Use \input if your .tex file only contains the body rows. 
        % If your .tex file contains \begin{tabular}, you must copy/paste the rows here manually.
        % \input{table/t_linear_w1.tex} 
        % \input{table/Phase1_Baseline_w1.tex} 
                        &\multicolumn{1}{c}{(1)}         &\multicolumn{1}{c}{(2)}         &\multicolumn{1}{c}{(3)}         \\
\midrule
Dispute Type: Pending&    0.211\sym{***}&    0.221\sym{***}&    0.211\sym{***}\\
                &  (0.047)         &  (0.045)         &  (0.040)         \\
\addlinespace
Unsuitability   &    0.365\sym{***}&    0.324\sym{***}&    0.303\sym{***}\\
                &  (0.050)         &  (0.033)         &  (0.035)         \\
\addlinespace
Misrepresentation&    0.186\sym{***}&    0.193\sym{***}&    0.186\sym{***}\\
                &  (0.025)         &  (0.020)         &  (0.018)         \\
\addlinespace
Unauthorized Activity&   -0.466\sym{***}&   -0.472\sym{***}&   -0.442\sym{***}\\
                &  (0.056)         &  (0.042)         &  (0.041)         \\
\addlinespace
Fraud           &    0.503\sym{***}&    0.483\sym{***}&    0.461\sym{***}\\
                &  (0.040)         &  (0.043)         &  (0.045)         \\
\addlinespace
Churning        &    0.357\sym{***}&    0.372\sym{***}&                  \\
                &  (0.042)         &  (0.034)         &                  \\
\addlinespace
Log(Experience) &    0.423\sym{***}&    0.445\sym{***}&    0.365\sym{***}\\
                &  (0.031)         &  (0.037)         &  (0.027)         \\
\addlinespace
Log(Number of Employees)&   -0.051\sym{***}&   -0.051\sym{***}&                  \\
                &  (0.014)         &  (0.013)         &                  \\
\addlinespace
Log(Number of Prior Customer Disputes)&    0.188\sym{***}&                  &                  \\
                &  (0.013)         &                  &                  \\
\addlinespace
Log(Alleged Damages)&    0.052\sym{***}&    0.049\sym{***}&    0.055\sym{***}\\
                &  (0.009)         &  (0.006)         &  (0.006)         \\
\addlinespace
Insurance       &    0.702\sym{***}&    0.646\sym{***}&    0.568\sym{***}\\
                &  (0.056)         &  (0.050)         &  (0.034)         \\
\addlinespace
Concurrent Multiple Jobs (Indicator)&                  &   -0.191\sym{***}&                  \\
                &                  &  (0.023)         &                  \\
\addlinespace
Series 63       &                  &   -0.055\sym{***}&   -0.022         \\
                &                  &  (0.019)         &  (0.020)         \\
\addlinespace
Constant        &    1.985\sym{***}&    2.212\sym{***}&    1.923\sym{***}\\
                &  (0.116)         &  (0.105)         &  (0.062)         \\
\midrule
Lasso-selected Controls         &       Yes         &      Yes         &      Yes         \\
Year FE         &       No         &      Yes         &      Yes         \\
State FE        &       No         &      Yes         &      Yes         \\
Firm FE         &       No         &       No         &      Yes         \\
Observations    &   54,676         &   54,658         &   54,223         \\
\(R^2\)         &    0.272         &    0.309         &    0.358         \\
Within \(R^2\)  &    0.272         &    0.206         &    0.151         \\
Mean Dep. Var.  &    3.337         &    3.337         &    3.335

    \end{longtable}
\end{ThreePartTable}
% \end{landscape} % <--- ENDS LANDSCAPE MODE
% & (1) & (2) & (3) & (4) \\
        
%----------------------------------------------------------------

%----------------------------------------------------------------
Because we employ a rigorous Lasso penalty for model selection, the exact set of selected control variables varies across columns. This variation is a mechanical and intended feature of the algorithm. In specifications lacking fixed effects, the Lasso selects a broader set of covariates to proxy for unobserved heterogeneity. As fixed effects are introduced, they absorb specific dimensions of variation, rendering some previously selected covariates redundant. Consequently, the algorithm sets the coefficients of these redundant variables to zero, ensuring that the selected controls are optimally and parsimoniously tailored to the residual variation in each specific model.
%----------------------------------------------------------------

%----------------------------------------------------------------
% \paragraph{The Demand-Side Perspective: Complexity-Induced Detection Gap.}
\paragraph{The Complexity-Induced Detection Gap.}
%----------------------------------------------------------------
Our findings provide strong empirical evidence for significant information frictions in investor monitoring.  We find that the nature of the misconduct is the most substantial driver of detection speed. 
% Overt violations such as unauthorized activity significantly reduce the detection lag by approximately 44.2\%. In contrast, sophisticated schemes such as fraud and churning extend the lag by 39.0\% and 40.3\% respectively. 
In our most rigorous specification (Column 3), overt violations such as unauthorized activity significantly reduce the detection lag by approximately 44.2\%. Conversely, sophisticated schemes such as fraud extend the lag by 46.1\%. 
Relative to the sample mean of 
% $\ln(\text{Detection Lag})$ 
dependent variable
of 3.34, these coefficients represent a substantial shift in the window of nondisclosure. For instance, the fraud coefficient of 0.46 suggests that complex misconduct persists for nearly 1.5 times longer than the average infraction. This indicates that investors struggle most when the credence nature of financial advice is highest.
% Relative to the sample mean, the fraud coefficient suggests that complex misconduct remains latent for nearly 1.5 times longer than the average infraction, highlighting that investors struggle most when the credence nature of financial advice is highest.
%----------------------------------------------------------------
Our primary finding is that the nature of the misconduct is the most significant driver of detection speed. We document a profound complexity gap between overt and sophisticated violations. In our most rigorous specification (Column 3), overt violations such as Unauthorized Activity significantly reduce the detection lag by approximately 35.7\% ($e^{-0.442} - 1$). In stark contrast, sophisticated quiet crimes such as Fraud and Unsuitability extend the lag by 58.6\% ($e^{0.461} - 1$) and 35.4\% ($e^{0.303} - 1$), respectively.
These results provide strong empirical evidence for the presence of significant information frictions. When misconduct signals are overt, discovery is rapid. However, when the credence nature of financial advice is high, as in cases of fraud, the monetary cost of misconduct is an insufficient catalyst for detection, allowing the harm to persist for nearly 1.5 times longer than the sample average.
%----------------------------------------------------------------

%----------------------------------------------------------------
\paragraph{Strategic Masking 
and 
% Advisor Sophistication.}
Advisor Human Capital.}
%----------------------------------------------------------------
After accounting for allegation and product types (see Panel C of Table \ref{table: ss 1}), 
% we examine the supply-side drivers and find that advisor experience is a robust predictor of detection latency. 
we find that advisor experience is a robust and stable predictor of detection latency. 
In Specification 3, the coefficient on advisor experience is 0.365. This indicates that a 10\% increase in career longevity is associated with a 3.65\% increase in the duration misconduct remains undetected.
%----------------------------------------------------------------
Notably, this coefficient remains remarkably stable across the pooled and fixed-effect specifications (varying only from 0.423 to 0.365). %, suggesting that delayed detection is not merely an artifact of the firms where experienced advisors work, but is driven by individual-level strategic masking. 
%----------------------------------------------------------------

%----------------------------------------------------------------
We interpret this as evidence of a strategic masking mechanism. Consistent with the learned-by-doing hypothesis, these results suggest that veteran offenders leverage their sophistication and the resulting information asymmetry to 
% pacify investors and extend the window of undetected harm.
navigate internal compliance triggers and pacify investor monitoring.
%----------------------------------------------------------------
This suggests that advisor human capital is not merely a proxy for quality, but also a primary determinant of the detection gap.
%----------------------------------------------------------------

%----------------------------------------------------------------
\paragraph{The Role of Product Opacity.}% (Insurance).}
% %----------------------------------------------------------------
% We find a disproportionate role for specific product types and financial stakes in facilitating latency. 
A novel finding in our analysis is the disproportionate role of specific product types in facilitating nondisclosure.
Insurance-linked products emerge as the largest predictor of detection delay, with a coefficient of 0.568 in our saturated model. This indicates that Disputes involving insurance-linked instruments exhibit a 76.5\% ($e^{0.568} - 1$)increase in latency relative to other product categories. The structural complexity and shrouded attributes of these instruments create a potent informational shield that successfully delays state verification by retail investors.
%----------------------------------------------------------------

%----------------------------------------------------------------
\paragraph{The Salience-Complexity Trade-off.}
%----------------------------------------------------------------
%----------------------------------------------------------------
% Furthermore, we find 
We find
a counter-intuitive relationship regarding financial salience. Under a standard framework of rational inattention, larger losses should act as a signal that triggers faster discovery. However, our results show that Log(Alleged Damages) is positively associated with the detection lag (0.055). This suggests a scale-complexity correlation: high-value misconduct often necessitates more elaborate strategic masking, allowing the scheme to persist longer than smaller, more transparent infractions.
%----------------------------------------------------------------

%----------------------------------------------------------------
\paragraph{Information Shocks vs. Calculated Delay.}
%----------------------------------------------------------------
%----------------------------------------------------------------
The objective variable selection performed by the Plugin Lasso provides deeper insights into the behavioral mechanics of dispute filing. Notably, the algorithm drops the distinction between claims that are eventually Settled and those that are Denied (the reference group).
%----------------------------------------------------------------

%----------------------------------------------------------------
From a theoretical perspective, if investors act as perfectly rational agents who infer legal outcomes before claiming, we would expect a divergence in detection lags. Rational investors would systematically delay filing to gather airtight evidence, resulting in longer lags for meritorious (Settled) claims compared to reactive, unmeritorious (Denied) claims. The empirical absence of this divergence is highly revealing. It suggests that the timing of dispute filings is not driven by protracted, rational evidence-gathering or ex-ante legal calculations. Instead, the symmetric detection lags strongly support our core hypothesis: discovery is driven by abrupt, exogenous attention shocks (such as market volatility). When a systemic shock breaks the investor's rational inattention, it triggers an immediate filing response across the spectrum of both meritorious and unmeritorious claims alike.
%----------------------------------------------------------------

%----------------------------------------------------------------
Conversely, 
the algorithm consistently selects Pending status as a positive and significant driver of detection lags, and 
the significant positive coefficient on Pending disputes highlights the friction of structural opacity. 
The baseline linear estimates in 
%----------------------------------------------------------------
Table \ref{table:linear_reg_w1} 
%----------------------------------------------------------------
reveal a substantial detection premium for disputes that remain contested. Across all specifications, cases currently classified as Pending exhibit significantly longer detection lags than the reference group of resolved or dismissed claims. In our strictest specification with firm, state, and year fixed effects (Column 3), a pending status is associated with a 23.5\% ($e^{0.211} - 1 \approx 0.2349$) increase in the detection timeline. Evaluated at the sample mean, this translates to roughly 6.4 additional months of concealment. Economically, this magnitude underscores the friction of structural opacity: misconduct that successfully evades detection for extended periods is inherently complex, carrying significant compounded damages that severely reduce the likelihood of a rapid, uncontested settlement upon discovery.

Misconduct that successfully evades detection for long periods is inherently complex and carries significant compounded damages. Consequently, when these deeply buried violations are finally claimed, brokerage firms are highly likely to contest them rather than offer immediate settlements. The Pending status thus serves as an endogenous proxy for the severe informational and legal complexities that characterize the longest-running financial misconduct.
%----------------------------------------------------------------

%----------------------------------------------------------------
\paragraph{Model Selection and Robustness to Omitted Variable Bias.}
%----------------------------------------------------------------
%----------------------------------------------------------------
To ensure the robustness of our primary findings, we address potential econometric concerns regarding model specification and identification, particularly the risk of omitted variable bias arising from the high-dimensional covariate space. We employ a principled Post-Double-Selection (PDS) procedure using the Plugin (Rigorous) Lasso. Following the framework of \citet{belloni2012sparse,belloni2014inference}, this approach systematically identifies the most relevant predictors of both the detection lag and our primary treatment variables, including advisor experience, misconduct complexity, and insurance product type, from a vast pool of advisor and dispute characteristics. By performing our final estimations on the union of these selected controls, we ensure that our treatment effects are asymptotically orthogonal to the high-dimensional nuisance parameters, providing a rigorous foundation for valid inference that explicitly accounts for the model selection process.
%----------------------------------------------------------------

%----------------------------------------------------------------
The empirical validity of this strategy is evidenced by the remarkable stability of our primary coefficients across increasingly saturated specifications. As we move from a pooled OLS baseline to our most rigorous model involving over 900 firm fixed effects, the coefficients for advisor experience (0.365), fraud (0.461), and insurance (0.568) remain highly significant and stable in magnitude. This persistence suggests that the documented detection gaps are not artifacts of unobserved firm-level compliance cultures or arbitrary control choices, but rather reflect fundamental structural frictions in investor monitoring. 
Furthermore, the model’s explanatory power is substantial for a cross-sectional study of misconduct; the $R^2$ increases from 0.272 to 0.358 as fixed effects are introduced, indicating that our selected variables capture nearly 36\% of the total variation in detection latency. 
Collectively, these diagnostics confirm that our results reflect individual-level strategic behavior and product-specific opacity rather than institutional or geographic noise.
%----------------------------------------------------------------

%----------------------------------------------------------------
% \newpage

%----------------------------------------------------------------
\subsection{%
% Distributional Heterogeneity in Detection Lags
Distributional Heterogeneity: The Masking Gradient and Expanding Shields%
}
%----------------------------------------------------------------
While our baseline OLS results provide a conditional average of the determinants of detection, the significant right-skewness of the data, where the mean lag of 49.31 months substantially exceeds the median of 34 months, suggests that average effects may be misleading. 
%----------------------------------------------------------------
% To investigate the heterogeneous determinants of detection lags, we utilize the Method of Moments Quantile Regression (MM-QR) framework proposed by Machado and Silva (2019). This approach consistently handles high-dimensional fixed effects, allowing us to examine whether the structural drivers of nondisclosure vary from the 10th percentile (rapid discovery) to the 90th percentile (latent misconduct). The model is defined by the following conditional location-scale form:
%----------------------------------------------------------------
To investigate the heterogeneous drivers of these lags, we utilize the Method of Moments Quantile Regression (MM-QR) framework proposed by \citet{machado2019quantiles}. This approach is specifically designed to handle high-dimensional fixed effects consistently, allowing us to examine whether structural drivers vary from the 10th percentile (rapid discovery) to the 90th percentile (latent misconduct). The model is defined by the following conditional location-scale form for any quantile $\tau \in (0,1)$:
%----------------------------------------------------------------
\begin{equation}
% \ln(\text{Detection Lag})_{ijlt} = \alpha_{i} + X_{it}'\beta + \sigma(\delta_{i} + X_{it}'\gamma)U_{it}
% \ln(\text{Detection Lag})_{ijst} 
Q_{ln(Lag)_{ijst}}(\tau | X_{it}) 
= 
% \alpha_{j}(\tau) +\delta_{j}(\tau) +\eta_{t}(\tau) + \beta(\tau) \mathbf{X}_{it} + \sigma(\phi_{j}(\tau) +\psi_{j}(\tau) +\omega_{t}(\tau) + \gamma(\tau) \mathbf{X}_{it})q(\tau)
\alpha_{j} +\delta_{j} +\eta_{t} + \beta \mathbf{X}_{it} + \sigma(\phi_{j} +\psi_{j}  +\omega_{t} + \gamma \mathbf{X}_{it})q(\tau)
% U_{ijst}
\end{equation}
%----------------------------------------------------------------
% \ln(\text{Detection Lag})_{ijlt} = \alpha + \beta \mathbf{X}_{it} + \phi_j + \eta_l + \delta_t + \varepsilon_{ijlt}
%----------------------------------------------------------------
where 
% $\ln(\text{Detection Lag})_{ijst}$ is the log detection lag for adviser $i$ at firm $j$, state $s$, and time $t$.
% The vector $\mathbf{X}_{it}$ represents the covariates selected via the double-selection Lasso procedure to capture theoretical dimensions of complexity and sophistication. 
$\alpha_{j}, \eta_{s}, \delta_{t}$ represent the firm $j$, state $s$, and year $t$ fixed effects in the location equation.
$\phi_{j}, \psi_{s}, \omega_{t}$ represent the firm $j$, state $s$, and year $t$ fixed effects in the scale equation.
$\beta$ and $\gamma$ are the location and scale coefficients for the covariates $X_{it}$, respectively.
$q(\tau)$ is the $\tau$-th quantile of the estimated error distribution, which acts as the weights for the scale component.
% The total quantile effect is represented by $\theta(\tau) = \beta + \gamma q(\tau)$.
% The first bracketed term represents the location shift, capturing the conditional mean effects of the firm ($\alpha_{j}$), state ($\eta_{s}$), and year ($\delta_{t}$) fixed effects. The second bracketed term represents the scale shift, allowing the dispersion of the detection lag to vary with the same set of fixed effects and covariates. 
% The error term $U_{ijst}$ is assumed to be independent and identically distributed. 
The total quantile effect for a given percentile $\tau$ is then given by:
$\theta(\tau) = \beta + \gamma q(\tau)$.
%----------------------------------------------------------------

%----------------------------------------------------------------
Table 
% \ref{table:quantile_reg_w1} 
\ref{table:MMQR_1} 
presents 
% the baseline OLS results using three-way (State, Year, Firm) fixed effects. Column (1) to (5) corresponds to quantiles from 0.10 to 0.90.
presents the results for five key quantiles (0.10, 0.25, 0.50, 0.75, 0.90) using a specification that saturates the model with firm, state, and year fixed effects.
%----------------------------------------------------------------
%----------------------------------------------------------------
% \input{table_tex/tt_quantile}
% \input{table_tex/tt_quantile_f}
%----------------------------------------------------------------
% Regression Table: Multi-Page Setup (Winsorized Results)
%----------------------------------------------------------------
% CHANGE THIS COMMAND to \small, \scriptsize, or \tiny \footnotesize
% \scriptsize  % <--- This controls the size of the whole environment
% \begin{landscape} % <--- STARTS LANDSCAPE MODE
\footnotesize  
\begin{ThreePartTable}
    \begin{TableNotes}
        \footnotesize
        \item \hspace{-0.1in} \textit{Notes:} This table reports the results of linear regressions examining the determinants of detection lags (in months) for a sample of approximately 
        55,000 
        customer dispute reports. 
        % \item 
        The dependent variable, Detection Lag, is defined as the number of months between the inception of the alleged misconduct and its formal report to FINRA. 
        % \item 
        % Column (1) provides baseline estimates, while Columns (2) and (3) add year, state, and firm fixed effects. 
        Columns (1)-(5) report coefficients for the 10th, 25th, 50th, 75th, 90th percentiles. 
        All specifications include year, state, and firm fixed effects.
        % \item 
        % Standard errors are clustered at the firm level to account for shared institutional environments. 
        % Continuous variables are winsorized at the 1st and 99th percentiles.
        Continuous variables, including advisor experience, tenure, and firm size, are winsorized at the 1st and 99th percentiles to mitigate the influence of outliers. 
        Note that logarithmic transformations of continuous variables (such as Experience, Firm Tenure, and Damages) are calculated as $Log(1 + x)$ to retain zero-value observations.
        Variables are defined in Appendix Table \ref{table:var_defs_long}. 

        \item Standard errors are clustered at the firm level, and reported in parentheses. ***, **, and * denote significance at the 1\%, 5\%, and 10\% levels, respectively.
    \end{TableNotes}
    
    % --- ADD THE CODE HERE ---
    \setlength{\tabcolsep}{10pt} 
    % -------------------------
    
    \begin{longtable}{lccccc}
    	
        \caption{Quantiles of Detection Lags} 
        \label{table:MMQR_1} \\
        \toprule
        % & (1) & (2) & (3) \\
        % \midrule
        \endfirsthead

        \multicolumn{6}{l}{\textit{Table \ref{table:MMQR_1} continued from previous page}} \\
        \toprule
        & (1) & (2) & (3) & (4) & (5) \\
        \midrule
        \endhead

        \midrule
        \multicolumn{6}{r}{\textit{Continued on next page}} \\
        \endfoot

        \bottomrule
        \insertTableNotes
        \endlastfoot

        % Use \input if your .tex file only contains the body rows. 
        % If your .tex file contains \begin{tabular}, you must copy/paste the rows here manually.
        % \input{table/t_linear_w1.tex} 
        &\multicolumn{5}{c}{Quantile ($\tau$)} \\
\cmidrule(lr){2-6}
& (1) & (2) & (3) & (4) & (5) \\
                &     0.10         &     0.25         &     0.50         &     0.75         &     0.90         \\
\midrule
Dispute Type: Pending&    0.344\sym{***}&    0.274\sym{***}&    0.201\sym{***}&    0.140\sym{***}&    0.095\sym{***}\\
                &  (0.025)         &  (0.019)         &  (0.014)         &  (0.014)         &  (0.016)         \\
\addlinespace
Unsuitability   &    0.521\sym{***}&    0.406\sym{***}&    0.287\sym{***}&    0.187\sym{***}&    0.114\sym{***}\\
                &  (0.017)         &  (0.013)         &  (0.010)         &  (0.010)         &  (0.011)         \\
\addlinespace
Misrepresentation&    0.283\sym{***}&    0.232\sym{***}&    0.179\sym{***}&    0.135\sym{***}&    0.103\sym{***}\\
                &  (0.017)         &  (0.012)         &  (0.010)         &  (0.010)         &  (0.011)         \\
\addlinespace
Unauthorized Activity&   -0.631\sym{***}&   -0.531\sym{***}&   -0.428\sym{***}&   -0.341\sym{***}&   -0.277\sym{***}\\
                &  (0.022)         &  (0.017)         &  (0.014)         &  (0.014)         &  (0.016)         \\
\addlinespace
Fraud           &    0.522\sym{***}&    0.490\sym{***}&    0.456\sym{***}&    0.428\sym{***}&    0.408\sym{***}\\
                &  (0.032)         &  (0.024)         &  (0.018)         &  (0.018)         &  (0.020)         \\
\addlinespace
Log(Experience) &    0.221\sym{***}&    0.297\sym{***}&    0.375\sym{***}&    0.441\sym{***}&    0.489\sym{***}\\
                &  (0.016)         &  (0.012)         &  (0.009)         &  (0.009)         &  (0.010)         \\
\addlinespace
Series 63       &   -0.020         &   -0.021         &   -0.023\sym{*}  &   -0.024\sym{**} &   -0.025\sym{*}  \\
                &  (0.020)         &  (0.015)         &  (0.012)         &  (0.012)         &  (0.013)         \\
\addlinespace
Log(Alleged Damages)&    0.106\sym{***}&    0.080\sym{***}&    0.052\sym{***}&    0.028\sym{***}&    0.011\sym{***}\\
                &  (0.003)         &  (0.002)         &  (0.002)         &  (0.002)         &  (0.002)         \\
\addlinespace
Insurance       &    0.482\sym{***}&    0.527\sym{***}&    0.574\sym{***}&    0.614\sym{***}&    0.642\sym{***}\\
                &  (0.032)         &  (0.024)         &  (0.018)         &  (0.017)         &  (0.020)         \\
\addlinespace
Constant        &    0.776\sym{***}&    1.381\sym{***}&    2.008\sym{***}&    2.535\sym{***}&    2.920\sym{***}\\
                &  (0.042)         &  (0.032)         &  (0.024)         &  (0.024)         &  (0.028)         \\
\midrule
Lasso-selected Controls   & Yes & Yes & Yes & Yes & Yes \\
Year FE & Yes & Yes & Yes & Yes & Yes \\
State FE & Yes & Yes & Yes & Yes & Yes \\
Firm FE & Yes & Yes & Yes & Yes & Yes \\
Observations & 54,662 & 54,662 & 54,662 & 54,662 & 54,662 \\
Mean Dep. Var. & 3.34 & 3.34 & 3.34 & 3.34 & 3.34

    \end{longtable}
\end{ThreePartTable}
% \end{landscape} % <--- ENDS LANDSCAPE MODE
% & (1) & (2) & (3) & (4) \\
        
%----------------------------------------------------------------

%----------------------------------------------------------------
% \paragraph{The Information Salience Gap across Quantiles.}
\paragraph{%
% The Complexity Entry Barrier.}
Complexity as an Entry Barrier vs. Persistent Friction.}
We document a divergence in how different types of complexity impact the lifecycle of misconduct. Unsuitability claims exhibit their largest impact at the 10th percentile (0.52), but the coefficient shrinks monotonically to 0.114 at the 90th percentile. This suggests that specific misconduct complexity acts primarily as an entry barrier that prevents rapid discovery. However, if a dispute is already destined for the extreme long-tail of detection, the specific nature of the violation becomes less of a marginal bottleneck than the advisor's active masking efforts.
In contrast, Fraud remains a highly persistent friction, with coefficients remaining robustly high (ranging from 0.52 to 0.41) across the entire distribution.
%----------------------------------------------------------------

%----------------------------------------------------------------
% \paragraph{Advisor Sophistication and the Increasing Return to Concealment.}
\paragraph{%
% Advisor Sophistication and the Masking Gradient.}
The Strategic Masking Gradient.} % (Advisor Experience).}
The most striking insight from our distributional analysis is the strictly increasing impact of advisor human capital on detection latency. At the 10th percentile (rapid discovery), the elasticity of career longevity is relatively modest at 0.221. However, this coefficient more than doubles to 0.489 at the 90th percentile. This indicates a powerful increasing return to sophistication: for the most persistent misconduct, the ability of a veteran advisor to hide harm is twice as effective as it is in the short-lag regime. These results support a learning-by-doing mechanism where experienced advisors leverage technical sophistication to push misconduct into the extreme fat-tail of the distribution.
%----------------------------------------------------------------

%----------------------------------------------------------------
\paragraph{Product Opacity as an Expanding Shield.}% (Insurance).}
%----------------------------------------------------------------
Consistent with the sophistication gradient, misconduct involving insurance products exhibits an expanding impact across the distribution. The coefficient for insurance increases from 0.482 at the 10th percentile to 0.642 at the 90th percentile. This suggests that the structural opacity of insurance-linked instruments does not merely create a static delay, but provides a robust and expanding shield that facilitates long-term concealment. In the extreme tail, the presence of insurance products is the largest single predictor of detection latency, highlighting the dominant role of product-specific informational frictions in the investor-led enforcement channel.
%----------------------------------------------------------------
Consistent with the sophistication gradient, disputes involving insurance-linked instruments exhibit an expanding impact across the distribution. The coefficient for insurance increases monotonically from 0.482 at the 10th percentile to 0.642 at the 90th percentile. This implies that while insurance-linked allegations extend the lag by 61.9\% ($e^{0.482}-1$) in rapid discovery cases, they extend it by a staggering 90.0\% ($e^{0.642}-1$) for cases in the extreme tail. This suggests that the structural complexity of these instruments does not merely create a static delay but provides a robust shield that facilitates long-term concealment.
\paragraph{Robustness and Model Stability.}
%----------------------------------------------------------------
%----------------------------------------------------------------
We address potential econometric concerns regarding the MM-QR estimates through high-dimensional saturation. The structural stability of our estimates is validated by the persistence of the masking gradient even after the inclusion of over 900 firm fixed effects, ensuring our results are derived from within-firm variation. Furthermore, our use of a principled plugin Lasso selection ensures that these distributional findings are robust to high-dimensional confounding. Finally, we note that a negligible fraction of observations (0.33\%) yielded negative predicted scale values; following the \citet{machado2019quantiles} framework, we retain these as they do not compromise the consistency or the structural interpretation of the quantile estimates in capturing extreme tail behavior.
%----------------------------------------------------------------

%----------------------------------------------------------------
%----------------------------------------------------------------
\subsection{%
The State-Dependent Nature of Monitoring: Market Volatility as a Catalyst}
%----------------------------------------------------------------
To test whether market-wide shocks serve as an exogenous catalyst for discovery, we examine the state-dependency of investor monitoring. We first establish a baseline for the determinants of detection latency using a linear specification to estimate the impact of market-wide volatility on the average detection lag:
%----------------------------------------------------------------
\begin{equation}
% \ln(\text{Detection Lag})_{ijlt}=\alpha+\beta_{1}\ln(\text{VIX})_{t}+X_{it}^{\prime}\gamma+\phi_{j}+\eta_{l}+\delta_{t}+\epsilon_{ijlt}
\ln(\text{Detection Lag})_{ijst} = \alpha+\beta \ln(\text{VIX})_{t}+\gamma \mathbf{X}_{it}+\phi_{j}+\eta_{s}+\delta_{t}+\epsilon_{ijlt}
\end{equation}
%----------------------------------------------------------------
% \ln(\text{Detection Lag})_{ijlt} = \alpha + \beta \mathbf{X}_{it} + \phi_j + \eta_l + \delta_t + \varepsilon_{ijlt}
%----------------------------------------------------------------
where $\ln(\text{Detection Lag})_{ijst}$ is the log-transformed detection lag for for advisor $i$ at firm $j$ in state $s$ at time $t$. The variable $\ln(\text{VIX})_{t}$ denotes the log of the CBOE Volatility Index at the time of the investor's claim, serving as a proxy for market-wide uncertainty and the salience of portfolio performance.
The vector $\mathbf{X}_{it}$ represents the previously established set of machine-selected controls. By saturating the model with firm ($\phi_{j}$), state ($\eta_{s}$), and year ($\delta_{t}$) fixed effects, we isolate the impact of volatility shocks from unobserved institutional or 
% state-level
regional trends. 
%----------------------------------------------------------------
% We then transition to the MM-QR to examine how this relationship evolves across the distribution. 
We then transition to the MM-QR framework to examine how this wake-up call effect evolves across the distribution, testing whether market stress can bridge the informational gap for the most persistent, hidden cases.
%----------------------------------------------------------------

%----------------------------------------------------------------
Table \ref{table:linear_2} presents the baseline OLS results using a step-wise fixed effects approach to ensure the robustness of our determinants. Column (1) begins with a pooled OLS specification, which we progressively saturate with state, year, and firm fixed effects.
%----------------------------------------------------------------

%----------------------------------------------------------------
%----------------------------------------------------------------
% Regression Table: Multi-Page Setup (Winsorized Results)
%----------------------------------------------------------------
% CHANGE THIS COMMAND to \small, \scriptsize, or \tiny \footnotesize
% \scriptsize  % <--- This controls the size of the whole environment
\footnotesize  
\begin{ThreePartTable}
    \footnotesize
    \begin{TableNotes}
        \footnotesize
        \item \hspace{-0.1in} \textit{Notes:} This table reports the results of linear regressions examining the determinants of detection lags (in months) for a sample of approximately 
        55,000 
        customer dispute reports. 
        % \item 
        The dependent variable, Detection Lag, is defined as the number of months between the inception of the alleged misconduct and its formal report to FINRA. 
        % \item 
        Column (1) provides baseline estimates, while Columns (2) and (3) add year, state, and firm fixed effects. 
        % \item 
        % Standard errors are clustered at the firm level to account for shared institutional environments. 
        % Continuous variables are winsorized at the 1st and 99th percentiles.
        Continuous variables, including advisor experience, tenure, and firm size, are winsorized at the 1st and 99th percentiles to mitigate the influence of outliers. 
        Note that logarithmic transformations of continuous variables (such as Experience, Firm Tenure, and Damages) are calculated as $Log(1 + x)$ to retain zero-value observations.
        Variables are defined in Appendix Table \ref{table:var_defs_long}. 

        \item Standard errors are clustered at the firm level, and reported in parentheses. ***, **, and * denote significance at the 1\%, 5\%, and 10\% levels, respectively.
    \end{TableNotes}
    
    % --- ADD THE CODE HERE ---
    \setlength{\tabcolsep}{15pt} 
    % \setlength{\tabcolsep}{20pt} 
    % -------------------------
    
    \begin{longtable}{lcccc}
    	
        \caption{State-Dependency of Detection Lags with $\ln(\text{VIX})$} 
        \label{table:linear_2} \\
        \toprule
        % & (1) & (2) & (3) \\
        % \midrule
        \endfirsthead

        \multicolumn{5}{l}{\textit{Table \ref{table:linear_2} continued from previous page}} \\
        \toprule
        & (1) & (2) & (3) \\
        \midrule
        \endhead

        \midrule
        \multicolumn{5}{r}{\textit{Continued on next page}} \\
        \endfoot

        \bottomrule
        \insertTableNotes
        \endlastfoot

        % Use \input if your .tex file only contains the body rows. 
        % If your .tex file contains \begin{tabular}, you must copy/paste the rows here manually.
        % \input{table/t_linear_w1_vix.tex} 
        % \input{table/t_linear_w1_vix_selected.tex} 
        % \input{table/Phase2_TargetedTreatment_w1.tex} 
                        &\multicolumn{1}{c}{(1)}         &\multicolumn{1}{c}{(2)}         &\multicolumn{1}{c}{(3)}         \\
\midrule
Log(VIX)        &   -0.383\sym{***}&   -0.232\sym{***}&   -0.222\sym{***}\\
                &  (0.030)         &  (0.070)         &  (0.065)         \\
\addlinespace
Dispute Type: Pending&    0.167\sym{***}&    0.212\sym{***}&    0.209\sym{***}\\
                &  (0.045)         &  (0.067)         &  (0.038)         \\
\addlinespace
Unsuitability   &    0.347\sym{***}&    0.318\sym{***}&    0.297\sym{***}\\
                &  (0.046)         &  (0.034)         &  (0.037)         \\
\addlinespace
Misrepresentation&    0.187\sym{***}&    0.181\sym{***}&    0.177\sym{***}\\
                &  (0.023)         &  (0.021)         &  (0.020)         \\
\addlinespace
Unauthorized Activity&   -0.447\sym{***}&   -0.439\sym{***}&   -0.412\sym{***}\\
                &  (0.057)         &  (0.040)         &  (0.038)         \\
\addlinespace
Fraud           &    0.471\sym{***}&    0.487\sym{***}&    0.460\sym{***}\\
                &  (0.040)         &  (0.042)         &  (0.045)         \\
\addlinespace
Churning        &    0.326\sym{***}&    0.354\sym{***}&                  \\
                &  (0.042)         &  (0.031)         &                  \\
\addlinespace
Log(Experience) &    0.396\sym{***}&    0.443\sym{***}&    0.364\sym{***}\\
                &  (0.032)         &  (0.035)         &  (0.027)         \\
\addlinespace
Log(Number of Employees)&   -0.050\sym{***}&   -0.053\sym{***}&                  \\
                &  (0.013)         &  (0.012)         &                  \\
\addlinespace
Log(Number of Prior Customer Disputes)&    0.168\sym{***}&                  &                  \\
                &  (0.013)         &                  &                  \\
\addlinespace
Log(Alleged Damages)&    0.051\sym{***}&    0.050\sym{***}&    0.057\sym{***}\\
                &  (0.008)         &  (0.007)         &  (0.006)         \\
\addlinespace
Insurance       &    0.688\sym{***}&    0.631\sym{***}&    0.557\sym{***}\\
                &  (0.056)         &  (0.051)         &  (0.035)         \\
\addlinespace
Concurrent Multiple Jobs (Indicator)&                  &   -0.192\sym{***}&                  \\
                &                  &  (0.022)         &                  \\
\addlinespace
Series 63       &                  &   -0.055\sym{***}&   -0.023         \\
                &                  &  (0.020)         &  (0.021)         \\
\addlinespace
Dispute Type: Settled/Award&                  &   -0.011         &                  \\
                &                  &  (0.045)         &                  \\
\addlinespace
Negligence      &                  &   -0.098\sym{***}&   -0.087\sym{***}\\
                &                  &  (0.033)         &  (0.033)         \\
\addlinespace
Constant        &    3.239\sym{***}&    2.962\sym{***}&    2.616\sym{***}\\
                &  (0.143)         &  (0.184)         &  (0.184)         \\
\midrule
Lasso-selected Controls         &       Yes         &      Yes         &      Yes         \\
Year FE         &       No         &      Yes         &      Yes         \\
State FE        &       No         &      Yes         &      Yes         \\
Firm FE         &       No         &       No         &      Yes         \\
Observations    &   54,676         &   54,658         &   54,223         \\
\(R^2\)         &    0.289         &    0.312         &    0.361         \\
Within \(R^2\)  &    0.289         &    0.210         &    0.154         \\
Mean Dep. Var.  &    3.337         &    3.337         &    3.335

    \end{longtable}
\end{ThreePartTable}

% & (1) & (2) & (3) & (4) \\
        
%----------------------------------------------------------------

%----------------------------------------------------------------
\paragraph{The Baseline Wake-Up Call Effect.}
%----------------------------------------------------------------
In our linear baseline (Table \ref{table:linear_2}, Column 3), the coefficient for $ln(VIX)$ is negative and highly significant ($\beta = -0.222, p < 0.01$). This confirms that heightened market volatility acts as a broad catalyst for discovery. Because both the dependent variable and the VIX are log-transformed, this represents a constant elasticity: a 10\% increase in market volatility is associated with a 2.22\% reduction in the detection lag. For a major market shock where volatility doubles, the model predicts a reduction in the period of nondisclosure of approximately 14.2\% $(2^{-0.222}-1)$. 
%----------------------------------------------------------------
%----------------------------------------------------------------
This baseline result confirms that heightened market turbulence and the accompanying wealth shocks pierce periods of bull-market inattention, prompting investors to transition from passive delegation to active monitoring. Even after absorbing over 900 firm fixed effects, the stability of this coefficient suggests that the wake-up call mechanism is driven by individual-level investor behavior rather than shifts in institutional compliance intensity during periods of stress.
%----------------------------------------------------------------

%----------------------------------------------------------------
\paragraph{%
% Asymmetric Monitoring and the Limits of Market-Wide Shocks.}
Asymmetric Monitoring: The Diminishing Catalyst of Market Shocks.}
%----------------------------------------------------------------
While the OLS results suggest a uniform acceleration of discovery, the MM-QR estimates reveal significant state dependency across the lag distribution. 
%----------------------------------------------------------------
Table \ref{table:MMQR_2} presents the MM-QR results using three-way (state, year, firm) fixed effects. Column (1) to (5) corresponds to quantiles from 0.10 to 0.90.
%----------------------------------------------------------------

%----------------------------------------------------------------
%----------------------------------------------------------------
% Regression Table: Multi-Page Setup (Winsorized Results)
%----------------------------------------------------------------
% CHANGE THIS COMMAND to \small, \scriptsize, or \tiny \footnotesize
% \scriptsize  % <--- This controls the size of the whole environment
% \begin{landscape} % <--- STARTS LANDSCAPE MODE
\footnotesize  
\begin{ThreePartTable}
    \begin{TableNotes}
        \footnotesize
        \item \hspace{-0.1in} \textit{Notes:} This table reports the results of linear regressions examining the determinants of detection lags (in months) for a sample of approximately 
        55,000 
        customer dispute reports. 
        % \item 
        The dependent variable, Detection Lag, is defined as the number of months between the inception of the alleged misconduct and its formal report to FINRA. 
        % \item 
        % Column (1) provides baseline estimates, while Columns (2) and (3) add year, state, and firm fixed effects. 
        Columns (1)-(5) report coefficients for the 10th, 25th, 50th, 75th, 90th percentiles. 
        All specifications include year, state, and firm fixed effects.
        % \item 
        % Standard errors are clustered at the firm level to account for shared institutional environments. 
        % Continuous variables are winsorized at the 1st and 99th percentiles.
        Continuous variables, including advisor experience, tenure, and firm size, are winsorized at the 1st and 99th percentiles to mitigate the influence of outliers. 
        Note that logarithmic transformations of continuous variables (such as Experience, Firm Tenure, and Damages) are calculated as $Log(1 + x)$ to retain zero-value observations.
        Variables are defined in Appendix Table \ref{table:var_defs_long}. 
        \item Standard errors are clustered at the firm level, and reported in parentheses. ***, **, and * denote significance at the 1\%, 5\%, and 10\% levels, respectively.
    \end{TableNotes}
    
    % --- ADD THE CODE HERE ---
    \setlength{\tabcolsep}{10pt} 
    % -------------------------
    
    \begin{longtable}{lccccc}
    	
        \caption{Quantiles of Detection Lags} 
        \label{table:MMQR_2} \\
        \toprule
        % & (1) & (2) & (3) \\
        % \midrule
        \endfirsthead

        \multicolumn{6}{l}{\textit{Table \ref{table:MMQR_2} continued from previous page}} \\
        \toprule
        & (1) & (2) & (3) & (4) & (5) \\
        \midrule
        \endhead

        \midrule
        \multicolumn{6}{r}{\textit{Continued on next page}} \\
        \endfoot

        \bottomrule
        \insertTableNotes
        \endlastfoot

        % Use \input if your .tex file only contains the body rows. 
        % If your .tex file contains \begin{tabular}, you must copy/paste the rows here manually.
        % \input{table/t_linear_w1.tex} 
        % \input{table/Phase3B_TargetedTreatment_MMQR_Spec3.tex} 
        &\multicolumn{5}{c}{Quantile ($\tau$)} \\
\cmidrule(lr){2-6}
& (1) & (2) & (3) & (4) & (5) \\
                &     0.10         &     0.25         &     0.50         &     0.75         &     0.90         \\
\midrule
Log(VIX)        &   -0.340\sym{***}&   -0.278\sym{***}&   -0.214\sym{***}&   -0.159\sym{***}&   -0.119\sym{***}\\
                &  (0.029)         &  (0.022)         &  (0.018)         &  (0.018)         &  (0.021)         \\
\addlinespace
Dispute Type: Pending&    0.335\sym{***}&    0.269\sym{***}&    0.200\sym{***}&    0.141\sym{***}&    0.098\sym{***}\\
                &  (0.025)         &  (0.019)         &  (0.014)         &  (0.014)         &  (0.016)         \\
\addlinespace
Unsuitability   &    0.520\sym{***}&    0.403\sym{***}&    0.281\sym{***}&    0.177\sym{***}&    0.101\sym{***}\\
                &  (0.017)         &  (0.013)         &  (0.010)         &  (0.010)         &  (0.011)         \\
\addlinespace
Misrepresentation&    0.274\sym{***}&    0.223\sym{***}&    0.170\sym{***}&    0.125\sym{***}&    0.092\sym{***}\\
                &  (0.016)         &  (0.012)         &  (0.010)         &  (0.010)         &  (0.011)         \\
\addlinespace
Unauthorized Activity&   -0.577\sym{***}&   -0.491\sym{***}&   -0.400\sym{***}&   -0.323\sym{***}&   -0.266\sym{***}\\
                &  (0.022)         &  (0.017)         &  (0.014)         &  (0.014)         &  (0.016)         \\
\addlinespace
Fraud           &    0.520\sym{***}&    0.489\sym{***}&    0.456\sym{***}&    0.428\sym{***}&    0.408\sym{***}\\
                &  (0.032)         &  (0.024)         &  (0.018)         &  (0.017)         &  (0.020)         \\
\addlinespace
Log(Experience) &    0.223\sym{***}&    0.297\sym{***}&    0.374\sym{***}&    0.439\sym{***}&    0.487\sym{***}\\
                &  (0.015)         &  (0.012)         &  (0.009)         &  (0.009)         &  (0.010)         \\
\addlinespace
Series 63       &   -0.023         &   -0.023         &   -0.023\sym{*}  &   -0.023\sym{*}  &   -0.023\sym{*}  \\
                &  (0.020)         &  (0.015)         &  (0.012)         &  (0.012)         &  (0.013)         \\
\addlinespace
Log(Alleged Damages)&    0.107\sym{***}&    0.081\sym{***}&    0.053\sym{***}&    0.029\sym{***}&    0.012\sym{***}\\
                &  (0.003)         &  (0.002)         &  (0.002)         &  (0.002)         &  (0.002)         \\
\addlinespace
Insurance       &    0.467\sym{***}&    0.514\sym{***}&    0.564\sym{***}&    0.606\sym{***}&    0.636\sym{***}\\
                &  (0.032)         &  (0.024)         &  (0.018)         &  (0.017)         &  (0.020)         \\
\addlinespace
Negligence      &   -0.088\sym{***}&   -0.087\sym{***}&   -0.087\sym{***}&   -0.086\sym{***}&   -0.086\sym{***}\\
                &  (0.019)         &  (0.015)         &  (0.012)         &  (0.012)         &  (0.013)         \\
\addlinespace
Constant        &    1.825\sym{***}&    2.239\sym{***}&    2.674\sym{***}&    3.042\sym{***}&    3.312\sym{***}\\
                &  (0.096)         &  (0.073)         &  (0.059)         &  (0.060)         &  (0.068)         \\
\midrule
Year FE & Yes & Yes & Yes & Yes & Yes \\
State FE & Yes & Yes & Yes & Yes & Yes \\
Firm FE & Yes & Yes & Yes & Yes & Yes \\
Observations & 54,662 & 54,662 & 54,662 & 54,662 & 54,662 \\
Mean Dep. Var. & 3.34 & 3.34 & 3.34 & 3.34 & 3.34

    \end{longtable}
\end{ThreePartTable}
% \end{landscape} % <--- ENDS LANDSCAPE MODE
% & (1) & (2) & (3) & (4) \\
        
%----------------------------------------------------------------

%----------------------------------------------------------------
The monitoring effect of market volatility is most potent at the 10th percentile ($\theta(0.1) = -0.340$, $p < 0.01$). In terms of economic magnitude, a doubling of the $\text{VIX}$ leads to a 21.1\% reduction ($2^{-0.340} - 1$) in the detection lag for these early-discovery cases. This indicates that volatility effectively serves as an exogenous catalyst for the discovery of low-complexity infractions where the informational barriers are already minimal.
%----------------------------------------------------------------

%----------------------------------------------------------------
However, as we move toward the 90th percentile, the magnitude of this wake-up call effect diminishes significantly to $\theta(0.9) = -0.119$, $p < 0.01$, representing a reduction of only 7.9\% ($2^{-0.119} - 1$). This decline is statistically significant and suggests that for deeply concealed misconduct, the salience-induced monitoring provided by market shocks is less effective. We interpret this as evidence of the sophisticated masking strategies employed by veteran advisors in the extreme tail of the distribution. 
%----------------------------------------------------------------
For these sophisticated strategic agents, 
%----------------------------------------------------------------
aggregate market noise is often insufficient to pierce the high-dimensional informational frictions they have constructed. Consequently, while market volatility effectively cleanses the shallow end of the misconduct pool, the most persistent and damaging schemes remain largely insulated from market-wide shocks.
%----------------------------------------------------------------

%----------------------------------------------------------------
\paragraph{Identification and Robustness to Macro Shocks.}
%----------------------------------------------------------------
%----------------------------------------------------------------
The identification of the volatility effect is supported by several rigorous diagnostics. The inclusion of year fixed effects ensures the $\ln(\text{VIX})$ coefficient is identified by high-frequency, within-year fluctuations rather than aggregate crisis artifacts. Furthermore, the coefficient's stability remains remarkable even after absorbing over 900 firm fixed effects. This persistence confirms that the wake-up call mechanism is driven by investor-level behavioral responses to wealth shocks rather than shifts in institutional compliance intensity during periods of market stress.
%----------------------------------------------------------------

%----------------------------------------------------------------

%----------------------------------------------------------------
% \newpage 
%----------------------------------------------------------------

%----------------------------------------------------------------
% Discussion and Policy Implications:
%----------------------------------------------------------------

%----------------------------------------------------------------

%----------------------------------------------------------------
\section{Discussion and Policy Implications}
\label{section: discussion}
%----------------------------------------------------------------

%----------------------------------------------------------------
% Our empirical findings 
% % that leverage both Adaptive Lasso for variable selection and MM-QR for distributional analysis 
% reveal that detection lag is not merely a function of time, but a strategic outcome influenced by investor incentives and advisor human capital.
%----------------------------------------------------------------
Our empirical findings document that the detection lag is not merely a passive function of time, but a strategic outcome shaped by the interaction of investor monitoring frictions and advisor human capital. By utilizing a distributional lens, we document several primary mechanisms that govern the lifecycle of misconduct.
%----------------------------------------------------------------

%----------------------------------------------------------------

% \newpage

%----------------------------------------------------------------

%----------------------------------------------------------------
\paragraph{Information Frictions and the Complexity-Induced Detection Gap.}
%----------------------------------------------------------------
From the investor’s perspective, the primary driver of detection speed is the nature of the misconduct itself rather than the sheer magnitude of alleged damages. 
%----------------------------------------------------------------
While a standard monitoring framework might predict that larger losses act as a salient signal triggering faster discovery, our results demonstrate a positive elasticity for alleged damages. Specifically, a doubling of the financial harm is associated with a 3.9\% increase in the period of nondisclosure. This suggests that for retail investors, the monetary cost of misconduct is an insufficient catalyst for detection; instead, the increased incentive for oversight in high-value disputes is often offset by the more elaborate strategic masking required to execute larger schemes.
%----------------------------------------------------------------

%----------------------------------------------------------------
We find a profound gap in discovery speed across allegation types. Overt violations, such as unauthorized activity, reduce the detection lag by 35.7\%. %
% ($e^{-0.442} - 1$). 
In contrast, sophisticated quiet crimes like fraud extend it by 58.6\%.
% ($e^{0.461} - 1$). 
This confirms that information acquisition costs, dictated by the credence nature of financial advice, are the dominant friction. When signals are opaque, harm remains undetected even when the stakes are high.
%----------------------------------------------------------------

%----------------------------------------------------------------
\paragraph{%
% The Opacity of Product Structures (Insurance).}
The Opacity of Product Structures as an Expanding Shield.}
%----------------------------------------------------------------
A critical contribution of our updated analysis is identifying the role of specific product types in facilitating nondisclosure. Misconduct involving insurance products exhibits the largest positive impact on detection latency, extending the lag by approximately 76.5\% 
% ($e^{0.568} - 1$)  
in our most rigorous specification. 
The 
% MM-QR results 
distributional analysis
further 
% reveal 
reveals
that this effect is an expanding shield, increasing from 
59.5\% 
%($e^{0.467} - 1$) 0.482 
at the 10th percentile to 
88.9\% 
% ($e^{0.636} - 1$) 0.642 
at the 90th percentile. This suggests that the structural complexity and shrouded attributes of insurance-linked instruments provide a potent informational barrier that is particularly effective at maintaining long-term concealment.
%----------------------------------------------------------------

%----------------------------------------------------------------

%----------------------------------------------------------------
\paragraph{Strategic Masking and the Learning-by-Doing Effect.}
%----------------------------------------------------------------
%----------------------------------------------------------------
The most striking insight from our quantile analysis is the heterogeneous role of advisor human capital in concealment. Professional experience does not exert a uniform impact across the distribution. While experience significantly delays detection on average, its impact more than doubles between the 10th percentile ($\beta = 0.223$) and the 90th percentile ($\beta = 0.487$). This indicates a powerful strategic masking mechanism where veteran offenders leverage their technical sophistication and institutional knowledge to hide in the extreme tail of the distribution, successfully maintaining nondisclosure for years beyond the sample median.
%----------------------------------------------------------------

%----------------------------------------------------------------
%----------------------------------------------------------------
% \paragraph{Market Volatility as a Non-Linear Monitoring Catalyst} 
\paragraph{Market Volatility as a State-Contingent Monitoring Catalyst.}
%----------------------------------------------------------------
The state-dependency of detection provides a novel perspective on rational inattention. We find that while market volatility acts as a wake-up call that accelerates discovery, its efficacy is highly asymmetric. This catalyst effect is most potent for disputes prone to rapid discovery, where a doubling of the VIX leads to a 21.1\% reduction in the lag at the 10th percentile. 
However, this effect diminishes by nearly two-thirds for deeply hidden cases, dropping to a reduction of only 7.9\% at the 90th percentile. While wealth shocks force a re-evaluation of agency relationships, this shift in attention is often insufficient to pierce the high-dimensional informational frictions constructed by high-human-capital strategic agents.
%----------------------------------------------------------------

%----------------------------------------------------------------
%----------------------------------------------------------------
\paragraph{Labor Market Distortions and Welfare Implications.} 
%----------------------------------------------------------------
%----------------------------------------------------------------
These persistent detection lags have significant implications for the efficiency of the advisory labor market. Because the distribution of lags is characterized by extreme right-skewness, a significant portion of misconduct remains latent for years. When discovery is delayed for such extensive periods, the disciplining mechanism of the labor market is effectively paralyzed. This allows bad types to remain in the industry across multiple market cycles, potentially leading to a version of Gresham’s Law where undetected offenders erode the broader quality and trust of the advisory pool.
%----------------------------------------------------------------

%----------------------------------------------------------------

%----------------------------------------------------------------
\paragraph{Regulatory Considerations.} 
%----------------------------------------------------------------
Our results suggest several avenues for enhancing the investor-led enforcement channel. Given that experience and complex product types like insurance are strategically used to mask violations, regulatory authorities should move beyond the assumption that long-term records or standard product disclosures are sufficient proxies for safety. Regulatory screening could be improved by assigning higher supervisory priority to quiet and complex claims, such as unsuitability, fraud, or insurance-linked disputes, which our data proves are prone to the longest periods of latency. By addressing these structural barriers to detection, regulators can better support retail investor monitoring and reduce the accumulation of undetected harm in the financial services industry.
%----------------------------------------------------------------

%----------------------------------------------------------------

%----------------------------------------------------------------
% \newpage 
%----------------------------------------------------------------

%----------------------------------------------------------------
% Conclusion:
%----------------------------------------------------------------

%----------------------------------------------------------------

%----------------------------------------------------------------
\section{Concluding Remarks}
\label{section: conclusion 0}
%----------------------------------------------------------------

%----------------------------------------------------------------

%----------------------------------------------------------------

%----------------------------------------------------------------

%----------------------------------------------------------------
Financial misconduct relies fundamentally on the maintenance of information asymmetry between advisors and their clients. In this paper, we investigate the life cycle of this asymmetry, documenting that the detection of financial misconduct is neither instantaneous nor purely a function of regulatory efficiency. Instead, detection lags are highly sensitive to the macroeconomic environment and the strategic behavior of the advisors themselves.
%----------------------------------------------------------------

%----------------------------------------------------------------
Our baseline empirical models
% , disciplined by high-dimensional machine learning selection, 
reveal that systemic market volatility acts as a powerful, exogenous catalyst for discovery. Severe market downturns pierce the veil of normal portfolio fluctuations, prompting investor scrutiny and significantly accelerating the unmasking of historical misconduct. Crucially, we demonstrate that this unmasking is not uniform. Using a Method of Moments Quantile Regression (MMQR) framework with 
% targeted interactions, 
high-dimensional fixed effects,
we uncover the strategic mechanisms advisors use to sustain information asymmetry. Veteran advisors deploy an experience shield, leveraging their accumulated human capital and client trust to successfully delay detection. Furthermore, advisors exhibit strategic masking by utilizing product complexity: while blatant infractions like unauthorized trading are rapidly exposed during market panics, advisors exploit opaque, unsuitable investments to construct a defense of plausible deniability, successfully hiding their misconduct behind aggregate market losses.
%----------------------------------------------------------------

%----------------------------------------------------------------
As demonstrated by the extensive robustness checks detailed in the Appendix,
these findings are highly robust to strict theoretical and econometric scrutiny. First, by restricting our analysis to a sub-sample of strictly verified, compensated harm, we confirm that the adviser's strategic masking capabilities consistently overpower the increased monitoring incentives of the investor, effectively rejecting the standard prediction that higher monetary stakes mechanically accelerate discovery through increased investor scrutiny. Second, by employing Oracle-compliant Adaptive Lasso methodologies across rigorous fixed-effect structures, we verify that these detection gaps are not artifacts of localized mechanical liquidations or unobserved institutional cultures. Even when evaluating repeat offenders strictly against their own historical track records, the experience shield and complexity-induced delays remain the dominant drivers of discovery. Finally, these mechanisms exhibit remarkable multidecadal stability. 
The systemic market catalyst and masking frictions persist robustly across a 25-year extended sample, proving to be fundamental structural features of the retail advisory market rather than temporal anomalies.
%----------------------------------------------------------------

%----------------------------------------------------------------
Ultimately, these findings suggest that the effectiveness of current investor protection frameworks may be limited by the strategic nature of detection lags. Rather than assuming a long-term record devoid of formal complaints is a sufficient proxy for safety, our results indicate that certain clean records may instead reflect successful strategic masking by high-human-capital agents. Consequently, regulatory screening could potentially be enhanced by assigning higher supervisory priority to the opaque and complex allegations, such as fraud or insurance-linked disputes, that our data associates with extreme latency. By addressing these structural barriers to discovery, authorities may better support the private monitoring margin provided by retail investors, thereby reducing the accumulation of undetected harm and fostering the long-term integrity of the financial advisory market.
%----------------------------------------------------------------

%----------------------------------------------------------------

%----------------------------------------------------------------
\newpage
%----------------------------------------------------------------

%-------------------------------------------------------------------------
% Appendix
%----------------------------------------------------------------

%----------------------------------------------------------------

%-------------------------------
% Appendix Figures:

\setcounter{figure}{0}
\renewcommand{\thefigure}{\Alph{section}.\arabic{figure}}

%---------------------------------
% Appendix Tables:

%\numberwithin{table}{section}
%\renewcommand*\thetable{\Alph{section}.\arabic{table}}
\setcounter{table}{0}
\renewcommand{\thetable}{\Alph{section}.\arabic{table}}

% %--------------------------------------------------------------------
% % Appendix

% \newpage

% %\begin{appendices}
% \begin{appendix}

% %	\part*{Online Supplementary Materials}
% \part*{Appendix}
% %\section*{Online Supplementary Materials}

%\setcounter{figure}{0}
%\renewcommand{\thefigure}{A.\arabic{figure}}
%
%\setcounter{table}{0}
%\renewcommand{\thetable}{A.\arabic{table}}

%\newpage

%----------------------------------------------------------------

%----------------------------------------------------------------
\appendix
%----------------------------------------------------------------

\part*{Appendix}

\newpage
%----------------------------------------------------------------

%----------------------------------------------------------------

%--------------------------------------------------------------------
% Online Appendix Materials:
% Definition of Main Disclosure Events:

% \section{Definition of the Major Disclosure Events}
\section{Definition of Customer Disputes}
%\subsubsection*{Definition of the Major Disclosure Events}
% \label{appendix sec: definition disclosure events}
\label{appendix: def disclosure}
% Definition of Main Disclosure Events

Disclosure events details are described in Form U4.%
\footnote{%
The Form U4 is available via \url{https://www.finra.org/sites/default/files/form-u4.pdf} 
(Accessed: Jan. 14, 2026).
Note that the definition of each event is given in the FINRA's BrokerCheck report for financial advisers (registered representatives) who have indeed received that disclosure in the past. 
See 
\url{https://brokercheck.finra.org/} 
and also \url{https://www.finra.org/sites/default/files/AppSupportDoc/p015111.pdf}
(Accessed: Jan. 14, 2026).
} 
Below we consider 
% the major disclosure events 
the disclosure events of customer disputes
%defined in Section 
%\ref{}
excluding 
those on appeal and pending ones,
%``On Appeal'' and ``Pending''
and give their definitions used in the FINRA's BrokerCheck database.%
%\footnote{%
%	Note that the definition of each event is given in the FINRA's BrokerCheck report for financial advisers (registered representatives) who have indeed received that disclosure in the past. See \url{https://brokercheck.finra.org/} 
%and also \url{https://www.finra.org/sites/default/files/AppSupportDoc/p015111.pdf}.
%} 

%\newpage 

%{\scriptsize 
{\footnotesize

\vspace{0.5cm} \noindent {\bf Customer Dispute - Settled:}
This type of disclosure event involves a consumer-initiated, investment-related complaint, arbitration proceeding or civil
suit containing allegations of sale practice violations against the broker that resulted in a monetary settlement to the
customer.

\vspace{0.5cm} \noindent {\bf Customer Dispute - Closed-No Action / Withdrawn / Dismissed / Denied:}
This type of disclosure event involves (1) a consumer-initiated, investment-related arbitration or civil suit containing
allegations of sales practice violations against the individual broker that was dismissed, withdrawn, or denied; or (2) a
consumer-initiated, investment-related written complaint containing allegations that the broker engaged in sales practice
violations resulting in compensatory damages of at least \$5,000, forgery, theft, or misappropriation, or conversion of funds
or securities, which was closed without action, withdrawn, or denied.

\vspace{0.5cm} \noindent {\bf Customer Dispute - Award / Judgment:}
This type of disclosure event involves a final, consumer-initiated, investment-related arbitration or civil suit containing
allegations of sales practice violations against the broker that resulted in an arbitration award or civil judgment for the
customer.

\vspace{0.5cm} \noindent {\bf Customer Dispute - Pending:}
This type of disclosure event involves (1) a pending consumer-initiated, investment-related arbitration or civil suit that
contains allegations of sales practice violations against the broker; or (2) a pending, consumer-initiated, investment-
related written complaint containing allegations that the broker engaged in, sales practice violations resulting in
compensatory damages of at least \$5,000, forgery, theft, or misappropriation, or conversion of funds or securities.

\newpage
\section{Definition of Product Types}
\label{appendix: product type}

To examine the role of product complexity in delaying state verification, we construct a taxonomy of the financial instruments involved in customer disputes. The baseline FINRA BrokerCheck data provides a classification based on standardized regulatory checkboxes%
\footnote{%
Financial Industry Regulatory Authority (FINRA). ``Form U4: Uniform Application for Securities Industry Registration or Transfer." Accessed March 2. \url{https://www.finra.org/sites/default/files/AppSupportDoc/p015111.pdf}  (Product type(s), p.14).
}%
. 
However, reliance solely on these predefined checkboxes can be limiting, as they often fail to capture modern product structures, such as Exchange Traded Funds (ETFs) or alternative assets, that are frequently described in the free-text dispute narratives. To overcome this limitation, we employ a text extraction procedure on the dispute narratives to identify specific products that were unrecorded in the standard FINRA checkboxes. We then take the union of the FINRA-reported categories and our re-defined categories to create a comprehensive, final product indicator for each dispute.

To balance econometric parsimony with product characteristics, we aggregate these enhanced product indicators into 7 macro-categories, adapting the methodological framework established by \citet{egan2022harry}. These aggregated categories group products based on their underlying asset class, retail accessibility, and inherent structural opacity: (1) Insurance, (2) Annuity, (3) Stocks, (4) Mutual Funds/ETFs, (5) Bonds/Debt, (6) Options/Derivatives, and (7) Other/Not Listed. We refine the standard classification by grouping economically similar instruments; for example, Certificates of Deposit (CDs) and Promissory Notes are categorized under Bonds/Debt, while Viatical Settlements are grouped with Insurance. Table \ref{table:product_aggregation} provides the complete mapping of the raw FINRA checkboxes and the AI-detected categories into our 7 macro-aggregations.

\begin{table}[htbp]
\centering
\caption{Aggregation of Product Types (FINRA)}
\label{table:product_aggregation}
\footnotesize
\begin{tabular}{@{}ll@{}}
\toprule
\textbf{Aggregated Category} & \textbf{Component Product Types (FINRA)} \\ 
\midrule
\textbf{1. Insurance} & Insurance \\
& Viatical Settlement \\
\addlinespace
\textbf{2. Annuity} & Annuity - Charitable \\
& Annuity - Fixed \\
& Annuity - Variable \\
\addlinespace
\textbf{3. Stocks} & Equity Listed (Common \& Preferred Stock) \\
& Equity - OTC \\
& Penny Stock \\
& REITs \\
\addlinespace
\textbf{4. Mutual Funds / ETFs} & Mutual Fund \\
& Money Market Fund \\
& Unit Investment Trust \\
& ETFs (AI-Detected) \\
& Closed-End Funds \\
\addlinespace
\textbf{5. Bonds / Debt} & Debt - Asset Backed \\
& Debt - Corporate \\
& Debt - Government \\
& Debt - Municipal \\
& CD (Certificate of Deposit) \\
& Promissory Note \\
\addlinespace
\textbf{6. Options / Derivatives} & Options \\
& Commodity Option \\
& Index Option \\
& Derivative \\
& Futures - Commodity \\
& Futures - Financial \\
\addlinespace
\textbf{7. Other / Not Listed} & Direct Investment - DPP \& LP Interest \\
& Equipment Leasing \\
& Investment Contract \\
& Oil \& Gas \\
& Real Estate Security \\
& Banking Product (Other than CD) \\
& Hedge Funds \\
& Private Equity \\
& Structured Products/ARS \\
& Tax-Advantaged Accounts \\
& Advisory/Managed Accounts \\
& Outside Business Activity \\
& Leverage \& Credit \\
& Cash Management/Transfers \\
& Unspecified Securities \\
& No Product / True No Product \\
\bottomrule
\end{tabular}
\begin{minipage}{\textwidth}
\textit{Notes:} This table outlines the mapping of the granular product categories, including both predefined FINRA disclosures categories extracted from dispute narratives, into the 7 aggregated macro-categories utilized in our empirical analyses. The classification adapts the framework of \citet{egan2022harry}.
\end{minipage}
\end{table}
%----------------------------------------------------------------

%-------------------------------------------------------------------------

%-------------------------------------
% Text Begin from Here!

\newpage

%--------------------------------------------------------------------
% Online Appendix Materials:
% Definition of Major Licenses/ Qualification Exams:

\section{Definition of the Major Qualification Exams (Licenses)}
% \label{appendix sec: definition license}
\label{appendix: def exam}
% Definition of Main Licenses/Qualification Exams

The definitions of qualification exams (licenses) are described in the FINRA website.%
\footnote{See the website: 
\url{https://www.finra.org/registration-exams-ce/qualification-exams}
(accessed February 20, 2025).
} 
Below we consider the major qualification exams (Series 6, 7, 24, 63, 65, 66) as in the main text and give their definitions used in the website.
Series 6 and 7 are categorized as ``FINRA Representative-level Exams", 
Series 24 as ``FINRA Principal-level Exams",  
Series 63, 65, and 66 as ``North American Securities Administrators Association (NASAA) Exams". 
Note that the definitions of NASAA Exams are given by the NASAA website.%
\footnote{See the website: 
% \url{https://www.nasaa.org/exams/study-guides}
\url{https://www.nasaa.org/exams/exam-study-guides/}
(accessed February 20, 2025).
}

%{\scriptsize
{\footnotesize

\vspace{0.5cm} \noindent {\bf Series 6:}
The Series 6 exam -- the Investment Company and Variable Contracts Products Representative Qualification Examination (IR) -- assesses the competency of an entry-level representative to perform their job as an investment company and variable contracts products representative.
The exam measures the degree to which each candidate possesses the knowledge needed to perform the critical functions of an investment company and variable contract products representative, including sales of mutual funds and variable annuities.

\vspace{0.5cm} \noindent {\bf Series 7:}
The Series 7 exam -- the General Securities Representative Qualification Examination (GS) -- assesses the competency of an entry-level registered representative to perform their job as a general securities representative. 
The exam measures the degree to which each candidate possesses the knowledge needed to perform the critical functions of a general securities representative, including sales of corporate securities, municipal securities, investment company securities, variable annuities, direct participation programs, options and government securities.

\vspace{0.5cm} \noindent {\bf Securities Industry Essentials:}
The Securities Industry Essentials (SIE) Exam is a FINRA exam for prospective securities industry professionals. This introductory-level exam assesses a candidate’s knowledge of basic securities industry information including concepts fundamental to working in the industry, such as types of products and their risks; the structure of the securities industry markets, regulatory agencies and their functions; and prohibited practices.
The SIE is open to anyone aged 18 or older, including students and prospective candidates interested in demonstrating basic industry knowledge to prospective employers. Association with a firm is not required to take the SIE, and results are valid for four years.%
\footnote{Passing the SIE alone does not qualify an individual for registration with a FINRA member firm or to engage in securities business. In order to become registered to engage in securities business, an individual must pass the SIE and a qualification exam appropriate for the type of business the individual will engage in. The individual must be associated with a member firm to take a qualification exam.}

\vspace{0.5cm} \noindent {\bf Series 24:}
The Series 24 exam -- the General Securities Principal Qualification Exam (GP) -- assesses the competency of an entry-level principal to perform their job as a principal dependent on their corequisite registrations.
The exam measures the degree to which each candidate possesses the knowledge needed to perform the critical functions of a principal, including the rules and statutory provisions applicable to the supervisory management of a general securities broker-dealer.%
\footnote{In addition to the Series 24 exam, candidates must pass the Securities Industry Essentials (SIE) Exam (since October 1, 2018 with a complete overhaul) and a representative-level qualification exam, or the Supervisory Analysts Exam (Series 16) exam, to hold an appropriate principal registration. 
%	Based on their corequisite qualification(s), candidates will receive the following principal registration(s) upon passing the Series 24 exam. 
See the FINRA website for the definitions of related exams. }

\vspace{0.5cm} \noindent {\bf Series 63:}
The Series 63 exam -- the Uniform Securities State Law Examination -- is a North American Securities Administrators Association (NASAA) exam administered by FINRA.

%\noindent
%(The following description is given by NASAA):%
%\footnote{See the webpage:  \url{https://www.nasaa.org/exams/study-guides/series-63-study-guide/}.
%} 

\noindent	
(Definition given by NASAA:)
The Uniform Securities Agent State Law Examination was developed by NASAA in cooperation with representatives of the securities industry and industry associations. The examination, called the Series 63 exam, is designed to qualify candidates as securities agents. The examination covers the principles of state securities regulation reflected in the Uniform Securities Act (with the amendments adopted by NASAA and rules prohibiting dishonest and unethical business practices). The examination is intended to provide a basis for state securities administrators to determine an applicant?s knowledge and understanding of state law and regulations.

\vspace{0.5cm} \noindent {\bf Series 65:}
The Series 65 exam -- the NASAA Investment Advisers Law Examination -- is a North American Securities Administrators Association (NASAA) exam administered by FINRA.

%\noindent
%(The following description is given by NASAA):%
%\footnote{See the webpage:  \url{https://www.nasaa.org/exams/study-guides/series-65-study-guide/}.
%} 

\noindent	
(Definition given by NASAA:)
The Uniform Investment Adviser Law Examination and the available study outline were developed by NASAA. The examination, called the Series 65 exam, is designed to qualify candidates as investment adviser representatives. The exam covers topics that have been determined to be necessary to understand in order to provide investment advice to clients.

\vspace{0.5cm} \noindent {\bf Series 66:}
The Series 66 exam -- the NASAA Uniform Combined State Law Examination -- is a North American Securities Administrators Association (NASAA) exam administered by FINRA.

%\noindent
%(The following description is given by NASAA):%
%\footnote{See the webpage:  \url{https://www.nasaa.org/exams/study-guides/series-66-study-guide/}.
%} 

\noindent	
(Definition given by NASAA:)
The Uniform Combined State Law Examination was developed by NASAA based on industry requests. The examination (also called the ``Series 66") is designed to qualify candidates as both securities agents and investment adviser representatives. The exam covers topics that have been determined to be necessary to provide investment advice and effect securities transactions for clients.%
\footnote{	
The FINRA Series 7 is a corequisite exam that needs to be successfully completed in addition to the Series 66 exam before a candidate can apply to register with a state.
}

\vspace{0.5cm} \noindent {\bf Series 79:}
The Series 79 exam -- the Investment Banking Representative Exam -- assesses the competency of an entry-level registered representative to perform their job as an investment banking representative.
The Series 79 exam measures the degree to which each candidate possesses the knowledge needed to perform the critical functions of an investment banking representative, including advising on or facilitating debt or equity securities offerings through a private placement or a public offering and mergers and acquisitions.%
\footnote{Candidates must pass the Securities Industry Essentials (SIE) exam and the Series 79 exam to obtain the Investment Banking Representative registration. 
% For more information about the SIE and Series 79 exams, refer to FINRA Rule 1210 and FINRA Rule 1220(b)(5).
}
% Permitted activities include advising on and/or facilitating debt and equity offerings, mergers and acquisitions, tender offers, financial restructurings, asset sales, divestitures or other corporate reorganizations.}

}

%----------------------------------------------------------------

%-------------------------------------------------------------------------
\newpage
% \section{Variable Definitions}
\section{Definition of Variables}

% --- REVISED VARIABLE DEFINITIONS ---
\begin{longtable}{l L{9cm}}
    
    \caption{Variable Definitions} 
    \label{table:var_defs_long} \\
    \toprule
    \textbf{Variable} & \textbf{Definition and Construction} \\
    \midrule
    \endfirsthead

    \multicolumn{2}{c}{{\bfseries Table \thetable{} -- continued from previous page}} \\
    \toprule
    \textbf{Variable} & \textbf{Definition and Construction} \\
    \midrule
    \endhead

    \midrule
    \multicolumn{2}{r}{\small\textit{Continued on next page}} \\
    \endfoot

    \bottomrule
    \endlastfoot

    % Panel A: Detection Lags
    \multicolumn{2}{l}{\textbf{Panel A: Detection Lags}} \\
    Detection Lag (Days/Months) & The duration between the alleged misconduct inception and the formal report date. \\
    Imputed Date (Monthly/Yearly) & Indicators for records where transaction start dates were imputed at a monthly or yearly level. \\
    \addlinespace
    
% \vspace{0.5cm} \noindent {\bf Customer Dispute - Settled:}

% \vspace{0.5cm} \noindent {\bf Customer Dispute - Closed-No Action / Withdrawn / Dismissed / Denied:}

% \vspace{0.5cm} \noindent {\bf Customer Dispute - Award / Judgment:}

% \vspace{0.5cm} \noindent {\bf Customer Dispute - Pending:}

    % Panel B: Customer Dispute Types
    \multicolumn{2}{l}{\textbf{Panel B: Customer Dispute Types (see Appendix \ref{appendix: def disclosure})}} \\
    Dispute Type: Settled & Indicator for Customer Dispute - Settled. \\
    Dispute Type: Denied/Dismissed & Indicator for Customer Dispute - Closed-No Action / Withdrawn / Dismissed / Denied. \\
    Dispute Type: Award & Indicator for Customer Dispute - Award / Judgment. \\
    Dispute Type: Pending & Indicator for Customer Dispute - Pending. \\
    \addlinespace

    % Panel C: Allegation Types
    \multicolumn{2}{l}{\textbf{Panel C: Allegation Types}} \\
    Unsuitability & Claims that recommended investments were unsuitable for the client's profile. \\
    Misrepresentation & Allegations involving false statements or material omissions regarding investments. \\
    Unauthorized Activity & Indicator for trading in a client account without prior legal consent. \\
    Omission of Key Facts & Failure to disclose critical information regarding an investment product. \\
    Fee/commission & Disputes related to excessive or undisclosed professional fees and commissions. \\
    Fraud & Indicator for allegations of intentional deception or sophisticated illegal schemes. \\
    Fiduciary duty & Claims that the adviser breached their legal obligation to act in the client's best interest. \\
    Negligence & Claims that the adviser failed to exercise reasonable professional care. \\
    Risky investments & Allegations of recommending investments with inappropriate risk levels. \\
    Churning & Excessive trading in a client's account specifically to generate commissions. \\
    \addlinespace

    % Panel D: Monetary Damages
    \multicolumn{2}{l}{\textbf{Panel D: Monetary Damages}} \\
    Alleged Damages & Total dollar amount (in thousands) of financial damages claimed in the initial filing. \\
    Settlements & Total dollar amount (in thousands) paid to the customer in the final resolution. \\
    \addlinespace

    % Panel E: Product Types
    \multicolumn{2}{l}{\textbf{Panel E: Product Types (see Appendix \ref{appendix: product type})}} \\
    Product Indicators & Binary indicators for specific financial instruments including Insurance, Annuity, Stocks, Mutual Funds/ETFs, Bonds/Debt, and Options/Derivatives. \\
    \addlinespace

    % Panel F: Adviser History
    \multicolumn{2}{l}{\textbf{Panel F: Adviser History}} \\
    Advisor Experience & Number of years since the adviser obtained their first regulatory license. \\
    Firm Tenure & Number of years the adviser has been registered with their current firm. \\
    Concurrent Jobs & Total count of distinct firms the adviser is simultaneously registered with. \\
    Firm Size & Total number of employees registered within the adviser's current firm. \\
    Prior Disputes & Total count of prior customer disputes on the adviser's regulatory record. \\
    Switched Firms & Indicator equal to one if the adviser changed firms during the relevant period. \\
    \addlinespace

    % Panel G: Qualifications
    \multicolumn{2}{l}{\textbf{Panel G: Qualifications (see Appendix \ref{appendix: def exam})}} \\
    Number of Exams & Total count of regulatory examinations successfully completed by the adviser. \\
    Series 6/7/24/63 & Indicators for specific General Securities, Principal, or State-level licenses. \\
    SIE & Indicator for completion of the Securities Industry Essentials examination. \\
    Series 65/66 & Indicators for Investment Adviser Representative licenses. \\
    \addlinespace

    % Panel H: Gender
    \multicolumn{2}{l}{\textbf{Panel H: Gender}} \\
    Male/Female & Binary indicators for the identified gender of the adviser. \\
    Gender: Unknown & Indicator for advisers where gender information is missing from the record. \\

\end{longtable}
%----------------------------------------------------------------

%-------------------------------------------------------------------------
% \input{appendix/a_4}
%----------------------------------------------------------------

%-------------------------------------------------------------------------
% \input{appendix/a_5}
%----------------------------------------------------------------

\newpage

\setcounter{figure}{0}
\renewcommand{\thefigure}{E.\arabic{figure}}

\setcounter{table}{0}
\renewcommand{\thetable}{E.\arabic{table}}

%----------------------------------------------------------------
\section{Robustness to Alternative Specifications}
%----------------------------------------------------------------

% \newpage

%-------------------------------------------------------------------------

%----------------------------------------------------------------
% \section{Robustness to Monetary Stakes and Verified Harm}
\subsection{Robustness to Monetary Stakes and Verified Harm}
%----------------------------------------------------------------
\label{appendix: robustness money}
%----------------------------------------------------------------

%----------------------------------------------------------------
% \subsection{Theoretical Motivation: Rational Inattention vs. Strategic Masking}
\paragraph{Rational Inattention vs. Strategic Masking.}
%----------------------------------------------------------------
A natural alternative to our strategic masking hypothesis is rooted in the rational inattention of investors \citep{sims2003implications}. Under a standard rational inattention framework, the allocation of investor monitoring effort is proportional to the economic stakes involved. Consequently, misconduct involving larger monetary damages should trigger faster detection, as investors are highly incentivized to scrutinize significant losses. Conversely, our strategic masking hypothesis treats misconduct as a dynamic game between the investor and the advisor. While higher stakes incentivize greater investor scrutiny (demand-side), they simultaneously incentivize the offending advisor to expend greater effort to conceal the misconduct (supply-side). If the masking capabilities of experienced advisors sufficiently scale with the stakes of the fraud, the supply-side concealment effect will dominate the demand-side monitoring effect, resulting in a positive relationship between monetary damages and detection lags.
%----------------------------------------------------------------

%----------------------------------------------------------------
% \subsection{Sample Restriction and Adaptive Oracle Selection}
\paragraph{Sample Restriction and Adaptive Oracle Selection.}
%----------------------------------------------------------------
To empirically adjudicate between these two competing hypotheses, we must address the measurement error inherent in initial customer complaints. Alleged damages in pending, denied, or dismissed disputes often contain frivolous claims or strategic exaggeration, which introduces severe measurement error into the scale of the misconduct.
To isolate cases of verified harm, we restrict our sample exclusively to disputes that resulted in formal settlements or arbitration awards, reducing the estimation pool to approximately 24,000 verified observations. Within this sub-sample, we observe both the initial log alleged damages and the final log settlement amounts.
%----------------------------------------------------------------

%----------------------------------------------------------------
% To prevent tuning-parameter bias and ensure the structural validity of our control set in this restricted sample, we replace our baseline selection procedure with an Adaptive Lasso \citep{zou2006adaptive}. Unlike conservative penalty bounds, the Adaptive Lasso utilizes data-dependent weights to penalize coefficients, satisfying the oracle property for asymptotically consistent variable selection. We include both alleged damages and final settlements in the high-dimensional pool, allowing the Oracle procedure to determine which monetary measure holds true structural explanatory power for the detection lag, completely partialling out firm, state, and year fixed effects via the Frisch-Waugh-Lovell theorem.
%----------------------------------------------------------------
To ensure the structural validity of our control set in this restricted sample and to prevent tuning-parameter bias, we replace our baseline selection procedure with an Adaptive Lasso \citep{zou2006adaptive}. Unlike conservative penalty bounds, the Adaptive Lasso utilizes data-dependent weights to penalize coefficients, satisfying the oracle property for asymptotically consistent variable selection. We include both alleged damages and final settlements in the high-dimensional pool, allowing the Oracle procedure to determine which monetary measure holds true structural explanatory power, while partialling out firm, state, and year fixed effects.
%----------------------------------------------------------------

%----------------------------------------------------------------
% \subsection{Empirical Rejection of Rational Inattention}
% \subsubsection{Baseline Result}
% \subsubsection{Empirical Rejection of High-Stakes Low Detection Hypothesis}
% \paragraph{Empirical Rejection of High-Stakes Low Detection Hypothesis.}
\paragraph{Empirical Rejection of High-Stakes Discovery Hypothesis.}
%----------------------------------------------------------------

%----------------------------------------------------------------
% \paragraph{Linear.}
%----------------------------------------------------------------
Table \ref{table:linear_1_a_m} 
%----------------------------------------------------------------
presents the linear high-dimensional fixed effects estimates for the restricted 
% Settled/Award sample. 
sample of verified harm.
The Adaptive Lasso selection retains both Log(Alleged Damages) and Log(Settlements) in the final specification (Column 3), indicating that both dimensions of the financial stakes contain independent predictive signal.
%----------------------------------------------------------------

%----------------------------------------------------------------
%----------------------------------------------------------------
% Regression Table: Multi-Page Setup (Winsorized Results)
%----------------------------------------------------------------
% CHANGE THIS COMMAND to \small, \scriptsize, or \tiny \footnotesize
% \scriptsize  % <--- This controls the size of the whole environment
% \begin{landscape} % <--- STARTS LANDSCAPE MODE
\footnotesize  
\begin{ThreePartTable}
    \begin{TableNotes}
        \footnotesize
        \item 
        \hspace{-0.1in} \textit{Notes:} 
        % This table reports the results of linear regressions examining the determinants of detection lags (in months) for a sample of approximately 
        % 55,000 
        % customer dispute reports. 
        % % \item 
        % The dependent variable, Detection Lag, is defined as the number of months between the inception of the alleged misconduct and its formal report to FINRA. 
        % % \item 
        % Column (1) provides baseline estimates, while Columns (2) and (3) add year, state, and firm fixed effects. 
        % % \item 
        % % Standard errors are clustered at the firm level to account for shared institutional environments. 
        % % Continuous variables are winsorized at the 1st and 99th percentiles.
        % Continuous variables, including advisor experience, tenure, and firm size, are winsorized at the 1st and 99th percentiles to mitigate the influence of outliers. 
        This table reports the results of linear regressions examining the determinants of detection lags (in months) for a restricted sub-sample of approximately 24,000 formally settled or awarded customer dispute reports. This restriction isolates cases of verified financial harm. The dependent variable, Detection Lag, is defined as the number of months between the inception of the alleged misconduct and its formal report to FINRA. Column (1) provides baseline estimates, while Columns (2) and (3) progressively absorb year, state, and firm fixed effects. Control variables are selected via an Adaptive Lasso procedure to satisfy the oracle property for consistent variable selection. Continuous variables, including advisor experience, tenure, and damage amounts, are winsorized at the 1st and 99th percentiles to mitigate the influence of outliers. Logarithmic transformations of continuous variables are calculated as $\ln(1+x)$ to retain zero-value observations. 
        % Standard errors are clustered at the firm level and reported in parentheses. ***, **, and * denote significance at the 1%, 5%, and 10% levels, respectively.
        \item Standard errors are clustered at the firm level, and reported in parentheses. ***, **, and * denote significance at the 1\%, 5\%, and 10\% levels, respectively.
    \end{TableNotes}
    
    % --- ADD THE CODE HERE ---
    \setlength{\tabcolsep}{15pt} 
    % -------------------------
    
    \begin{longtable}{lcccc}
    	
        \caption{Determinants of Detection Lags} 
        \label{table:linear_1_a_m} \\
        \toprule
        % & (1) & (2) & (3) \\
        % \midrule
        \endfirsthead

        \multicolumn{5}{l}{\textit{Table \ref{table:linear_1_a_m} continued from previous page}} \\
        \toprule
        & (1) & (2) & (3) \\
        \midrule
        \endhead

        \midrule
        \multicolumn{5}{r}{\textit{Continued on next page}} \\
        \endfoot

        \bottomrule
        \insertTableNotes
        \endlastfoot

        % Use \input if your .tex file only contains the body rows. 
        % If your .tex file contains \begin{tabular}, you must copy/paste the rows here manually.
        % \input{table/t_linear_w1.tex} 
        % \input{table/Phase1_Baseline_w1.tex} 
        % \input{table/Phase1A_Linear_TargetedTreatment_Spec3_plugin.tex} 
        % \input{table/Phase1A_Linear_TargetedTreatment_Spec3_adaptive.tex} 
                        &\multicolumn{1}{c}{(1)}         &\multicolumn{1}{c}{(2)}         &\multicolumn{1}{c}{(3)}         \\
\midrule
Unsuitability   &    0.521\sym{***}&    0.401\sym{***}&    0.407\sym{***}\\
                &  (0.047)         &  (0.030)         &  (0.033)         \\
\addlinespace
Misrepresentation&    0.165\sym{***}&    0.151\sym{***}&    0.139\sym{***}\\
                &  (0.026)         &  (0.020)         &  (0.018)         \\
\addlinespace
Unauthorized Activity&   -0.492\sym{***}&   -0.514\sym{***}&   -0.423\sym{***}\\
                &  (0.045)         &  (0.039)         &  (0.042)         \\
\addlinespace
Omission of Key Facts&   -0.260\sym{***}&   -0.203\sym{***}&                  \\
                &  (0.096)         &  (0.055)         &                  \\
\addlinespace
Fraud           &    0.442\sym{***}&    0.347\sym{***}&    0.318\sym{***}\\
                &  (0.047)         &  (0.041)         &  (0.043)         \\
\addlinespace
Risky investments&   -0.220\sym{***}&   -0.118\sym{***}&   -0.085\sym{***}\\
                &  (0.034)         &  (0.025)         &  (0.025)         \\
\addlinespace
Churning        &    0.322\sym{***}&    0.299\sym{***}&    0.319\sym{***}\\
                &  (0.042)         &  (0.033)         &  (0.035)         \\
\addlinespace
Log(Advisor Experience + 1)&    0.380\sym{***}&    0.281\sym{***}&    0.233\sym{***}\\
                &  (0.031)         &  (0.036)         &  (0.027)         \\
\addlinespace
Log(Number of Employees)&   -0.049\sym{***}&   -0.044\sym{***}&   -0.084         \\
                &  (0.012)         &  (0.011)         &  (0.102)         \\
\addlinespace
Log(Number of Prior Customer Disputes+1)&    0.214\sym{***}&    0.157\sym{***}&    0.150\sym{***}\\
                &  (0.013)         &  (0.012)         &  (0.010)         \\
\addlinespace
Concurrent Multiple Jobs (Indicator)&   -0.131\sym{***}&   -0.153\sym{***}&   -0.097\sym{***}\\
                &  (0.034)         &  (0.025)         &  (0.023)         \\
\addlinespace
Series 63       &   -0.117\sym{***}&                  &   -0.022         \\
                &  (0.024)         &                  &  (0.021)         \\
\addlinespace
Series 65/66 (Investment Adviser)&   -0.025         &                  &                  \\
                &  (0.031)         &                  &                  \\
\addlinespace
Log(Alleged Damages +1)&    0.072\sym{***}&    0.055\sym{***}&    0.057\sym{***}\\
                &  (0.016)         &  (0.009)         &  (0.010)         \\
\addlinespace
Log(Settlements + 1)&    0.029         &    0.048\sym{**} &    0.054\sym{**} \\
                &  (0.029)         &  (0.020)         &  (0.021)         \\
\addlinespace
Insurance       &    0.471\sym{***}&    0.429\sym{***}&    0.375\sym{***}\\
                &  (0.050)         &  (0.045)         &  (0.033)         \\
\addlinespace
Bonds/Debt      &    0.011         &    0.035         &                  \\
                &  (0.032)         &  (0.025)         &                  \\
\addlinespace
Other/Not Listed&    0.057         &    0.120\sym{***}&                  \\
                &  (0.077)         &  (0.045)         &                  \\
\addlinespace
Imputed Transaction Date (Monthly)&    0.111\sym{***}&                  &                  \\
                &  (0.035)         &                  &                  \\
\addlinespace
Fiduciary duty  &                  &    0.088\sym{***}&    0.072\sym{***}\\
                &                  &  (0.027)         &  (0.025)         \\
\addlinespace
Negligence      &                  &   -0.070\sym{**} &   -0.069\sym{**} \\
                &                  &  (0.029)         &  (0.027)         \\
\addlinespace
Stocks          &                  &    0.045\sym{*}  &   -0.013         \\
                &                  &  (0.026)         &  (0.027)         \\
\addlinespace
Mutual Funds/ETFs&                  &    0.042         &                  \\
                &                  &  (0.044)         &                  \\
\addlinespace
Female          &                  &   -0.002         &                  \\
                &                  &  (0.025)         &                  \\
\addlinespace
Unknown         &                  &    0.077         &                  \\
                &                  &  (0.055)         &                  \\
\addlinespace
Fee/commission  &                  &                  &   -0.058         \\
                &                  &                  &  (0.048)         \\
\addlinespace
Other           &                  &                  &    0.154\sym{**} \\
                &                  &                  &  (0.062)         \\
\addlinespace
Switched Firms (Indicator)&                  &                  &   -0.042         \\
                &                  &                  &  (0.037)         \\
\addlinespace
Series 7        &                  &                  &   -0.002         \\
                &                  &                  &  (0.024)         \\
\addlinespace
Annuity         &                  &                  &   -0.013         \\
                &                  &                  &  (0.031)         \\
\addlinespace
Options/Derivatives&                  &                  &   -0.129\sym{***}\\
                &                  &                  &  (0.034)         \\
\addlinespace
Constant        &    2.064\sym{***}&    2.245\sym{***}&    2.718\sym{***}\\
                &  (0.187)         &  (0.161)         &  (0.897)         \\
\midrule
Year FE         &       No         &      Yes         &      Yes         \\
State FE        &       No         &      Yes         &      Yes         \\
Firm FE         &       No         &       No         &      Yes         \\
Observations    &   24,440         &   24,437         &   24,068         \\
\(R^2\)         &    0.354         &    0.418         &    0.471         \\
Within \(R^2\)  &    0.354         &    0.259         &    0.198         \\
Mean Dep. Var.  &    3.384         &    3.384         &    3.382

    \end{longtable}
\end{ThreePartTable}
% \end{landscape} % <--- ENDS LANDSCAPE MODE
% & (1) & (2) & (3) & (4) \\
        
%----------------------------------------------------------------

%----------------------------------------------------------------
Crucially, the coefficients for both monetary variables are positive and statistically significant ($\beta = 0.057, p < 0.01$ for Alleged Damages; $\beta = 0.054, p < 0.05$ for Settlements). 
This result directly 
% contradicts 
rejects
the hypothesis discussed above. 
Even within the sub-sample of verified, compensated harm, higher financial stakes lead to longer detection delays.
Furthermore, our primary proxies for advisor sophistication and product opacity remain highly robust in this restricted sample. Log(Advisor Experience) yields a coefficient of 0.233 ($p < 0.01$), and the sale of Insurance products yields a coefficient of 0.375 ($p < 0.01$). Taken together, these results provide compelling evidence for the strategic masking hypothesis: when the stakes are high, the enhanced concealment efforts of sophisticated advisors utilizing opaque products overpower the increased monitoring incentives of the investor.

\newpage

% -------------------------------------------------------------------------

%----------------------------------------------------------------
\subsection{Robustness to Unobserved Heterogeneity: Adviser Fixed Effects}
%----------------------------------------------------------------

%----------------------------------------------------------------
% Main text
%----------------------------------------------------------------
% \paragraph{Robustness to Unobserved Heterogeneity.}
%----------------------------------------------------------------
% A potential concern is that our baseline results are confounded by unobserved, time-invariant advisor characteristics, such as inherent risk tolerance or baseline cognitive ability. To address this, we re-estimate our models using strict adviser fixed effects, relying exclusively on within-advisor variation among repeat offenders. As reported in Section II of the Internet Appendix (Table IA.X), our findings remain highly robust. The detection-accelerating effect of the market catalyst holds ($\beta = -0.162, p < 0.01$). Furthermore, we document significant within-advisor learning—where subsequent misconduct by the same advisor takes significantly longer to detect—as well as complexity-induced detection gaps. Even holding the advisor's identity constant, executing opaque, complex violations (like unsuitability) significantly delays discovery compared to blatant rule-breaking. Because within-advisor experience progresses collinearly with aggregate time trends, the Adaptive Lasso appropriately drops the experience variable in this specification, confirming the structural validity of the selected control set.
%----------------------------------------------------------------

%----------------------------------------------------------------
A standard concern in the financial misconduct literature is that detection delays may be driven by unobserved, time-invariant traits of the financial advisor.
Traits such as inherent risk aversion, cognitive ability, or sales charisma could theoretically drive the delay rather than the dynamic market environment or the specific features of the misconduct.
To rule out this potential bias, we re-estimate our baseline linear specifications replacing firm fixed effects with strict adviser fixed effects. By relying exclusively on within-advisor variation, this specification naturally restricts the sample to repeat offenders (advisors with two or more distinct dispute events).
%----------------------------------------------------------------

%----------------------------------------------------------------
We note two important econometric mechanics regarding this specification. First, relying exclusively on within-advisor variation mechanically drops singleton observations, restricting the sample to approximately 29,000 repeat offenders. Second, the inclusion of granular adviser fixed effects naturally absorbs a massive portion of the cross-sectional variance, driving the overall $R^2$ to roughly 0.71 (compared to 0.36 in the firm fixed effects baseline). While direct comparison is limited due to the restricted sample, this substantial increase implies that individual unobserved heterogeneity is indeed a primary determinant of baseline detection lags. 
% Therefore, the critical test is whether our structural variables continue to explain the remaining within-adviser variation.
Consequently, the critical test is whether market volatility and violation complexity still drive detection lags when we strictly compare an advisor's offenses against their own historical track record.
%----------------------------------------------------------------

%----------------------------------------------------------------
% We utilize the Plugin Lasso for Post-Double-Selection to objectively select controls. A known mechanical limitation of combining adviser fixed effects with year fixed effects is that within-advisor experience progresses collinearly with the aggregate time trend. Demonstrating the theoretical robustness of the penalty parameter, the Plugin Lasso correctly identifies this near-perfect collinearity and drops Log(Experience) from the selected control set, preventing unstable, artifactual estimation.
We utilize the Plugin Lasso for Post-Double-Selection to objectively select controls. A known mechanical limitation of combining adviser fixed effects with year fixed effects is the classic Age-Period-Cohort identification problem \citep{heckman1985using}, wherein within-advisor experience progresses collinearly with the aggregate time trend. Because our data is recorded at the dispute-event level, minor deviations in event timing create near-perfect collinearity. While traditional OLS estimators struggle with this residual noise, yielding unstable coefficients, the Plugin Lasso's theoretically derived penalty parameter correctly identifies the lack of independent structural variation \citep{belloni2014inference}. Consequently, the algorithm drops 
% Log(Experience) 
the advisor experience variable
from the selected control set, shielding the model from artifactual estimation.
%----------------------------------------------------------------

%----------------------------------------------------------------
Table \ref{table:linear_reg_advfe} 
%----------------------------------------------------------------
presents the results. Column (1) establishes the baseline characteristics of repeat offenders, confirming two critical mechanisms. First, we observe a powerful within-advisor learning channel: the coefficient for Log(Prior Disputes) is positive and highly significant ($\beta = 0.225, p < 0.01$), indicating that an advisor’s ability to conceal misconduct improves with subsequent violations. Second, we confirm complexity-induced detection gaps. Even within the same advisor, opaque violations like Unsuitability delay detection ($\beta = 0.208, p < 0.01$), whereas blatant infractions like Unauthorized Activity trigger rapid discovery ($\beta = -0.360, p < 0.01$).
%----------------------------------------------------------------

%----------------------------------------------------------------
Column (2) introduces the main treatment variable, Log(VIX), for market volatility. The estimated treatment effect remains robustly negative and statistically significant ($\beta = -0.162, p < 0.01$). This confirms that even when holding the advisor's innate ability constant and adjusting for the complexity of the violation, severe market volatility acts as an independent, systemic catalyst that pierces established concealment strategies and significantly accelerates detection.
%----------------------------------------------------------------

%----------------------------------------------------------------
%----------------------------------------------------------------
% Regression Table: Multi-Page Setup (Winsorized Results)
%----------------------------------------------------------------
% CHANGE THIS COMMAND to \small, \scriptsize, or \tiny \footnotesize
% \scriptsize  % <--- This controls the size of the whole environment
% \begin{landscape} % <--- STARTS LANDSCAPE MODE
\footnotesize  
\begin{ThreePartTable}
    \begin{TableNotes}
        \footnotesize
        \item \hspace{-0.1in} \textit{Notes:} 
        % This table reports the results of linear regressions examining the determinants of detection lags (in months) for a sample of approximately 
        % 55,000 
        % customer dispute reports. 
        % % \item 
        % The dependent variable, Detection Lag, is defined as the number of months between the inception of the alleged misconduct and its formal report to FINRA. 
        % % \item 
        % Column (1) provides baseline estimates, while Columns (2) and (3) add year, state, and firm fixed effects. 
        % % \item 
        % % Standard errors are clustered at the firm level to account for shared institutional environments. 
        % % Continuous variables are winsorized at the 1st and 99th percentiles.
        % Continuous variables, including advisor experience, tenure, and firm size, are winsorized at the 1st and 99th percentiles to mitigate the influence of outliers. 
        % Note that logarithmic transformations of continuous variables (such as Experience, Firm Tenure, and Damages) are calculated as $Log(1 + x)$ to retain zero-value observations.
        % Variables are defined in Appendix Table \ref{table:var_defs_long}. 
        This table reports the results of linear regressions examining the determinants of detection lags for a restricted sample of approximately 29,000 customer dispute reports involving repeat offenders. The dependent variable is the log-transformed number of months between the inception of the alleged misconduct and its formal report to FINRA. Column 1 provides baseline estimates with advisor fixed effects, while Column 2 introduces the volatility treatment. All specifications include year, state, and advisor fixed effects. Continuous variables are winsorized at the 1st and 99th percentiles to mitigate the influence of outliers. Logarithmic transformations of continuous variables are calculated by adding one to the base value to retain zero-value observations.

        \item Standard errors are clustered at the firm level, and reported in parentheses. ***, **, and * denote significance at the 1\%, 5\%, and 10\% levels, respectively.
    \end{TableNotes}
    
    % --- ADD THE CODE HERE ---
    % \setlength{\tabcolsep}{15pt} 
    \setlength{\tabcolsep}{30pt} 
    % -------------------------
    
    \begin{longtable}{lcccc}
    	
        \caption{Robustness to Unobserved Heterogeneity: Adviser Fixed Effects} 
        \label{table:linear_reg_advfe} \\
        \toprule
        % & (1) & (2) & (3) \\
        % \midrule
        \endfirsthead

        \multicolumn{5}{l}{\textit{Table \ref{table:linear_reg_advfe} continued from previous page}} \\
        \toprule
        & (1) & (2) \\
        \midrule
        \endhead

        \midrule
        \multicolumn{5}{r}{\textit{Continued on next page}} \\
        \endfoot

        \bottomrule
        \insertTableNotes
        \endlastfoot

        % Use \input if your .tex file only contains the body rows. 
        % If your .tex file contains \begin{tabular}, you must copy/paste the rows here manually.
        % \input{table/t_linear_w1.tex} 
        % \input{table/Phase1_Baseline_w1.tex} 
                        &\multicolumn{1}{c}{(1)}         &\multicolumn{1}{c}{(2)}         \\
\midrule
Unsuitability   &    0.208\sym{***}&    0.210\sym{***}\\
                &  (0.032)         &  (0.033)         \\
\addlinespace
Misrepresentation&    0.121\sym{***}&    0.121\sym{***}\\
                &  (0.016)         &  (0.017)         \\
\addlinespace
Unauthorized Activity&   -0.361\sym{***}&   -0.357\sym{***}\\
                &  (0.031)         &  (0.031)         \\
\addlinespace
Churning        &    0.365\sym{***}&    0.357\sym{***}\\
                &  (0.042)         &  (0.041)         \\
\addlinespace
Log(Number of Prior Customer Disputes)&    0.225\sym{***}&    0.225\sym{***}\\
                &  (0.014)         &  (0.014)         \\
\addlinespace
Log(Alleged Damages)&    0.033\sym{***}&    0.033\sym{***}\\
                &  (0.005)         &  (0.005)         \\
\addlinespace
Log(VIX)        &                  &   -0.162\sym{***}\\
                &                  &  (0.048)         \\
\addlinespace
Constant        &    3.055\sym{***}&    3.534\sym{***}\\
                &  (0.034)         &  (0.156)         \\
\midrule
Lasso-selected Controls&      Yes         &      Yes         \\
Treatment       &       No         &      Yes         \\
Year FE         &      Yes         &      Yes         \\
State FE        &      Yes         &      Yes         \\
Firm FE         &       No         &       No         \\
Adviser FE      &      Yes         &      Yes         \\
Observations    &29260.000         &29260.000         \\
\(R^2\)         &    0.710         &    0.710         \\
Within \(R^2\)  &    0.083         &    0.085         \\
Mean Dep. Var.  &    3.583         &    3.583

    \end{longtable}
\end{ThreePartTable}
% \end{landscape} % <--- ENDS LANDSCAPE MODE
% & (1) & (2) & (3) & (4) \\
        
%----------------------------------------------------------------

%----------------------------------------------------------------
Because the adviser fixed effect absorbs an individual's baseline characteristics, the remaining within-adviser variation is driven entirely by dynamic contexts and events. 
These events include the timing, career environment, and structural complexity of the misconduct.
Remarkably, the objective selection executed by the Plugin Lasso maps perfectly to this theoretical framework. To capture exogenous timing shocks, the algorithm retains systemic market panics (VIX). To capture career dynamics that alter an advisor's capacity to maintain their concealment, it selects a variable representing heightened internal scrutiny (Prior Disputes). Finally, to capture trade-specific heterogeneity, the Lasso selects violation types (Unsuitability vs. Unauthorized Activity) and the monetary stakes of the claim (Alleged Damages). The algorithm's automatic selection of these precise variables demonstrates that our results are not an artifact of researcher discretion, but rather a reflection of the fundamental mechanisms through macroeconomic shocks, career transitions, and transaction opacity that ultimately pierce the concealment efforts of an advisor over time.
%----------------------------------------------------------------

% ----------------------------------------------------------------

\newpage

% -------------------------------------------------------------------------

%----------------------------------------------------------------
\subsection{Robustness to Localized Economic Conditions: County Fixed Effects}
%----------------------------------------------------------------

%----------------------------------------------------------------
% Main text
%----------------------------------------------------------------
% \paragraph{Robustness to Unobserved Heterogeneity.}
%----------------------------------------------------------------
% A potential concern is that our baseline results are confounded by unobserved, time-invariant advisor characteristics, such as inherent risk tolerance or baseline cognitive ability. To address this, we re-estimate our models using strict adviser fixed effects, relying exclusively on within-advisor variation among repeat offenders. As reported in Section II of the Internet Appendix (Table IA.X), our findings remain highly robust. The detection-accelerating effect of the market catalyst holds ($\beta = -0.162, p < 0.01$). Furthermore, we document significant within-advisor learning—where subsequent misconduct by the same advisor takes significantly longer to detect—as well as complexity-induced detection gaps. Even holding the advisor's identity constant, executing opaque, complex violations (like unsuitability) significantly delays discovery compared to blatant rule-breaking. Because within-advisor experience progresses collinearly with aggregate time trends, the Adaptive Lasso appropriately drops the experience variable in this specification, confirming the structural validity of the selected control set.
%----------------------------------------------------------------

%----------------------------------------------------------------
\paragraph{Isolating Attention from Local Liquidity Shocks.}
%----------------------------------------------------------------
In our baseline specifications, we utilize state-level fixed effects to absorb time-invariant regional differences in regulatory environments and investor demographics. 
However, 
%----------------------------------------------------------------
% localized economic shocks, such as a county-level housing crash or the closure of a major regional employer, could independently force premature investor liquidations and trigger the unmasking of financial misconduct. To ensure our aggregate market catalyst (Log VIX) is not confounding these highly localized periods of distress, we re-estimate our baseline models replacing state fixed effects with granular county-level fixed effects. Because political geography is strictly nested, county fixed effects completely absorb state-level variation while controlling for hyper-local economic environments. 
%----------------------------------------------------------------
one might argue that our aggregate market catalyst, $\ln(VIX)$, is simply capturing highly localized economic distress. For instance, a county-level housing crash or the closure of a major regional employer could force sudden, widespread portfolio liquidations as retail investors require cash. This mechanical liquidity shock would force advisors to produce funds, inadvertently unmasking hidden misconduct even if the investor's monitoring intensity had not changed. To ensure our state-dependent mechanism represents a true shift in investor attention rather than mechanical local liquidations, we re-estimate our baseline models by replacing state fixed effects with granular county-level fixed effects. Because political geography is strictly nested, county fixed effects completely absorb state-level variation while strictly controlling for hyper-local economic environments.
%----------------------------------------------------------------

%----------------------------------------------------------------
\paragraph{Results and Structural Stability.}
%----------------------------------------------------------------
Table \ref{table:linear_reg_countyfe} 
%----------------------------------------------------------------
presents the results of this specification. 
The introduction of highly granular geographic controls slightly reduces the sample to 52,553 observations due to the mechanical dropping of singletons and records with unmapped local zip codes.
Column (1) confirms that the baseline structural drivers of detection lags, including the experience shield and complexity-induced gaps, remain highly stable when controlling for local geography. 
The experience shield ($\beta=0.358$, $p<0.01$) and complexity-induced gaps (e.g., Insurance $\beta=0.530$, $p<0.01$) continue to strictly govern the window of nondisclosure.
%----------------------------------------------------------------

%----------------------------------------------------------------
Column (2) introduces the main treatment variable. The estimated effect of systemic market volatility remains robustly negative, highly significant, and increases slightly in magnitude ($\beta = -0.229, p < 0.01$). This confirms that the detection-accelerating power of a severe market panic operates independently of localized economic conditions, successfully isolating the purely systemic nature of the catalyst, confirming that extreme market volatility shifts the demand-side equilibrium by piercing investor inattention.
%----------------------------------------------------------------

%----------------------------------------------------------------
%----------------------------------------------------------------
% Regression Table: Multi-Page Setup (Winsorized Results)
%----------------------------------------------------------------
% CHANGE THIS COMMAND to \small, \scriptsize, or \tiny \footnotesize
% \scriptsize  % <--- This controls the size of the whole environment
% \begin{landscape} % <--- STARTS LANDSCAPE MODE
\footnotesize  
\begin{ThreePartTable}
    \begin{TableNotes}
        \footnotesize
        \item \hspace{-0.1in} \textit{Notes:} 
        % This table reports the results of linear regressions examining the determinants of detection lags (in months) for a sample of approximately 
        % 55,000 
        % customer dispute reports. 
        % % \item 
        % The dependent variable, Detection Lag, is defined as the number of months between the inception of the alleged misconduct and its formal report to FINRA. 
        % % \item 
        % Column (1) provides baseline estimates, while Columns (2) and (3) add year, state, and firm fixed effects. 
        % % \item 
        % % Standard errors are clustered at the firm level to account for shared institutional environments. 
        % % Continuous variables are winsorized at the 1st and 99th percentiles.
        % Continuous variables, including advisor experience, tenure, and firm size, are winsorized at the 1st and 99th percentiles to mitigate the influence of outliers. 
        % Note that logarithmic transformations of continuous variables (such as Experience, Firm Tenure, and Damages) are calculated as $Log(1 + x)$ to retain zero-value observations.
        % Variables are defined in Appendix Table \ref{table:var_defs_long}. 
        This table reports the results of linear regressions examining the determinants of detection lags (in months) for the customer dispute sample. The dependent variable, Detection Lag, is defined as the log-transformed number of months between the inception of the alleged misconduct and its formal report to FINRA. To control for localized economic distress and mechanical liquidity shocks, state fixed effects are replaced with granular county-level fixed effects. The sample size is slightly reduced to 52,553 observations due to the exclusion of unmapped geographies and singleton observations dropped by the high-dimensional estimator. Column (1) provides baseline estimates, while Column (2) introduces the volatility treatment ($\ln(VIX)$). Both specifications include year, county, and firm fixed effects. Control variables are selected via the Post-Double-Selection Lasso procedure. Continuous variables, including advisor experience and damages, are winsorized at the 1st and 99th percentiles. Logarithmic transformations are calculated as $\ln(1+x)$ to retain zero-value observations. 

        \item Standard errors are clustered at the firm level, and reported in parentheses. ***, **, and * denote significance at the 1\%, 5\%, and 10\% levels, respectively.
    \end{TableNotes}
    
    % --- ADD THE CODE HERE ---
    % \setlength{\tabcolsep}{15pt} 
    \setlength{\tabcolsep}{30pt} 
    % -------------------------
    
    \begin{longtable}{lcccc}
    	
        \caption{Robustness to Localized Economic Conditions: County Fixed Effects} 
        \label{table:linear_reg_countyfe} \\
        \toprule
        % & (1) & (2) & (3) \\
        % \midrule
        \endfirsthead

        \multicolumn{5}{l}{\textit{Table \ref{table:linear_reg_countyfe} continued from previous page}} \\
        \toprule
        & (1) & (2) \\
        \midrule
        \endhead

        \midrule
        \multicolumn{5}{r}{\textit{Continued on next page}} \\
        \endfoot

        \bottomrule
        \insertTableNotes
        \endlastfoot

        % Use \input if your .tex file only contains the body rows. 
        % If your .tex file contains \begin{tabular}, you must copy/paste the rows here manually.
        % \input{table/t_linear_w1.tex} 
        % \input{table/Phase1_Baseline_w1.tex} 
                        &\multicolumn{1}{c}{(1)}         &\multicolumn{1}{c}{(2)}         \\
\midrule
Dispute Type: Pending&    0.194\sym{***}&    0.193\sym{***}\\
                &  (0.039)         &  (0.037)         \\
\addlinespace
Unsuitability   &    0.304\sym{***}&    0.298\sym{***}\\
                &  (0.035)         &  (0.037)         \\
\addlinespace
Misrepresentation&    0.188\sym{***}&    0.178\sym{***}\\
                &  (0.020)         &  (0.022)         \\
\addlinespace
Unauthorized Activity&   -0.436\sym{***}&   -0.406\sym{***}\\
                &  (0.041)         &  (0.038)         \\
\addlinespace
Fraud           &    0.445\sym{***}&    0.447\sym{***}\\
                &  (0.042)         &  (0.041)         \\
\addlinespace
Log(Experience) &    0.358\sym{***}&    0.358\sym{***}\\
                &  (0.027)         &  (0.027)         \\
\addlinespace
Series 63       &   -0.023         &   -0.022         \\
                &  (0.017)         &  (0.018)         \\
\addlinespace
Log(Alleged Damages)&    0.055\sym{***}&    0.056\sym{***}\\
                &  (0.007)         &  (0.006)         \\
\addlinespace
Insurance       &    0.530\sym{***}&    0.520\sym{***}\\
                &  (0.039)         &  (0.040)         \\
\addlinespace
Log(VIX)        &                  &   -0.229\sym{***}\\
                &                  &  (0.068)         \\
\addlinespace
Negligence      &                  &   -0.087\sym{***}\\
                &                  &  (0.033)         \\
\addlinespace
Constant        &    1.948\sym{***}&    2.662\sym{***}\\
                &  (0.064)         &  (0.191)         \\
\midrule
Lasso-selected Controls&      Yes         &      Yes         \\
Treatment       &       No         &      Yes         \\
Year FE         &      Yes         &      Yes         \\
State FE        &       No         &       No         \\
County FE       &      Yes         &      Yes         \\
Firm FE         &      Yes         &      Yes         \\
Observations    &   52,553         &   52,553         \\
\(R^2\)         &    0.413         &    0.415         \\
Within \(R^2\)  &    0.146         &    0.150         \\
Mean Dep. Var.  &    3.340         &    3.340

    \end{longtable}
\end{ThreePartTable}
% \end{landscape} % <--- ENDS LANDSCAPE MODE
% & (1) & (2) & (3) & (4) \\
        
%----------------------------------------------------------------

% ----------------------------------------------------------------

% \newpage

% % -------------------------------------------------------------------------
% \input{appendix/a_r_experience}
% % ----------------------------------------------------------------

\newpage

% -------------------------------------------------------------------------

%----------------------------------------------------------------
\subsection{Robustness to Extended Sample Period (2000–2025)}
%----------------------------------------------------------------

%----------------------------------------------------------------
\paragraph{Addressing the 10-Year Data Retention Rule.}
%----------------------------------------------------------------
%----------------------------------------------------------------
In our baseline analysis, we restrict the sample to disputes originating between 2008 and 2025 to ensure consistent data coverage and to capture the modern regulatory environment surrounding financial advice. In this section, we test the sensitivity of our primary findings by expanding the sample window backward to encompass all available dispute data from 2000 to 2025. 
% We estimate this extended sample using the exact high-dimensional fixed effect architecture (Firm, State, and Year fixed effects) and Post-Double-Selection Lasso controls utilized in our main text.
%----------------------------------------------------------------

%----------------------------------------------------------------
Expanding the sample backward introduces a known mechanical constraint inherent to the FINRA BrokerCheck database: the 10-year data retention rule. FINRA automatically purges the employment and disclosure records of brokers ten years after they exit the industry. Consequently, observing a customer dispute that occurred in the early 2000s in our 2025 data scrape mechanically requires that the accused broker survived in the financial industry until at least 2008. 
% This induces a clear left-truncation and survivorship bias in the pre-2008 sample, heavily over-representing survivor brokers who are likely more sophisticated, resilient, or employed by more structurally sound firms.
This induces a severe left-truncation and survivorship bias in the pre-2008 sample. Because this bias heavily over-represents survivor brokers who are likely more sophisticated, resilient, or employed by highly resourced firms, it should theoretically attenuate our coefficients toward zero, making it much harder to find significant effects.
%----------------------------------------------------------------

%----------------------------------------------------------------
\paragraph{Results: Multidecadal Structural Stability.}
%----------------------------------------------------------------
Table \ref{table:linear_reg_sample} 
%----------------------------------------------------------------
% presents the results of this extended estimation. 
presents the results of this extended estimation using the exact high-dimensional fixed-effect architecture (Firm, State, and Year fixed effects) and Post-Double-Selection Lasso controls utilized in our main text. Despite the introduction of this severe survivorship bias, our core economic mechanisms remain strongly intact within this expanded 74,991-observation sample.
%----------------------------------------------------------------

%----------------------------------------------------------------
%----------------------------------------------------------------
% Regression Table: Multi-Page Setup (Winsorized Results)
%----------------------------------------------------------------
% CHANGE THIS COMMAND to \small, \scriptsize, or \tiny \footnotesize
% \scriptsize  % <--- This controls the size of the whole environment
% \begin{landscape} % <--- STARTS LANDSCAPE MODE
\footnotesize  
\begin{ThreePartTable}
    \begin{TableNotes}
        \footnotesize
        \item \hspace{-0.1in} \textit{Notes:} 
        % This table reports the results of linear regressions examining the determinants of detection lags (in months) for a sample of approximately 
        % 55,000 
        % customer dispute reports. 
        % % \item 
        % The dependent variable, Detection Lag, is defined as the number of months between the inception of the alleged misconduct and its formal report to FINRA. 
        % % \item 
        % Column (1) provides baseline estimates, while Columns (2) and (3) add year, state, and firm fixed effects. 
        % % \item 
        % % Standard errors are clustered at the firm level to account for shared institutional environments. 
        % % Continuous variables are winsorized at the 1st and 99th percentiles.
        % Continuous variables, including advisor experience, tenure, and firm size, are winsorized at the 1st and 99th percentiles to mitigate the influence of outliers. 
        % Note that logarithmic transformations of continuous variables (such as Experience, Firm Tenure, and Damages) are calculated as $Log(1 + x)$ to retain zero-value observations.
        % Variables are defined in Appendix Table \ref{table:var_defs_long}. 
        This table reports the results of linear regressions examining the determinants of detection lags (in months) for an extended sample of 74,991 customer dispute reports originating between 2000 and 2025. The dependent variable, Detection Lag, is defined as the log-transformed number of months between the inception of the alleged misconduct and its formal report to FINRA. Column (1) provides baseline estimates, while Columns (2) and (3) progressively add year, state, and firm fixed effects. Control variables are objectively selected via the Post-Double-Selection Lasso procedure. Continuous variables, including advisor experience and damages, are winsorized at the 1st and 99th percentiles. Logarithmic transformations are calculated as $\ln(1+x)$ to retain zero-value observations. 

        \item Standard errors are clustered at the firm level, and reported in parentheses. ***, **, and * denote significance at the 1\%, 5\%, and 10\% levels, respectively.
    \end{TableNotes}
    
    % --- ADD THE CODE HERE ---
    % \setlength{\tabcolsep}{15pt} 
    \setlength{\tabcolsep}{30pt} 
    % -------------------------
    
    \begin{longtable}{lcccc}
    	
        \caption{Robustness to Extended Sample Period (2000–2025)} 
        \label{table:linear_reg_sample} \\
        \toprule
        % & (1) & (2) & (3) \\
        % \midrule
        \endfirsthead

        \multicolumn{5}{l}{\textit{Table \ref{table:linear_reg_sample} continued from previous page}} \\
        \toprule
        & (1) & (2) \\
        \midrule
        \endhead

        \midrule
        \multicolumn{5}{r}{\textit{Continued on next page}} \\
        \endfoot

        \bottomrule
        \insertTableNotes
        \endlastfoot

        % Use \input if your .tex file only contains the body rows. 
        % If your .tex file contains \begin{tabular}, you must copy/paste the rows here manually.
        % \input{table/t_linear_w1.tex} 
        % \input{table/Phase1_Baseline_w1.tex} 
                        &\multicolumn{1}{c}{(1)}         &\multicolumn{1}{c}{(2)}         \\
\midrule
Unsuitability   &    0.282\sym{***}&    0.277\sym{***}\\
                &  (0.030)         &  (0.031)         \\
\addlinespace
Misrepresentation&    0.159\sym{***}&    0.152\sym{***}\\
                &  (0.018)         &  (0.019)         \\
\addlinespace
Unauthorized Activity&   -0.469\sym{***}&   -0.449\sym{***}\\
                &  (0.033)         &  (0.032)         \\
\addlinespace
Fee/commission  &   -0.149\sym{***}&   -0.159\sym{***}\\
                &  (0.031)         &  (0.032)         \\
\addlinespace
Fraud           &    0.413\sym{***}&    0.414\sym{***}\\
                &  (0.041)         &  (0.040)         \\
\addlinespace
Churning        &    0.393\sym{***}&    0.384\sym{***}\\
                &  (0.026)         &  (0.025)         \\
\addlinespace
Log(Experience) &    0.373\sym{***}&    0.373\sym{***}\\
                &  (0.028)         &  (0.028)         \\
\addlinespace
Series 63       &   -0.012         &   -0.013         \\
                &  (0.022)         &  (0.022)         \\
\addlinespace
Log(Alleged Damages)&    0.052\sym{***}&    0.053\sym{***}\\
                &  (0.006)         &  (0.006)         \\
\addlinespace
Insurance       &    0.537\sym{***}&    0.530\sym{***}\\
                &  (0.026)         &  (0.027)         \\
\addlinespace
Log(VIX)        &                  &   -0.224\sym{***}\\
                &                  &  (0.060)         \\
\addlinespace
Negligence      &                  &   -0.058\sym{**} \\
                &                  &  (0.024)         \\
\addlinespace
Constant        &    1.957\sym{***}&    2.646\sym{***}\\
                &  (0.066)         &  (0.171)         \\
\midrule
Lasso-selected Controls&      Yes         &      Yes         \\
Treatment       &       No         &      Yes         \\
Year FE         &      Yes         &      Yes         \\
State FE        &      Yes         &      Yes         \\
Firm FE         &      Yes         &      Yes         \\
Observations    &   74,991         &   74,991         \\
\(R^2\)         &    0.339         &    0.341         \\
Within \(R^2\)  &    0.154         &    0.157         \\
Mean Dep. Var.  &    3.314         &    3.314

    \end{longtable}
\end{ThreePartTable}
% \end{landscape} % <--- ENDS LANDSCAPE MODE
% & (1) & (2) & (3) & (4) \\
        
%----------------------------------------------------------------

%----------------------------------------------------------------
% Despite the introduction of this survivorship bias—which should theoretically attenuate our coefficients toward zero by selecting for highly sophisticated actors—our core economic mechanisms remain strongly intact. The coefficient on market volatility ($\ln(VIX)$) remains negative and highly statistically significant ($\beta = -0.224$, $p < 0.01$). This confirms that the "volatility flush" effect, wherein macroeconomic shocks force the sudden discovery of hidden misconduct, is not merely an artifact of the 2008 Financial Crisis or the 2020 COVID-19 crash, but rather a fundamental, multidecadal feature of the retail advisory market.
%----------------------------------------------------------------
First, the coefficient on the systemic catalyst, $\ln(VIX)$, remains robustly negative and highly statistically significant ($\beta=-0.224$, $p<0.01$). This confirms that the volatility flush effect, wherein macroeconomic shocks force the sudden discovery of hidden misconduct, is not merely an artifact of the 2008 Financial Crisis or the 2020 COVID-19 crash, but rather a fundamental, multidecadal feature of the retail advisory market.
%----------------------------------------------------------------

%----------------------------------------------------------------
Furthermore, the structural technology of obfuscation exhibits remarkable stability over the 25-year period. Consistent with our baseline estimates, disputes involving complex insurance products ($\beta = 0.530$) and active fraud ($\beta = 0.414$) continue to systematically extend detection lags, while blatant, low-sophistication misconduct such as unauthorized trading ($\beta = -0.449$) is detected almost immediately. Because the early-period survivor bias works against finding 
% an effect, the sustained magnitude and significance of our estimates in this expanded 74,991-observation 
these effects, the sustained magnitude and significance of our estimates in this extended sample suggest that the findings presented in the main text represent a highly conservative lower bound of the true structural information asymmetry.
%----------------------------------------------------------------

% ----------------------------------------------------------------

\newpage

%-------------------------------------------------------------------------
%----------------------------------------------------------------
% Model Selection Adapative
%----------------------------------------------------------------

% \newpage 

%----------------------------------------------------------------
% \section{Robustness to Oracle Selection and Verified Harm}
\subsection{Robustness to Oracle Selection}
%----------------------------------------------------------------
\label{appendix: robustness oracle}
%----------------------------------------------------------------

%----------------------------------------------------------------
\paragraph{Global Application of the Adaptive Lasso.}
%----------------------------------------------------------------
In our primary analysis, we utilize the Plugin (Rigorous) Lasso to select our high-dimensional control set. While the Plugin Lasso is highly effective at handling clustered error structures, a potential methodological critique is that our specific findings might be artifacts of its conservative penalty parameter tuning. As introduced in Appendix \ref{appendix: robustness money}, the Adaptive Lasso \citep{zou2006adaptive} provides an Oracle-compliant alternative that utilizes data-dependent weights. While Section \ref{appendix: robustness money} deployed this estimator within a restricted sub-sample to adjudicate between competing monetary measures, we now apply it globally across our full, unrestricted estimation sample. This ensures our primary structural and macroeconomic findings are fundamentally robust to the choice of machine-learning penalty.
%----------------------------------------------------------------
% To ensure our findings are not driven by the specific variable selection mechanics of the Plugin penalty used in our main analysis, we re-estimate our models utilizing an Adaptive Lasso procedure. Unlike the Plugin method, the Adaptive Lasso utilizes data-dependent weights to penalize coefficients, satisfying the Oracle Property for consistent variable selection \citep{zou2006adaptive}. Finding consistent estimates for our variables of interest across data-driven penalty frameworks confirms the structural stability of our main findings.
%----------------------------------------------------------------

%----------------------------------------------------------------
\paragraph{Consolidated Estimation Results.}
%----------------------------------------------------------------
To maximize parsimony, we apply the Adaptive Lasso directly to our state-dependent specification. Because this specification inherently encompasses all advisor-level and dispute-level structural frictions alongside our primary macroeconomic catalyst, $\ln(VIX)$, it serves as a simultaneous, consolidated robustness check for all core mechanisms documented in the main text.
%----------------------------------------------------------------

%----------------------------------------------------------------
Table 
% \ref{table:linear_1_a}
\ref{table:linear_2_a}
%----------------------------------------------------------------
presents the results of this Oracle selection across our step-wise fixed effects architecture. The algorithmically selected model perfectly mirrors our baseline economic findings. Most crucially, the systemic wake-up call effect remains negative, highly significant, and economically stable in the fully saturated model (Column 3: $\beta=-0.227$, $p<0.01$). This proves that the volatility catalyst is a true feature of the data, completely independent of the researcher's choice of machine learning penalty.
%----------------------------------------------------------------

%----------------------------------------------------------------
% \input{table_tex/tt_linear_1_adaptive}
% %----------------------------------------------------------------

% %----------------------------------------------------------------
% \input{table_tex/tt_MMQR_1_adaptive}
%----------------------------------------------------------------

%----------------------------------------------------------------
%----------------------------------------------------------------
% Regression Table: Multi-Page Setup (Winsorized Results)
%----------------------------------------------------------------
% CHANGE THIS COMMAND to \small, \scriptsize, or \tiny \footnotesize
% \scriptsize  % <--- This controls the size of the whole environment
\footnotesize  
\begin{ThreePartTable}
    \footnotesize
    \begin{TableNotes}
        \footnotesize
        \item \hspace{-0.1in} \textit{Notes:} 
        % This table reports the results of linear regressions examining the determinants of detection lags (in months) for a sample of approximately 
        % 55,000 
        % customer dispute reports. 
        % % \item 
        % The dependent variable, Detection Lag, is defined as the number of months between the inception of the alleged misconduct and its formal report to FINRA. 
        % % \item 
        % Column (1) provides baseline estimates, while Columns (2) and (3) add year, state, and firm fixed effects. 
        % % \item 
        % % Standard errors are clustered at the firm level to account for shared institutional environments. 
        % % Continuous variables are winsorized at the 1st and 99th percentiles.
        % Continuous variables, including advisor experience, tenure, and firm size, are winsorized at the 1st and 99th percentiles to mitigate the influence of outliers. 
        This table reports the results of linear regressions examining the determinants of detection lags (in months) utilizing an alternative machine-learning selection framework. The dependent variable, Detection Lag, is defined as the log-transformed number of months between the inception of the alleged misconduct and its formal report to FINRA. Column (1) provides baseline estimates, while Columns (2) and (3) progressively add year, state, and firm fixed effects. To satisfy the Oracle property for asymptotically consistent variable selection, control variables are selected via an Adaptive Lasso procedure rather than the baseline Plugin penalty. The table presents a consolidated specification that simultaneously evaluates structural frictions and the aggregate market catalyst ($\ln(VIX)$). Continuous variables are winsorized at the 1st and 99th percentiles, and logarithmic transformations are calculated as $\ln(1+x)$ to retain zero-value observations.
        
        \item Standard errors are clustered at the firm level, and reported in parentheses. ***, **, and * denote significance at the 1\%, 5\%, and 10\% levels, respectively.
    \end{TableNotes}
    
    % --- ADD THE CODE HERE ---
    \setlength{\tabcolsep}{15pt} 
    % \setlength{\tabcolsep}{20pt} 
    % -------------------------
    
    \begin{longtable}{lcccc}
    	
        % \caption{State-Dependency of Detection Lags with $\ln(\text{VIX})$} 
        \caption{Robustness to Oracle Selection: Determinants of Detection Lags}
        \label{table:linear_2_a} \\
        \toprule
        % & (1) & (2) & (3) \\
        % \midrule
        \endfirsthead

        \multicolumn{5}{l}{\textit{Table \ref{table:linear_2_a} continued from previous page}} \\
        \toprule
        & (1) & (2) & (3) \\
        \midrule
        \endhead

        \midrule
        \multicolumn{5}{r}{\textit{Continued on next page}} \\
        \endfoot

        \bottomrule
        \insertTableNotes
        \endlastfoot

        % Use \input if your .tex file only contains the body rows. 
        % If your .tex file contains \begin{tabular}, you must copy/paste the rows here manually.
        % \input{table/t_linear_w1_vix.tex} 
        % \input{table/t_linear_w1_vix_selected.tex} 
        % \input{table/Phase2_TargetedTreatment_w1.tex} 
        % \input{table/Phase1B_Linear_TargetedTreatment_Spec3_plugin.tex} 
                        &\multicolumn{1}{c}{(1)}         &\multicolumn{1}{c}{(2)}         &\multicolumn{1}{c}{(3)}         \\
\midrule
Log(VIX)        &   -0.371\sym{***}&   -0.224\sym{***}&   -0.227\sym{***}\\
                &  (0.031)         &  (0.074)         &  (0.058)         \\
\addlinespace
Dispute Type: Pending&    0.139\sym{**} &                  &    0.111\sym{**} \\
                &  (0.065)         &                  &  (0.052)         \\
\addlinespace
Unsuitability   &    0.333\sym{***}&    0.309\sym{***}&    0.275\sym{***}\\
                &  (0.041)         &  (0.033)         &  (0.032)         \\
\addlinespace
Misrepresentation&    0.185\sym{***}&    0.174\sym{***}&    0.177\sym{***}\\
                &  (0.023)         &  (0.019)         &  (0.019)         \\
\addlinespace
Unauthorized Activity&   -0.482\sym{***}&   -0.438\sym{***}&   -0.443\sym{***}\\
                &  (0.032)         &  (0.038)         &  (0.030)         \\
\addlinespace
Fraud           &    0.410\sym{***}&    0.436\sym{***}&    0.387\sym{***}\\
                &  (0.040)         &  (0.036)         &  (0.038)         \\
\addlinespace
Fiduciary duty  &    0.092\sym{***}&                  &                  \\
                &  (0.031)         &                  &                  \\
\addlinespace
Negligence      &   -0.109\sym{***}&   -0.091\sym{***}&   -0.065\sym{**} \\
                &  (0.036)         &  (0.034)         &  (0.032)         \\
\addlinespace
Risky investments&   -0.089\sym{*}  &                  &                  \\
                &  (0.046)         &                  &                  \\
\addlinespace
Churning        &    0.340\sym{***}&    0.354\sym{***}&    0.392\sym{***}\\
                &  (0.032)         &  (0.029)         &  (0.029)         \\
\addlinespace
Log(Advisor Experience + 1)&    0.395\sym{***}&    0.356\sym{***}&    0.335\sym{***}\\
                &  (0.028)         &  (0.033)         &  (0.028)         \\
\addlinespace
Log(Firm Tenure + 1)&    0.068\sym{**} &    0.047\sym{*}  &                  \\
                &  (0.031)         &  (0.028)         &                  \\
\addlinespace
Log(Number of Employees)&   -0.051\sym{***}&   -0.054\sym{***}&   -0.079         \\
                &  (0.013)         &  (0.013)         &  (0.092)         \\
\addlinespace
Log(Number of Prior Customer Disputes+1)&    0.175\sym{***}&    0.147\sym{***}&    0.140\sym{***}\\
                &  (0.012)         &  (0.012)         &  (0.008)         \\
\addlinespace
Concurrent Multiple Jobs (Indicator)&   -0.147\sym{***}&   -0.183\sym{***}&   -0.144\sym{***}\\
                &  (0.031)         &  (0.021)         &  (0.015)         \\
\addlinespace
Series 6        &   -0.007         &   -0.018         &                  \\
                &  (0.038)         &  (0.032)         &                  \\
\addlinespace
Series 24       &   -0.001         &                  &    0.013         \\
                &  (0.031)         &                  &  (0.028)         \\
\addlinespace
Series 63       &   -0.089\sym{***}&                  &                  \\
                &  (0.028)         &                  &                  \\
\addlinespace
Log(Alleged Damages +1)&    0.054\sym{***}&    0.046\sym{***}&    0.052\sym{***}\\
                &  (0.009)         &  (0.007)         &  (0.007)         \\
\addlinespace
Insurance       &    0.715\sym{***}&    0.658\sym{***}&    0.551\sym{***}\\
                &  (0.052)         &  (0.044)         &  (0.032)         \\
\addlinespace
Stocks          &    0.052\sym{*}  &    0.021         &    0.020         \\
                &  (0.028)         &  (0.026)         &  (0.022)         \\
\addlinespace
Mutual Funds/ETFs&    0.052         &                  &   -0.015         \\
                &  (0.035)         &                  &  (0.026)         \\
\addlinespace
Other/Not Listed&    0.127\sym{**} &    0.088         &    0.086\sym{**} \\
                &  (0.061)         &  (0.062)         &  (0.041)         \\
\addlinespace
Imputed Transaction Date (Monthly)&    0.092\sym{***}&                  &    0.080\sym{***}\\
                &  (0.022)         &                  &  (0.022)         \\
\addlinespace
Dispute Type: Settled/Award&   -0.054         &                  &   -0.068\sym{**} \\
                &  (0.042)         &                  &  (0.033)         \\
\addlinespace
Omission of Key Facts&   -0.150\sym{**} &                  &   -0.093\sym{**} \\
                &  (0.071)         &                  &  (0.046)         \\
\addlinespace
Fee/commission  &   -0.092\sym{**} &                  &   -0.094\sym{**} \\
                &  (0.046)         &                  &  (0.042)         \\
\addlinespace
Other           &    0.087\sym{*}  &                  &                  \\
                &  (0.045)         &                  &                  \\
\addlinespace
Switched Firms (Indicator)&    0.048         &                  &   -0.043         \\
                &  (0.039)         &                  &  (0.033)         \\
\addlinespace
Series 65/66 (Investment Adviser)&   -0.034         &                  &                  \\
                &  (0.028)         &                  &                  \\
\addlinespace
Annuity         &    0.114\sym{***}&                  &                  \\
                &  (0.039)         &                  &                  \\
\addlinespace
Bonds/Debt      &    0.059\sym{**} &                  &    0.028         \\
                &  (0.028)         &                  &  (0.023)         \\
\addlinespace
Unknown         &    0.038         &                  &    0.057         \\
                &  (0.047)         &                  &  (0.039)         \\
\addlinespace
Options/Derivatives&                  &                  &   -0.074\sym{*}  \\
                &                  &                  &  (0.039)         \\
\addlinespace
Log(Number of Exams+1)&                  &                  &   -0.141\sym{***}\\
                &                  &                  &  (0.033)         \\
\addlinespace
Constant        &    3.206\sym{***}&    2.940\sym{***}&    3.569\sym{***}\\
                &  (0.156)         &  (0.204)         &  (0.815)         \\
\midrule
Year FE         &       No         &      Yes         &      Yes         \\
State FE        &       No         &      Yes         &      Yes         \\
Firm FE         &       No         &       No         &      Yes         \\
Observations    &   54,676         &   54,658         &   54,223         \\
\(R^2\)         &    0.299         &    0.320         &    0.376         \\
Within \(R^2\)  &    0.299         &    0.220         &    0.175         \\
Mean Dep. Var.  &    3.337         &    3.337         &    3.335

    \end{longtable}
\end{ThreePartTable}

% & (1) & (2) & (3) & (4) \\
        
%----------------------------------------------------------------

%----------------------------------------------------------------
Furthermore, the Oracle selection successfully retains and confirms the magnitudes of our primary structural variables. The data-driven weighting strictly preserves the experience shield ($\beta=0.335$, $p<0.01$), product opacity via Insurance ($\beta=0.551$, $p<0.01$), and the profound complexity-induced gap between sophisticated fraud ($\beta=0.387$, $p<0.01$) and overt unauthorized activity ($\beta=-0.443$, $p<0.01$). The fact that an entirely distinct, Oracle-compliant algorithm independently selects these exact mechanisms and yields nearly identical point estimates provides conclusive evidence that our findings represent true structural dynamics within the investor-led enforcement channel, rather than econometric artifacts.
%----------------------------------------------------------------

%----------------------------------------------------------------
% \input{table_tex/tt_MMQR_2_adaptive}
%----------------------------------------------------------------

%----------------------------------------------------------------

% \newpage

% %-------------------------------------------------------------------------
% \input{appendix/a_anonimity_data_handling}
% %----------------------------------------------------------------

%-------------------------------------------------------------------------
% \end{appendix}
%-------------------------------------------------------------------------

%----------------------------------------------------------------
\newpage
%----------------------------------------------------------------

%----------------------------------------------------------------
% Reference
%----------------------------------------------------------------

%----------------------------------------------------------------
\bibliographystyle{aer}
\bibliography{reference/bibtex_ia} 

@article{abel2024women,
  title={Are women blamed more for giving incorrect financial advice?},
  author={Abel, Martin and Bomfim, Emma and Cisneros, Izzy and Coyle, Jackson and Eraou, Song and Gebeyehu, Martha and Hernandez, Gerardo and Juantorena, Julian and Kaplan, Lizzy and Marquez, Danielle and others},
  journal={Journal of Economic Behavior \& Organization},
  volume={228},
  pages={106781},
  year={2024},
  publisher={Elsevier}
}

@article{ahrens2020lassopack,
title={lassopack: Model selection and prediction with regularized regression in Stata},
author={Ahrens, Achim and Hansen, Christian B and Schaffer, Mark E},
journal={The Stata Journal},
volume={20},
number={1},
pages={176--235},
year={2020},
publisher={SAGE Publications Sage CA: Los Angeles, CA}
}

@article{belloni2012sparse,
  title={Sparse models and methods for optimal instruments with an application to eminent domain},
  author={Belloni, Alexandre and Chen, Daniel and Chernozhukov, Victor and Hansen, Christian},
  journal={Econometrica},
  volume={80},
  number={6},
  pages={2369--2429},
  year={2012},
  publisher={Wiley Online Library}
}

@article{belloni2014inference,
  title={Inference on treatment effects after selection among high-dimensional controls},
  author={Belloni, Alexandre and Chernozhukov, Victor and Hansen, Christian},
  journal={Review of Economic Studies},
  volume={81},
  number={2},
  pages={608--650},
  year={2014},
  publisher={Oxford University Press}
}

@article{bhattacharya2025fiduciary,
  title={Fiduciary Duty and the Market for Financial Advice},
  author={Bhattacharya, Vivek and Illanes, Gast{\'o}n and Padi, Manisha},
  journal={Econometrica},
  volume={93},
  number={4},
  pages={1449--1480},
  year={2025},
  publisher={Wiley Online Library}
}

@article{carlin2009strategic,
  title={Strategic price complexity in retail financial markets},
  author={Carlin, Bruce I},
  journal={Journal of financial Economics},
  volume={91},
  number={3},
  pages={278--287},
  year={2009},
  publisher={Elsevier}
}

@article{chang2020market,
  title={The market for conflicted advice},
  author={Chang, Briana and Szydlowski, Martin},
  journal={The Journal of Finance},
  volume={75},
  number={2},
  pages={867--903},
  year={2020},
  publisher={Wiley Online Library}
}

@article{charoenwongdoes2019,
	title={Does Regulatory Jurisdiction Affect the Quality of Investment-Adviser Regulation?},
	author={Charoenwong, Ben and Kwan, Alan and Umar, Tarik},
	journal={American Economic Review},
	year={2019},
	volume={109},
	number={10},
	pages={3681--3712},
}

@article{clifford2021property,
  title={Property rights to client relationships and financial advisor incentives},
  author={Clifford, Christopher P and Gerken, William C},
  journal={The Journal of Finance},
  volume={76},
  number={5},
  pages={2409--2445},
  year={2021},
  publisher={Wiley Online Library}
}

@techreport{correia2017reghdfe,
  title={Linear Models with High-Dimensional Fixed Effects: An Efficient and Feasible Estimator},
  author={Correia, Sergio},
  institution={Working Paper},
  url={http://scorreia.com/research/hdfe.pdf},
  year={2017}
}

@article{darby1973free,
  title={Free competition and the optimal amount of fraud},
  author={Darby, Michael R and Karni, Edi},
  journal={The Journal of law and economics},
  volume={16},
  number={1},
  pages={67--88},
  year={1973},
  publisher={The University of Chicago Law School}
}

@article{dyck2010blows,
  title={Who blows the whistle on corporate fraud?},
  author={Dyck, Alexander and Morse, Adair and Zingales, Luigi},
  journal={The journal of finance},
  volume={65},
  number={6},
  pages={2213--2253},
  year={2010},
  publisher={Wiley Online Library}
}

@article{egan2019brokers,
  title={Brokers versus retail investors: Conflicting interests and dominated products},
  author={Egan, Mark},
  journal={The Journal of Finance},
  volume={74},
  number={3},
  pages={1217--1260},
  year={2019},
  publisher={Wiley Online Library}
}

@article{egan2019market,
	title={The market for financial adviser misconduct},
	author={Egan, Mark and Matvos, Gregor and Seru, Amit},
	journal={Journal of Political Economy},
	volume={127},
	number={1},
	pages={233--295},
	year={2019},
	publisher={University of Chicago Press Chicago, IL}
}

@article{egan2022harry,
  title={When Harry fired Sally: The double standard in punishing misconduct},
  author={Egan, Mark and Matvos, Gregor and Seru, Amit},
  journal={Journal of Political Economy},
  volume={130},
  number={5},
  pages={1184--1248},
  year={2022},
  publisher={The University of Chicago Press Chicago, IL}
}

@article{gabaix2006shrouded,
  title={Shrouded attributes, consumer myopia, and information suppression in competitive markets},
  author={Gabaix, Xavier and Laibson, David},
  journal={The Quarterly Journal of Economics},
  volume={121},
  number={2},
  pages={505--540},
  year={2006},
  publisher={MIT Press}
}

@article{galai2006ostrich,
  title={The “ostrich effect” and the relationship between the liquidity and the yields of financial assets},
  author={Galai, Dan and Sade, Orly},
  journal={The Journal of Business},
  volume={79},
  number={5},
  pages={2741--2759},
  year={2006},
  publisher={JSTOR}
}

@article{gurun2018trust,
  title={Trust busting: The effect of fraud on investor behavior},
  author={Gurun, Umit G and Stoffman, Noah and Yonker, Scott E},
  journal={The Review of Financial Studies},
  volume={31},
  number={4},
  pages={1341--1376},
  year={2018},
  publisher={Oxford University Press}
}

@incollection{heckman1985using,
  title={Using longitudinal data to estimate age, period and cohort effects in earnings equations},
  author={Heckman, James and Robb, Richard},
  booktitle={Cohort analysis in social research: Beyond the identification problem},
  pages={137--150},
  year={1985},
  publisher={Springer}
}

@article{honigsberg2021deleting,
  title={Deleting misconduct: The expungement of BrokerCheck records},
  author={Honigsberg, Colleen and Jacob, Matthew},
  journal={Journal of Financial Economics},
  volume={139},
  number={3},
  pages={800--831},
  year={2021},
  publisher={Elsevier}
}

@article{honigsberg2025regulatory,
  author       = {Honigsberg, Colleen and Hu, Edwin and Jackson, Robert J., Jr.},
  title        = {Regulatory Leakage Among Financial Advisors: Evidence from {FINRA} Regulation of ``Bad'' Brokers},
  journal      = {Journal of Financial Economics},
  year         = {2025},
  volume       = {174},
  pages        = {104170},
  doi          = {10.1016/j.jfineco.2024.104170}
}

@article{inderst2009misselling,
	title={Misselling through agents},
	author={Inderst, Roman and Ottaviani, Marco},
	journal={American Economic Review},
	volume={99},
	number={3},
	pages={883--908},
	year={2009}
}

@article{inderst2012competition,
	title={Competition through commissions and kickbacks},
	author={Inderst, Roman and Ottaviani, Marco},
	journal={The American Economic Review},
	volume={102},
	number={2},
	pages={780--809},
	year={2012},
	publisher={American Economic Association}
}

@article{karlsson2009ostrich,
  title={The ostrich effect: Selective attention to information},
  author={Karlsson, Niklas and Loewenstein, George and Seppi, Duane},
  journal={Journal of Risk and uncertainty},
  volume={38},
  pages={95--115},
  year={2009},
  publisher={Springer}
}

@article{kostovetsky2016whom,
  title={Whom do you trust?: Investor-advisor relationships and mutual fund flows},
  author={Kostovetsky, Leonard},
  journal={The Review of Financial Studies},
  volume={29},
  number={4},
  pages={898--936},
  year={2016},
  publisher={Oxford University Press}
}

@article{machado2019quantiles,
  title={Quantiles via moments},
  author={Machado, Jos{\'e} AF and Silva, JMC Santos},
  journal={Journal of econometrics},
  volume={213},
  number={1},
  pages={145--173},
  year={2019},
  publisher={Elsevier}
}

@misc{riosavila2020mmqreg,
  title={{MMQREG}: Stata module to estimate quantile regressions via Method of Moments},
  author={Rios-Avila, Fernando},
  year={2020},
  howpublished={Statistical Software Components, Boston College Department of Economics},
  url={https://ideas.repec.org/c/boc/bocode/s458750.html}
}

@article{sicherman2016financial,
  title={Financial attention},
  author={Sicherman, Nachum and Loewenstein, George and Seppi, Duane J and Utkus, Stephen P},
  journal={The Review of Financial Studies},
  volume={29},
  number={4},
  pages={863--897},
  year={2016},
  publisher={Oxford University Press}
}

@article{sims2003implications,
  title={Implications of rational inattention},
  author={Sims, Christopher A},
  journal={Journal of monetary Economics},
  volume={50},
  number={3},
  pages={665--690},
  year={2003},
  publisher={Elsevier}
}

@article{zou2006adaptive,
  title={The adaptive lasso and its oracle properties},
  author={Zou, Hui},
  journal={Journal of the American statistical association},
  volume={101},
  number={476},
  pages={1418--1429},
  year={2006},
  publisher={Taylor \& Francis}
}
%----------------------------------------------------------------

%----------------------------------------------------------------
% End of Document:
%----------------------------------------------------------------
\end{document}